\documentclass[reprint,amsmath,amssymb,aps,prb,superscriptaddress]{revtex4-1}

\pdfoutput=1

\usepackage{hyperref,url}
\usepackage{amsmath}
\usepackage{amsfonts}
\usepackage{mathtools}
\usepackage[capitalise]{cleveref}
\usepackage{placeins}
\usepackage{mhchem}
\hypersetup{breaklinks,colorlinks=true,linkcolor=blue,citecolor=blue,urlcolor=blue,filecolor=blue}
\usepackage{graphicx}
\usepackage{dcolumn}

\newcommand\mat\mathbf
\newcommand\tr{\operatorname{tr}}

\begin{document}

\author{Joonho Lee}
\email{jl5653@columbia.edu}
\affiliation{Department of Chemistry, Columbia University, New York, NY 10027, USA.}
\author{Fionn D. Malone}
\email{fionn.malone@gmail.com}
\affiliation{Quantum Simulations Group, Lawrence Livermore National Laboratory, 7000 East Avenue, Livermore, CA 94551 USA.}
\author{Miguel A. Morales}
\email{moralessilva2@llnl.gov}
\affiliation{Quantum Simulations Group, Lawrence Livermore National Laboratory, 7000 East Avenue, Livermore, CA 94551 USA.}
\author{David R. Reichman}
\email{drr2103@columbia.edu}
\affiliation{Department of Chemistry, Columbia University, New York, NY 10027, USA.}

\title{Spectral Functions from Auxiliary-Field Quantum Monte Carlo without Analytic Continuation: The Extended Koopmans' Theorem Approach}
%\title{EKT-AFQMC}
\begin{abstract}
We explore the extended Koopmans' theorem (EKT)
within the phaseless auxiliary-field quantum Monte Carlo (AFQMC) method.
The EKT allows for the direct calculation of electron addition and removal spectral functions using reduced density matrices of the $N$-particle system, and avoids the need for analytic continuation.
The lowest level of EKT with AFQMC, called EKT1-AFQMC, is benchmarked using small molecules, 14-electron and 54-electron uniform electron gas supercells, and diamond at the $\Gamma$-point. Via comparison with numerically exact results (when possible) and coupled-cluster methods, 
we find that EKT1-AFQMC can reproduce the qualitative features of spectral functions for Koopmans-like charge excitations
with errors in peak locations of less than 0.25 eV in a finite basis. 
We also note the numerical difficulties that arise in the EKT1-AFQMC eigenvalue problem, especially when
back-propagated quantities are very noisy.  We show how a systematic higher order EKT approach can correct errors in EKT1-based theories with respect to the satellite region of the spectral function.
Our work will be of use for the  study of low-energy charge excitations and spectral functions in correlated molecules and solids where AFQMC can be reliably performed.
\end{abstract}
\maketitle
\section{Introduction}
%We are often interested in the ground state of an {\it ab-initio} Hamiltonian,
%\begin{equation}
%\hat{\mathcal H} = 
%\sum_{pq} h_{pq} \hat{a}_p^\dagger\hat{a}_q + \frac12 \sum_{pqrs} \langle pq | rs \rangle \hat{a}_p^\dagger\hat{a}_q^\dagger\hat{a}_s\hat{a}_r
%\end{equation}
%Within a projector Quantum Monte Carlo (QMC) approach such as auxiliary-field QMC (AFQMC), 
%the infamous fermionic minus sign problem can be removed by imposing a constraint along the imaginary-time propagation of each walker.
The dynamical response to external perturbation is
one of the most powerful means of experimentally probing molecules and materials.
Examples include angle-resolved photoemission spectroscopy,\cite{lu2012angle} electron energy loss spectroscopy,\cite{egerton2008electron} and inelastic neutron scattering,\cite{andreani2005measurement} each of which encodes the excitation spectrum of a many-body system.
The theoretical description of such experiments can be modeled (in the linear response regime) by considering many-body Green's functions.\cite{alexanderfetter2003,onida2002electronic}
For example, differential cross sections from direct and inverse photoemission experiments can be related to the (retarded) single-particle Green's function.\cite{cederbaum1977theoretical}
In a general sense, these observables are connected to spectral functions describing electron removal and addition via the single-particle Green's function.\cite{cederbaum1977theoretical,hedin1998transition}

Given the above facts, the theoretical description of dynamical response properties have been dominated by 
Green's function-based approaches
mainly due to the direct access
to the spectral function that they afford.\cite{onida2002electronic,duchemin2020robust}
Among Green's function methods, the G$_0$W$_0$ approach\cite{hybertsen1986electron} is perhaps the most widely used approximation to Hedin's equations for the description of quasi-particle spectra in solids.\cite{hedin1965new}
This approach has provided numerous valuable insights for the interpretation of experimental data.\cite{onida2002electronic,reining2018gw}
Despite these successes, G$_0$W$_0$ has several deficiencies,
such as a notable dependence
on the input Green's function G$_0$ and the screened Coulomb operator W$_0$,\cite{rinke2005combining, fuchs2007quasiparticle,marom2012benchmark,bruneval2012ionization}
a poor description of satellite region of the spectral function,\cite{langreth1970singularities,aryasetiawan1996multiple,guzzo2011valence,lischner2013physical} and the absence of certain conserving properties.\cite{stan2009levels}
Attempts have been made to address these deficiencies
via (partially) self-consistent GW approaches,\cite{von1996self,holm1998fully,faleev2004all,van2006quasiparticle,shishkin2007self,rostgaard2010fully,koval2014fully} as well as with the incorporation of vertex corrections\cite{bobbert1994lowest,schindlmayr1998systematic,romaniello2009self,chen2015accurate,maggio2017gw} including via the use of cumulant-based approaches.\cite{holm1997self,kas2014cumulant,lischner2014satellite,caruso2016gw,mayers2016descr}

There has also been a sizable effort to construct
spectral functions based on wavefunction methods.
These approaches include the use of
matrix product states (MPS),\cite{Jeckelmann2002Jul,Schollwock2006Feb}
algebraic diagrammatic constructions,\cite{Schirmer1982Nov,Schirmer1983Sep,Tarantelli1989Feb,Wenzel2014Oct,Dreuw2015Jan,Sokolov2018Nov}
and
coupled-cluster Green's function (or equation-of-motion coupled-cluster) methods.\cite{Monkhorst1977Jan,Nooijen1992Mar,Nooijen1993Oct,Peng2016Dec,McClain2016Jun,McClain2017Mar,Furukawa2018May,Peng2018Mar,Shee2019Nov}
These and related approaches have
distinct strengths and weaknesses
in terms of both cost and accuracy and continue to be actively pursued.

Another useful path to the description of spectral information is based on projector quantum Monte Carlo (PQMC) approaches.\cite{Blankenbecler1983Mar,Becca2017Nov}
PQMC methods provide a highly accurate means to simulate
the ground state properties of correlated solids.\cite{foulkes_rmp}
Unlike the aforementioned wavefunction-based approaches, PQMC methods do not provide
direct access to real-time and real-frequency Green's functions.
This is a direct consequence of 
the imaginary-time propagation at the heart of all PQMC approaches.
A popular way around this hurdle is to first obtain
the imaginary-time Green's function 
and then
% <<<<<<< HEAD
% analytically continue this to obtain a real-time Green's function.\cite{Silver1990Feb,Gubernatis1991Sep,Jarrell1996May,Motta2014Jan,Motta2015Oct,Otsuki2017Jun,Bertaina2017Mar}
% Imaginary-time Green's functions from PQMC approaches
% have been shown to be accurate\cite{Silver1990Feb,Gubernatis1991Sep,Jarrell1996May,Motta2014Jan,Motta2015Oct,Otsuki2017Jun,Bertaina2017Mar}
% so a general hope is to preserve this accuracy as much as possible during the analytic continuation process.
% Unfortunately, analytic continuation is numerically ill-conditioned and
% exhibits difficulties in resolving
% sharp features of the true spectral functions
% even if the exact imaginary-time Green's function is used as an input.[]
% =======
perform analytical continuation to obtain the real-frequency Green's function.\cite{SilverMaxEnt1990,GubernatisMaxEnt1991,Jarrell1996May,Motta2014Jan,Motta2015Oct,Otsuki2017Jun,Bertaina2017Mar}
%Imaginary-time Green's functions from PQMC approaches
%have been shown to be accurate []
%so a general hope is to preserve this accuracy as much as possible during the analytic continuation process.
Unfortunately, analytic continuation is numerically ill-conditioned, and the methods to perform analytic continuation such as the maximum entropy method\cite{SilverMaxEnt1990,GubernatisMaxEnt1991}
can exhibit difficulties in resolving
sharp features in the real-frequency spectral function
even if high quality imaginary-time Green's functions are used as input.\cite{Reichman2009Aug,Goulko2017Jan,DornheimDynamicalStructure2018}
% >>>>>>> 62a69a9ec7fb27348ebf137aa6528612d01a2189
Therefore, it is highly desirable to develop an alternative means to obtain spectral functions which can
work with PQMC methods without sacrificing its ground state accuracy.
There have been approaches based on diffusion Monte Carlo (DMC) \cite{Ceperley1988Nov} and
the Krylov-projected
full-configuration QMC (KP-FCIQMC)\cite{Blunt2015Jul,Blunt2018Aug} where one samples a low-energy Hamiltonian matrix and solves an eigenvalue problem to obtain a low-energy spectrum.
%solve an eigenvalue problem within the low-energy manifold.
Similar to its ground state counterpart, the
excited states from a DMC-based approach would be biased due to the fixed node error.\cite{Ceperley1988Nov}
On the other hand, KP-FCIQMC is numerically exact but scales exponentially in system size analogously to its ground state counterpart, FCIQMC.\cite{Booth2009Aug}
Therefore, the scope of KP-FCIQMC has been limited to small systems.\cite{Blunt2015Jul,Blunt2018Aug}
Furthermore, because one is trying to obtain excited state information from imaginary-time propagation in these approaches, the higher-lying excited states are exponentially harder to obtain. This makes it challenging for these approaches to estimate high energy spectral information.
%Lastly, we note
%that
%there are quantum Monte Carlo (QMC) approaches
%which directly work in real-time such as those based on path-integral approaches for open quantum systems [].

The approach that we will examine in this work
is called the extended Koopmans' theorem (EKT).\cite{Day1975,Smith1975,Pickup1975,Morrison1975Jan,Ellenbogen1977Jun,Morrison1992,Morrison1992a,Sundholm1993,Cioslowski1997,Kent1998,Olsen1998May,Pernal2001Mar,Farnum2004,Pernal2005Aug,Ernzerhof2009,Vanfleteren2009,Piris2012May,Piris2013Jan,Bozkaya2013,Welden2015,Bozkaya2018,Pavlyukh2018,Pavlyukh2019}
The EKT generalizes the Koopmans' theorem in Hartree-Fock (HF) theory for arbitrary many-body wavefunctions.
Its working ingredients are reduced density matrices (RDMs) for an $N$-particle system
and it produces approximate
($N$-1)-particle and ($N$+1)-particle {\it wavefunctions} even without 
the $N$-particle ground state wavefunction.
Due to this desirable feature, EKT methods have been widely used as a means to obtain spectral information
for approaches for which one has access neither to real-time Green's functions nor wavefunctions.
Examples include direct RDM-based methods,\cite{Farnum2004} density matrix functional theory,\cite{Pernal2005Aug} natural orbital functional methods,\cite{Piris2012May,Piris2013Jan} and second-order Green's function methods.\cite{Welden2015}
EKT has also been explored with wavefunction methods such as configuration interaction methods,\cite{Morrison1992a} M{\o}ller-Plesset perturbation methods,\cite{Cioslowski1997,Bozkaya2013} and coupled-cluster methods.\cite{Bozkaya2018}
It is also a promising way for any QMC method to compute excited state and spectral information if the necessary RDMs can be constructed.
%It is not the first time for one to apply the EKT approach to QMC approaches.
The EKT has also been used to obtain the quasi-particle band structure of silicon,\citep{Kent1998} ionization potentials and electron affinities of atoms,\citep{Zheng2016Jul} and the Fermi velocity of graphene using VMC.\citep{Zheng2016Jul} Lastly, the EKT has been combined with DMC to study similar systems.\citep{Zheng2016Jul}

A PQMC approach that can be readily combined with the EKT is the phaseless auxiliary-field QMC (ph-AFQMC) method.\cite{Zhang1995May,Zhang1997Mar,zhang2003quantum}
ph-AFQMC
has emerged as a flexible, accurate and scalable 
many-body method.
It imposes an approximate gauge boundary condition (i.e., the phaseless constraint) on the imaginary-time evolution of Slater determinant walkers, completely removing the Fermionic phase problem.\cite{zhang2003quantum}
While the resulting energy is neither exact nor a variational upper bound to the exact ground state energy,\cite{carlson1999issues}
many benchmark studies have
demonstrated the accuracy of ph-AFQMC and its related variants.\cite{leblanc2015solutions,zheng2017stripe,motta2017towards,zhang_nio,motta2019ground,lee_2019_UEG,Lee2020benzene,williams2020direct,malone2020gpu,lee2020utilizing,qin2020absence,Malone2020Oct}
Furthermore, with recent advances in local energy evaluation techniques in ph-AFQMC,\cite{malone_isdf,lee2020stochastic}
the cost for obtaining each statistical sample scales cubically with system size,
which renders it less expensive than many other many-body methods.
With the advent of the back-propagation (BP) method in ph-AFQMC,\cite{Zhang1997Mar,Purwanto2004Nov,motta_back_prop}
with some additional effort one can compute
pure estimators for any operator, including those that do not commute with the Hamiltonian.
Therefore, one can compute the relevant input for the EKT directly from ph-AFQMC using the BP algorithm. This is the direction we persue in this work.
%We also use a brute-force selected configuration interaction approach called heat-bath CI (HCI) []
%to obtain ``exact'' EKT data for benchmark purposes.

This paper is organized as follows. We first 
present the general framework of the EKT, its most common form EKT-1, and its extension, EKT-3.
We then discuss how to obtain the relevant input for EKT-1 using BP and ph-AFQMC.
We assess the accuracy of EKT-1-AFQMC on a variety of small molecules and the uniform electron gas model.
We further show the qualitative failure of EKT-1 for the core spectra of the 14-electron uniform electron gas model and
illustrate the drastic improvement upon this result from using EKT-3 on the same model. 
We also apply EKT1-AFQMC to the 54-electron uniform electron gas model and diamond at the $\Gamma$-point.
We conclude and summarize our most important findings.

\section{Theory}
\subsection{The Extended Koopmans' Theorem}
In order to compute quasi-particle gaps and spectral functions, 
one must compute ionization potential and electron attachment energies along with the associated wavefunctions (or at least squared amplitudes for spectral weights).
While we focus in this work on electron removal processes, we keep our presentation of theory general so that it is also applicable to electron addition processes.

In the EKT approach, we consider wavefunctions
\begin{equation}
|\Psi_\nu^{N\pm1}\rangle  = \hat{O}_\nu^{\pm} | \Psi_0^N\rangle ,
\label{eq:psi_np}
\end{equation}
where the electron addition operator $\hat{O}_\nu^{+}$ is
\begin{equation}
\hat{O}_\nu^{+} = 
\sum_{p} (c_+)_{p}^\nu \hat{a}_{p} ^\dagger 
\label{eq:ea}
\end{equation}
for 1-particle excitations,
and the electron removal operator $\hat{O}_\nu^{-}$ is
\begin{equation}
\hat{O}_\nu^{-} = 
\sum_{p} (c_-)_{p}^\nu \hat{a}_{p}  
%\sum_{pqr} (c_-)_{pqr}^\nu \hat{a}_{r}^\dagger \hat{a}_{q} \hat{a}_{p} + \cdots
\label{eq:ip}
\end{equation}
for 1-hole excitations.
We obtain the linear coefficients $\mathbf c_\pm$ by minimizing the following variational energy expression:
\begin{equation}
\Delta E_\nu^{\pm} = E_\nu^{(N\pm1)} - E_0^{(N)} = 
\frac{\langle \Psi_0^N | ({\hat{O}_\nu^{\pm}})^\dagger [\hat{\mathcal H}, \hat{O}_\nu^\pm] |\Psi_0^N\rangle}
{\langle \Psi_0^N | ({\hat{O}_\nu^{\pm}})^\dagger \hat{O}_\nu^\pm | \Psi_0^N \rangle} ,
\label{eq:def_npm1}
\end{equation}
where we have defined 
\begin{equation}
E_\nu^{(N\pm1)} = 
\frac{\langle \Psi_0^N | ({\hat{O}_\nu^{\pm}})^\dagger \hat{\mathcal H} \hat{O}_\nu^\pm |\Psi_0^N\rangle}
{\langle \Psi_0^N | ({\hat{O}_\nu^{\pm}})^\dagger \hat{O}_\nu^\pm | \Psi_0^N \rangle} ,
\label{eq:enpm}
\end{equation}
and assumed 
\begin{equation}
\hat{\mathcal H}| \Psi_0^N \rangle = E_0^{(N)} | \Psi_0^N \rangle .
\end{equation}
We refer this approach to as EKT-1.
The excitation levels in \cref{eq:ip} and \cref{eq:ea} can be systematically increased to achieve a greater accuracy in principle
at the expense of greater computational costs.\cite{Farnum2004,Pavlyukh2018}
The next level of theory would incorporate 2h1p and 2p1h excitations instead of \cref{eq:ip} and \cref{eq:ea}, respectively:
\begin{equation}
\hat{O}_\nu^{+} = 
 \sum_{pqr} (c_+)_{pqr}^\nu \hat{a}_{r}\hat{a}_{q} ^\dagger\hat{a}_{p}^\dagger
 \label{eq:ea2}
 \end{equation}
 and
\begin{equation}
\hat{O}_\nu^{-} = 
\sum_{pqr} (c_-)_{pqr}^\nu \hat{a}_{r}^\dagger \hat{a}_{q} \hat{a}_{p}
\label{eq:ip2}
 \end{equation} 
These operators include EKT-1 excitations because when $r=q$ we recover the 1h and 1p excitations, as in \cref{eq:ip} and \cref{eq:ea}.
We refer this higher level of theory to as EKT-3.
%For the purpose of this paper, we focus on the simplest EKT theory which involves only 1h and 1p excitations.

%We provide more implementation details of EKT-1 and EKT-3.

\subsubsection{EKT-1}
We consider the following Lagrangian for 1h and 1p excitations
\begin{align}
\mathcal L[\mathbf c^\nu] = 
\langle \Psi_0^N | ({\hat{O}_\nu^{\pm}})^\dagger [\hat{\mathcal H}, \hat{O}_\nu^\pm] |\Psi_0^N\rangle
-\epsilon^\nu_\pm((\mathbf c^\nu)^\dagger \mathbf S_{\pm} \mathbf c^\nu - 1) ,
\label{eq:lag}
\end{align}
where 
$\mathbf{S}_\pm$ is
a pertinent metric matrix for normalization
and $\epsilon^\nu_\pm$ is a Lagrange multiplier.
We note that
\begin{equation}
\epsilon_\pm^\nu = \pm \Delta E_\nu^{\pm} .
\end{equation}
The normalization of $|\Psi_\nu^{N\pm1}\rangle$ is ensured by the constraint in \cref{eq:lag}.
The stationary condition of \cref{eq:lag} with respect to $(\mathbf c_\pm^\nu)^\dagger$ then leads to a
generalized eigenvalue equation, 
\begin{equation}
\mathbf F_{\pm} \mathbf c_\pm^\nu = \epsilon_\pm^\nu \mathbf S_{\pm} \mathbf c_\pm^\nu\label{eq:gen_eig}
\end{equation}
where the generalized Fock matrix is defined as (assuming that $|\Psi_0^N\rangle$ is normalized)
\begin{equation}
(\mathbf F_{-})_{pq} = 
\langle \Psi_0^N |  
\hat{a}_p^\dagger
[\hat{\mathcal H},\hat{a}_q]
| \Psi_0^N \rangle ,
\label{eq:fm}
\end{equation}
and
\begin{equation}
(\mathbf F_{+})_{pq} = 
\langle \Psi_0^N |  
\hat{a}_p
[\hat{\mathcal H},\hat{a}_q^\dagger]
| \Psi_0^N \rangle ,
\label{eq:fp}
\end{equation}
and the corresponding metric matrix $\mathbf S_\pm$ is
\begin{equation}
\mathbf S_- = \mathbf P ,
\end{equation}
and
\begin{equation}
\mathbf S_+ = \mathbf I - \mathbf P^T .
\end{equation}
Here, $\mathbf P$ is the one-body reduced density matrix (1-RDM),
\begin{equation}
P_{pq} = \langle \Psi_0^N | \hat{a}_p^\dagger \hat{a}_q | \Psi_0^N \rangle .
\end{equation}
The electron attachment and ionization potential simply follow
$\epsilon_+=-\text{EA}$ and $\epsilon_-=\text{IP}$ (assuming $\nu$ corresponds to the lowest energy state).
Then the quasiparticle gap is given as $\Delta E_\text{qp} = \epsilon_+ + \epsilon_-$.
We note that these Fock matrices are not Hermitian unless $|\Psi_0^N\rangle$ is an exact eigenstate of $\hat{\mathcal H}$.

To provide more detailed expressions, let us define a generic {\it ab-initio} Hamiltonian,
\begin{equation}
\hat{\mathcal H}
=
\hat{\mathcal H}_1
+
\hat{\mathcal H}_2 ,
\label{eq:ham}
\end{equation}
with
\begin{align}
\hat{\mathcal H}_1
&=
\sum_{pq}
h_{pq}a_p^\dagger \hat{a}_q ,\\
\hat{\mathcal H}_2
&=
\frac12
\sum_{pqrs}
\langle pq | rs \rangle
\hat{a}_p^\dagger
\hat{a}_q^\dagger
\hat{a}_s
\hat{a}_r ,
\label{eq:h2chem}
\end{align}
where $h_{pq}$ is the one-body Hamiltonian matrix element and $\langle pq | rs \rangle$ is the two-electron integral tensor in Dirac notation.
Substituting \cref{eq:ham} into \cref{eq:fm} and \cref{eq:fp}, it can be shown that $\mathbf F_{\pm}$ can be evaluated with the 1-RDM and the two-body RDM (2-RDM):
\begin{equation}
(\mathbf F_-)_{pq} = - \sum_q h_{qr}(\mathbf S_-)_{pr} 
+ \frac12 \sum_{trs} \langle tq || rs\rangle \Gamma_{pt}^{rs} ,
\label{eq:fmcomp}
\end{equation}
and
\begin{align}\nonumber
(\mathbf F_+)_{pq} =& \sum_q h_{qr}(\mathbf S_+)_{pr} 
+ \frac12 \sum_{trs} \langle rt || qs\rangle \Gamma_{rt}^{sp}\\
&+ \sum_{rs} (\mathbf S_-)_{rs} \langle pr||qs\rangle ,
\end{align}
where the 2-RDM $\Gamma$ is 
\begin{equation}
\Gamma_{pt}^{rs} = 
\langle \Psi_0^N |  
\hat{a}_p^\dagger
\hat{a}_t^\dagger
\hat{a}_s
\hat{a}_r
| \Psi_0^N \rangle ,
\label{eq:2-RDM}
\end{equation}
and the antisymmetrized two-electron integral tensor is defined as
\begin{equation}
\langle pq || rs \rangle
=
\langle pq | rs \rangle
-
\langle pq | sr \rangle .
\end{equation}
%In passing we note that including higher excitations in \cref{eq:ip} and \cref{eq:ea} requires higher-order density matrices.
%Increasing the excitation level by one naturally requires 1-order-higher density matrices, which will then increase the overall cost and storage.

\subsubsection{EKT-3}
For many solid state systems, including 1h or 1p excitations only is not sufficient as the restriction to such excitations would not be capable of describing satellite peaks.\cite{hedin1980effects}
A straightforward way to obtain satellite peaks in addition to dominant quasiparticle peaks is to include higher-order excitations in the EKT ansatz.
Thus, EKT-3 is the next level in this hierarchy that can be attempted.
Although EKT-3 has been mentioned in literature\cite{Farnum2004,Pavlyukh2018,Pavlyukh2019} and approximately implemented (neglecting the opposite spin term) before,\cite{Farnum2004}
to the best of our knowledge this work presents the
first complete implementation of EKT-3 along with numerical results.

The corresponding generalized Fock operator for the IP problem reads
\begin{align}
(\mathbf{F}_-)_{pqr,stu} &= \langle \Psi_0^N | \hat{a}^\dagger_p\hat{a}^\dagger_q\hat{a}_r [\hat{H}, \hat{a}_u^\dagger \hat{a}_t\hat{a}_s] |\Psi_0^N\rangle .
\end{align}
Using the SecondQuantizationAlgebra package\cite{sqa} developed by Neuscamman and others,\cite{Neuscamman2009Mar,Saitow2013Jul}
we derived a complete spin-orbital equation of the generalized Fock operator:
\begin{align}\nonumber
(\mathbf{F}_-)_{ijk,lmn}
&=
-h_{kn}\Gamma_{ij}^{ml}
-\sum_a (h_{la}\Gamma_{ij}^{am} \delta_{kn}
+h_{ma}\delta_{kn} \Gamma_{ij}^{al}
\\ \nonumber
&
+h_{la}\Gamma_{ijn}^{amk}
-h_{ma}\Gamma_{ijn}^{alk} 
+h_{na} \Gamma_{ija}^{mlk})\\ \nonumber
&
+ \frac12
\sum_{ab}(
\langle lm || ab \rangle
\delta_{kn}
\Gamma_{ij}^{ba}
-
\langle kl || ab \rangle
\Gamma_{ijn}^{bam}\\\nonumber
&+
\langle km || ab \rangle
\Gamma_{ijn}^{bal}
- 2
\langle ka || nb \rangle
\Gamma_{ija}^{bml}
-
\langle lm || ab \rangle
\Gamma_{ijn}^{bak})\\ \nonumber
&+\frac12\sum_{abc}(
% (   0.50000) h2pqrs(m,a,b,c) kdelta(k,n) cre(i) cre(j) cre(a) des(l) des(b) des(c) 
\langle ma || bc \rangle \delta_{kn} \Gamma_{ija}^{cbl}
% (  -0.50000) h2pqrs(l,a,b,c) kdelta(k,n) cre(i) cre(j) cre(a) des(m) des(b) des(c) 
-\langle la || bc \rangle \delta_{kn} \Gamma_{ija}^{cbm}\\ \nonumber
% (  -0.50000) h2pqrs(l,a,b,c) cre(i) cre(j) cre(n) cre(a) des(k) des(m) des(b) des(c) 
&-\langle la || bc \rangle \Gamma_{ijna}^{cbmk}
% (   0.50000) h2pqrs(m,a,b,c) cre(i) cre(j) cre(n) cre(a) des(k) des(l) des(b) des(c) 
+ \langle ma || bc \rangle \Gamma_{ijna}^{cblk}\\ 
% (  -0.50000) h2pqrs(n,a,b,c) cre(i) cre(j) cre(b) cre(c) des(k) des(l) des(m) des(a) 
&- \langle na || bc \rangle \Gamma_{ijbc}^{amlk}) ,
\label{eq:fmekt3}
\end{align}
where the three-body RDM (3-RDM) is
\begin{equation}
\Gamma_{ijk}^{npq} = 
\langle \Psi_0^N |  
\hat{a}_i^\dagger
\hat{a}_j^\dagger
\hat{a}_k^\dagger
\hat{a}_q
\hat{a}_p
\hat{a}_n
| \Psi_0^N \rangle ,
\end{equation}
 and the four-body RDM (4-RDM) is
\begin{equation}
\Gamma_{ijkl}^{mnpq} = 
\langle \Psi_0^N |  
\hat{a}_i^\dagger
\hat{a}_j^\dagger
\hat{a}_k^\dagger
\hat{a}_l^\dagger
\hat{a}_q
\hat{a}_p
\hat{a}_n
\hat{a}_m
| \Psi_0^N \rangle .
\end{equation}
The pertinent metric, $\mathbf S$, for this generalized eigenvalue problem is
\begin{align} \nonumber
S_{pqr,stu} &= \langle \Psi_0^N | \hat{a}^\dagger_p\hat{a}^\dagger_q\hat{a}_r  \hat{a}_u^\dagger \hat{a}_t\hat{a}_s |\Psi_0^N\rangle \\ 
&=
\delta_{ur}\Gamma_{pq}^{st}
-\Gamma_{pqu}^{str} .
\end{align}
The storage requirement of the 4-RDM scales as $\mathcal O(N^8)$ and
it becomes prohibitively expensive for more than 16 orbitals.
To circumvent this problem, we approximate the 4-RDM via a cumulant expansion.
The cumulant approximation to the 4-RDM has been used in multi-reference perturbation theory and configuration interaction methods previously.\cite{Zgid2009May,Saitow2013Jul}
In essence, the 4-RDM is approximately constructed from
four classes of terms:
(1) 1-RDM$\times$1-RDM$\times$1-RDM$\times$1-RDM,
(2) 2-RDM$\times$1-RDM$\times$1-RDM
(3) 2-RDM$\times$2-RDM,
and 
(4) 1-RDM$\times$3-RDM.
Interested readers are referred to Ref. \citenum{Zgid2009May} for more details.
To construct the cumulant terms we wrote a Python code based on the Fortran code presented in Ref. \citenum{Saitow2013Jul}.
For the systems we have investigated, we have found the error of the
cumulant approximation is insignificant
and we present results with the reconstructed cumulant 4-RDM later in this work.
We further note that Mazziotti and co-workers have used a cumulant expansion for both 3- and 4-RDMs in their EKT-3 calculations.\cite{Farnum2004}
%\begin{equation}
%\end{equation}

Practical implementations may be achieved using spin-orbital expressions where
we consider two spin-blocks in $\mathbf{c}_-$ ($(\alpha\alpha\alpha)$ and $(\alpha\beta\beta)$) for removing an $\alpha$ electron:
\begin{align}\nonumber
\hat{O}_\nu^{-} &= 
\sum_{p_\alpha q_\alpha r_\alpha} (c_-)_{p_\alpha q_\alpha r_\alpha}^\nu \hat{a}_{r_\alpha}^\dagger \hat{a}_{q_\alpha} \hat{a}_{p_\alpha}\\
&+\sum_{p_\alpha q_\beta r_\beta} (c_-)_{p_\alpha q_\beta r_\beta}^\nu \hat{a}_{r_\beta}^\dagger \hat{a}_{q_\beta} \hat{a}_{p_\alpha} .
 \end{align} 
 Consequently, this leads to four distinct spin-blocks for $\mathbf F$: $(\alpha\alpha\alpha\alpha\alpha\alpha)$, $(\alpha\beta\beta\alpha\alpha\alpha)$, 
 $(\alpha\alpha\alpha\alpha\beta\beta)$, and  $(\alpha\beta\beta\alpha\beta\beta)$.

%The steep scaling due to the terms involving 4-RDMs limits the applicability of this method to small systems.
%Therefore, we further employ a ``diagonal'' approximation for $\mathbf{F}_-$,
%\begin{equation}
%(\mathbf{F}_-)_{ijk,lmn} = \delta_{ijk,lmn} (\mathbf{F}_-)_{ijk,ijk}
%\end{equation}
%The diagonal element is 

%\subsection{Spin-free formulation of EKT}
%For spin-restricted orbital, it is often computationally advantageous to work with spin-free operators.
%The unitary group generator $\hat{E}_{pq}$ is defined as
%\begin{equation}
%\hat{E}_{pq} = a_{p_\alpha}^\dagger a_{q_\alpha} + a_{p_\beta}^\dagger a_{q_\beta}
%\end{equation}
%We use this operator to reduce the complexity of EKT-2h1p.

%Then the spin-free {\it ab-initio} Hamiltonian is given as
%\begin{align}
%\hat{H}
%&=
%\sum_{pq} h_{pq}\hat{E}_{pq}
%+\frac12
%\sum_{pqrs} \langle pq|rs\rangle(\hat{E}_{pr}\hat{E}_{qs} - \delta_{qr} \hat{E}_{ps}) \\
%&=
%\sum_{pq} t_{pq}\hat{E}_{pq}
%+\frac12
%\sum_{pqrs} \langle pq|rs\rangle\hat{E}_{pr}\hat{E}_{qs}
%\end{align}
%where
%\begin{equation}
%t_{pq} = h_{pq} - \frac12\sum_r \langle pr | r q \rangle %\hat{E}_{pq}
%\end{equation}
%
%\begin{equation}
%(\mathbf F_{-})_{pq} = 
%\langle \Psi_0^N |  
%\hat{a}_p^\dagger
%[\hat{\mathcal H},\hat{a}_q]
%| \Psi_0^N \rangle
%\end{equation}
%\begin{equation}
%(\mathbf F_-)_{pq} = - \sum_q h_{qr}(\tilde{\gamma}_-)_{pr} 
%- \sum_{trs} (ts|qr) \tilde{\Gamma}_{ptrs}
%\end{equation}
%where $\tilde{\gamma}_-$ and $\tilde{\Gamma}$ are spin-summed 1-RDM and 2-RDM, respectively.

\subsection{Spectral Functions from EKT}
We write the retarded single-particle Green's function in a finite basis as\cite{alexanderfetter2003}
\begin{equation}
    iG^R_{pq}(t,t') = \theta(t-t') \langle \Psi_0^N|\{\hat{a}_p(t),\hat{a}^{\dagger}_q(t')\}|\Psi_0^N\rangle,
\end{equation}
where $\theta(t)$ is the Heaviside step function. %, $\hat{\psi}(\bx,t)$ is an electron field operator, and $\bx=(\br,s)$ is a compound spatial and spin index.
Assuming the $\hat{\mathcal{H}}$ is time independent, we can write the Green's function in the frequency domain as
%the Lehman representation of 
\begin{align}\nonumber
    G^R_{pq}(\omega+i\eta) &=
	\langle \Psi^{N}_0|\hat{a}_p
\frac{1}
{
	\omega-(\hat{\mathcal{H}}-E_0^{(N)})+i\eta
}
\hat{a}_q^{\dagger}|\Psi_0^N\rangle\\
&+
\langle \Psi^{N}_0|
\hat{a}_q^{\dagger}
\frac{1}
{
	\omega-(\hat{\mathcal{H}}-E_0^{(N)})+i\eta
}
\hat{a}_p|\Psi_0^N\rangle ,
%&\sum_\nu
%\frac{
%	\langle \Psi^{N}_0|\hat{a}_p|\Psi_\nu^{N+1}\rangle
%	\langle \Psi^{N+1}_\nu|\hat{a}_q^{\dagger}|\Psi_0^N\rangle
%}
%{
%	\omega-(E_\nu^{N+1}-E_0^N)+i\eta
%}\\
%& + \sum_\nu
%\frac{
%	\langle \Psi^{N}_0|\hat{a}_q^{\dagger}|\Psi_\nu^{N-1}\rangle
%	\langle \Psi^{N-1}_\nu|\hat{a}_p|\Psi_0^N\rangle
%}
%{
%	\omega+(E_\nu^{N-1}-E_0^{N})+i\eta
%},
\label{eq:GR}
\end{align}
where $\eta$ is a small positive constant and $E_0^N$ is the ground-state energy of $N$-particle system.

The EKT approach offers a systematically improvable way to approximate the evaluation of \cref{eq:GR}.
This is because one can form projection operators on the subspace of EKT excitations, 
\begin{equation}
\hat{P}_\pm = \sum_{\mu\nu}|\Psi_\mu^{N\pm1}\rangle(S_{\mu\nu}^{\pm})^{-1}\langle \Psi_\nu^{N\pm1} | 
\label{eq:proj}
\end{equation}
where
$|\Psi_\nu^{N\pm1}\rangle$ are approximate wavefunctions obtained via the EKT as defined in \cref{eq:psi_np}
and $\mathbf S^{\pm}$ is a metric in the pertinent space.
Using \cref{eq:proj}, we obtain an approximate $\mathbf G^R$,
\begin{align}\nonumber
    G^R_{pq}(\omega+i\eta) &\simeq
	\langle \Psi^{N}_0|\hat{a}_p
\hat{P}_+
\frac{1}
{
	\omega-(\hat{\mathcal{H}}-E_0^{(N)})+i\eta
}
\hat{P}_+
\hat{a}_q^{\dagger}|\Psi_0^N\rangle\\
&+
\langle \Psi^{N}_0|
\hat{a}_q^{\dagger}
\hat{P}_-
\frac{1}
{
	\omega-(\hat{\mathcal{H}}-E_0^{(N)})+i\eta
}
\hat{P}_-
\hat{a}_p|\Psi_0^N\rangle
\label{eq:GR2}
\end{align}
This approximation can be systematically improved as higher-order excitations are included in \cref{eq:ip} and \cref{eq:ea}.
It is exact when $|\Psi_\nu^{N\pm1}\rangle$ spans the entire $({N\pm1})$-particle Hilbert space.
Substituting \cref{eq:proj} into \cref{eq:GR2} and using $\hat{P}_\pm\hat{\mathcal H}\hat{P}_\pm|\Psi_\nu^{N\pm1}\rangle = E_\nu^{N\pm1}|\Psi_\nu^{N\pm1}\rangle$
(from \cref{eq:enpm}),
we obtain the (approximate) Lehmann representation of the Green's function
\begin{align}\nonumber
G^R_{pq}(\omega)& = \sum_\nu
\frac{
	\langle \Psi^{N}_0|\hat{a}_p|\tilde{\Psi}_\nu^{N+1}\rangle
	\langle \tilde{\Psi}^{N+1}_\nu|\hat{a}_q^{\dagger}|\Psi_0^N\rangle
}
{
	\omega-(E_\nu^{(N+1)}-E_0^{(N)})+i\eta
}\\
& + \sum_\nu
\frac{
	\langle \Psi^{N}_0|\hat{a}_q^{\dagger}|\tilde{\Psi}_\nu^{N-1}\rangle
	\langle \tilde{\Psi}^{N-1}_\nu|\hat{a}_p|\Psi_0^N\rangle
}
{
	\omega+(E_\nu^{(N-1)}-E_0^{{(N)}})+i\eta
} ,
\label{eq:GR3}
\end{align}
where we orthogonalized the eigenvectors by
\begin{equation}
|\tilde{\Psi}^{N\pm1}_\nu\rangle
=
\sum_\mu |{\Psi}^{N\pm1}_\mu\rangle
( S^{\pm})^{-1/2}_{\mu\nu}.
\end{equation}
Details concerning the numerical implementation of this orthogonalization procedure are given in \cref{sec:numdet}.

A spectral function in a finite basis set can  be computed from
\begin{align}\nonumber
&A_{pq}(\omega) = -\frac{1}{\pi} \lim_{\eta\rightarrow 0^+} \mathrm{Im} \left[G_{pq}^R(\omega+i\eta)\right] \\ 
%&=
%\sum_\nu
%	\langle \Psi^{N}_0|\hat{a}_p|\Psi_\nu^{N+1}\rangle
%	\langle \Psi^{N+1}_\nu|\hat{a}_q^{\dagger}|\Psi_0^N\rangle\\ 
%& \times	\delta\left(\omega-(E_\nu^{(N+1)}-E_0^{(N}))\right)
%	\label{eq:spec_addition}\\ \nonumber
%& + \sum_\nu
%	\langle \Psi^{N}_0|\hat{a}_q^{\dagger}|\Psi_\nu^{N-1}\rangle
%	\langle \Psi^{N-1}_\nu|\hat{a}_p|\Psi_0^N\rangle\\ \nonumber
%	&\times\delta\left(\omega+(E_\nu^{(N-1)}-E_0^{(N)})\right)
%	\label{eq:spec_remov},\\
& = A_{pq}^{>}(\omega) + A^{<}_{pq}(\omega),
\label{eq:ImG}
\end{align}
where
\begin{align}\nonumber
 A_{pq}^{>}(\omega)
& =
\sum_\nu
	\langle \Psi^{N}_0|\hat{a}_p|\Psi_\nu^{N+1}\rangle
	\langle \Psi^{N+1}_\nu|\hat{a}_q^{\dagger}|\Psi_0^N\rangle\\ 
& \times	\delta\left(\omega-(E_\nu^{(N+1)}-E_0^{(N}))\right) ,
	\label{eq:spec_addition}
\end{align}
and
\begin{align}
	\nonumber
 A^{<}_{pq}(\omega)
&= \sum_\nu
	\langle \Psi^{N}_0|\hat{a}_q^{\dagger}|\Psi_\nu^{N-1}\rangle
	\langle \Psi^{N-1}_\nu|\hat{a}_p|\Psi_0^N\rangle\\ 
	&\times\delta\left(\omega+(E_\nu^{(N-1)}-E_0^{(N)})\right) ,
	\label{eq:spec_remov}
\end{align}
where $ {\mathbf A}^{>}$  and $ {\mathbf A}^{<}$ are the addition and removal single-particle spectral functions, which describe inverse and direct photoemission experiments, respectively, in the sudden approximation.

%\subsubsection{EKT-1}
Using the definition of \cref{eq:psi_np},
\cref{eq:spec_addition} and \cref{eq:spec_remov}
can be expressed in terms of directly computable quantities, $\tilde{\mathbf c}^\nu$ (orthogonalized eigenvectors) and $\mathbf P$ in the case of EKT-1:
\begin{equation}
A^{>}_{pq}(\omega)
=
\sum_{rs} (T_+)_{rs}(\omega) \left(\delta_{pr} - P_{rp}\right)
\left(\delta_{sq} - P_{qs}\right),
\end{equation}
and
\begin{equation}
A^{<}_{pq}(\omega)
=
\sum_{rs} (T_-)_{rs}(\omega) P_{qr} P_{sp} ,
\end{equation}
where the state-averaged one-particle transition density matrix $\mathbf T_{\pm}$ is defined as
\begin{equation}
\mathbf T_{\pm} (\omega)
=
\sum_\nu
\tilde{\mathbf c}^\nu_\pm
(\tilde{\mathbf c}^\nu_\pm)^\dagger
\delta\left(\omega\mp(E_\nu^{(N\pm1)}-E_0^{(N)}))\right) .
\end{equation}
Similarly, for EKT-3, 2-RDM naturally arises in the evaluation of the spectral functions.
The working equation for IP states is as follows:%in \cref{eq:ekteA}.
\begin{align} \nonumber
A^{<}_{pq}(\omega)
&=
\sum_\nu \sum_{ijklmn} \Gamma^{ij}_{qk}\Gamma^{pn}_{lm}
\tilde{c}_{ijk}^\nu (\tilde{c}_{lmn}^\nu)^* \\
&\times \delta\left(\omega+(E_\nu^{(N-1)}-E_0^{(N)})\right).
\label{eq:ekteA}
\end{align}
It is straightforward to find similar equations for the EA states.

Furthremore, we note that
the density-of-state (DOS) is simply defined as
\begin{equation}
g(\omega) = \frac{\tr(\mathbf A(\omega))}M
\end{equation}
where $M$ is the number of single-particle basis functions.
We note that for solid-state applications single-particle basis carries
an additional index for the crystalline momentum $\mathbf k$ which can be straightforwardly incorporated into the above formalism.
In such applications, it is useful to compute the momentum-dependent DOS, $g(\mathbf k, \omega)$.

%The Dyson's equation for retarded Green's functions follows
%\begin{equation}
%G^R(\mathbf k, \omega)
%=
%G_0^R(\mathbf k, \omega)
%+ G_0^R(\mathbf k, \omega)
%\Sigma^R(\mathbf k, \omega)
%G^R(\mathbf k, \omega)
%\label{eq:dyson}
%\end{equation}
%where $G_0^R(\mathbf k, \omega)$ is the Hartree-Fock retarded Green's function.
%This equation can be used to obtain a self-energy using the interacting Green's function computed from EKT.
%Inserting \cref{eq:psi_np} and \cref{eq:def_npm1} into the definition of the spectral functions above and by expanding the field operators in terms of the single-particle basis
%\begin{equation}
%	\hat{\psi}(\bx) = \sum_p \phi_{p\sigma}(\br) \haop_{p\sigma},
%\end{equation}
%we have that
%\begin{equation}
%	A^{>}(\bx,\bx',\omega) = \sum_{pq\sigma\sigma'}\phi_{p\sigma}(\br)\phi^{*}_{q\sigma'}(\br') A^{>}_{pq\sigma\sigma'}(\omega),
%\end{equation}
%where (FDM: need to reconcile spin indices throughout (double delta too, $\pm <, >$))
%\begin{equation}
%\begin{split}
%		A^{>}_{pq\sigma\sigma'}(\omega) &=\\
%	&\sum_\nu \sum_{sr}
%	c_r^\nu c^{*\nu}_s
%	(\delta_{pr}-P_{rp}^{\sigma\sigma})
%	(\delta_{sq}-P_{qs}^{\sigma'\sigma'})\\
%	&\times
%	\delta\left(\omega-\epsilon_\nu^{+}\right),
%\end{split}
%\end{equation}
%and similarly
%\begin{equation}
%\begin{split}
%		A^{<}_{pq\sigma\sigma'}(\omega) =
%	\sum_\nu \sum_{sr}
%	c_r^\nu c^{*\nu}_s
%	P_{qr}^{\sigma'\sigma'}
%	P_{sp}^{\sigma\sigma}\delta\left(\omega+\epsilon_\nu^{-}\right).
%\end{split}
%\end{equation}

\subsection{Phaseless Auxiliary-Field Quantum Monte Carlo}
While the ph-AFQMC formalism has been presented before in detail,\cite{Motta2019} we review the essence of the algorithm 
to provide a self-contained description.
The imaginary propagation is given as
\begin{equation}
|\Psi_0\rangle 
\propto
\lim_{\tau\rightarrow \infty}    
\exp{\left(-\tau \hat{\mathcal H}\right)} |\Phi_0\rangle
= 
\lim_{\tau\rightarrow \infty}    
|\Psi(\tau)\rangle,
\label{eq:exact}
\end{equation}
where $\tau$ is the imaginary time,
$|\Psi_0\rangle$ is the exact ground state of a Hamiltonian $\hat{\mathcal H}$
and $|\Phi_0\rangle$ is an initial starting wavefunction with a non-zero overlap with $|\Psi_0\rangle$. 
We assume no special structure in the underlying Hamiltonian and work with the generic {\it ab-initio} Hamiltonians of \cref{eq:ham}.

In ph-AFQMC, this imaginary-time propagation 
is stochastically implemented. 
One discretizes the imaginary time $\tau$ with a time step of $\Delta \tau$ such that for $N$ time steps we have $\tau = N \Delta\tau$.
Using the Trotter approximation and the Hubbard-Stratonovich transformation,\cite{Hubbard1959Jul,Hirsch1983} a single time step many-body propagator can be written in integral form,
\begin{equation}
\exp(-\Delta\tau \hat{\mathcal H}) \: = 
\int d^{N_\alpha}\mathbf{x}~
p(\mathbf{x})
\hat{B}(\Delta \tau, \mathbf x),
\label{eq:HS}
\end{equation}
where $p(\mathbf{x})$ is the standard normal distribution, $\mathbf x$ is a vector of $N_\alpha$ auxiliary fields and $\hat{B}$ is defined as
\begin{equation}
\hat{B}(\Delta \tau, \mathbf x) = e^{-\frac{\Delta\tau}{2} \hat{\mathcal H}_1}
e^{-\sqrt{\Delta\tau} \mathbf{x}\cdot\hat{\mathbf{v}}}
e^{-\frac{\Delta\tau}{2} \hat{\mathcal H}_1} + \mathcal O (\Delta \tau^3),
\label{eq:B}
\end{equation}
where $\hat{\mathbf v}$ is defined from 
\begin{equation}
    \hat{\mathcal H}_2 = -\frac12 \sum_\alpha^{N_\alpha} \hat{v}_\alpha^2.
\end{equation}
The computation of the integral in \cref{eq:HS}
is carried out via Monte Carlo sampling where each walker samples an instance of $\mathbf x$.

The global wavefunction is, with importance sampling, represented as
a linear combination of walker wavefunctions:
\begin{equation}
|\Psi(\tau)\rangle = 
\sum_i
w_i (\tau)
\frac{|\psi_i(\tau)\rangle}
{\langle\Psi_T | \psi_i(\tau)\rangle} ,
\label{eq:walkers}
\end{equation}
where
$w_i$ is the weight of the $i$-th walker,
$|\psi_i(\tau)\rangle$
is the single Slater determinant of the $i$-th walker, and
$|\Psi_T\rangle$ is the trial wavefunction.
At each time step, each walker samples a set of $\mathbf x$, forms $\hat{B}(\Delta\tau,\mathbf x)$, and updates its wavefunction by applying $\hat{B}(\Delta\tau,\mathbf x)$ to it.
Practical implementations employ the so-called ``optimal'' force bias which shifts the Gaussian distribution,\cite{zhang2003quantum}
\begin{equation}
    \mathbf{\bar{x}}_i(\Delta \tau, \tau) = -\sqrt{\Delta\tau}
\frac{\langle
    \Psi_T | \hat{\mathbf{v}}' | \psi_i(\tau)
    \rangle}{
    \langle
    \Psi_T | \psi_i(\tau)
    \rangle
    }.
\end{equation}
With the optimal force bias, a single time step propagation can be summarized with two equations
\begin{align}
w_i(\tau+\Delta\tau) &= I_\text{ph}(\mathbf{x}_i,\mathbf{\bar{x}}_i,\tau,\Delta\tau) \times w_i(\tau), \\
|\psi_i(\tau+\Delta\tau)\rangle &= \hat{B}(\Delta\tau, \mathbf{x}_i-\mathbf{\bar{x}}_i) |\psi_i(\tau)\rangle ,
\end{align}
where the phaseless importance function in hybrid form is defined as 
\begin{equation}
I_\text{ph}(\mathbf{x}_i,\mathbf{\bar{x}}_i,\tau,\Delta\tau) = |I (\mathbf{x}_i,\mathbf{\bar{x}}_i,\tau,\Delta\tau)|\times
\text{max}(0, \cos(\theta_i(\tau))) ,
\label{eq:ph}
\end{equation}
with
\begin{equation}
    I(\mathbf{x}_i,\mathbf{\bar{x}}_i,\tau,\Delta\tau) = S_i(\tau, \Delta\tau)
    e^{\mathbf{x}_i\cdot\mathbf{\bar{x}}_i-\mathbf{\bar{x}}_i\cdot\mathbf{\bar{x}}_i/2},
 \label{eq:import}
\end{equation}
and
\begin{equation}
S_i(\tau, \Delta\tau) = \frac{\langle
    \Psi_T |
    \hat{B}(\Delta\tau, \mathbf{x}_i-\mathbf{\bar{x}}_i) | \psi_i(\tau)
    \rangle}{
    \langle
    \Psi_T | \psi_i(\tau) .
    \rangle},
 \label{eq:ovl}
\end{equation}
With this specific walker update instruction, all walker weights in \cref{eq:walkers} remain real and positive
and thereby it completely eliminates the fermionic phases problem.

In ph-AFQMC, the simplest way to evaluate the expectation value of an operator $\hat{O}$ is using the mixed estimator
\begin{align}
\langle \hat{O} \rangle_{\mathrm{mixed}}
&\coloneqq
\frac{\langle \Psi_T | \hat{O} | \Psi(\tau)\rangle}
{\langle \Psi_T | \Psi(\tau)\rangle}
\\
&=
\frac{
\sum_i w_i(\tau)
    \frac{
        \langle \Psi_T | \hat{O} | \psi_i(\tau) \rangle
    }
    {
        \langle \Psi_T | \psi_i(\tau)\rangle
    }
}
{
    \sum_i w_i(\tau) .
}
\end{align}
The mixed estimator is an unbiased estimator only for operators that commute with the Hamiltonian.
For operators that do not commute with the Hamiltonian, the mixed estimator can introduce significant biases due to the approximate trial wavefunctions that can be practically used.
To overcome this do we use the back-propagation algorithm\cite{Zhang1995May,Purwanto2004Nov,motta_back_prop,Motta2019} and write
\begin{align}
\langle {\hat{O}} &\rangle \approx  
\lim_{\kappa\rightarrow\infty} 
\frac{\langle \Psi_T | e^{-\kappa \hat{\mathcal{H}}} \hat{O} | \Psi(\tau)\rangle}
{\langle \Psi_T | e^{-\kappa \hat{\mathcal{H}}} | \Psi(\tau)\rangle}
\\
&=
\lim_{\kappa\rightarrow\infty} 
\frac{
\sum_i w_i(\tau+\kappa)
    \frac{
        \langle \psi_i(\kappa) | \hat{O} | \psi_i(\tau) \rangle
    }
    {
        \langle \psi_i(\kappa) | \psi_i(\tau)\rangle
    }
}
{
    \sum_i w_i(\tau+\kappa).
}\label{eq:bp_obs}
\end{align}
%For example, to compute the 1-RDM we accumulate for each walker
%\begin{equation}
%P_{pq} = 
%\lim_{\tau\rightarrow\infty}
%\lim_{\kappa\rightarrow\infty}
%P_{pq}(\kappa, \tau)
%\end{equation}
%where
%\begin{equation}
%P_{pq}(\kappa, \tau) = 
%\frac{\langle 
%\Psi (\kappa)
%|
%\hat{a}_p^\dagger\hat{a}_q
%| \Psi(\tau)\rangle}{
%\langle 
%\Psi (\kappa)
%| \Psi(\tau)\rangle
%}
%\label{eq:rdm}
%\end{equation}
%with $| \Psi(\tau)\rangle$ being the walker wavefunction at imaginary time $\tau$.
To summarize, we propagate $|\Psi\rangle$ until $\kappa + \tau$, storing the walker wavefunction at time $\tau$. We can then split the propagation into $\kappa$ back-propagation and $\tau$ forward-propagation as in \cref{eq:bp_obs}.
The back propagated wavefunction is constructed by applying a walker's propagators to the trial wavefunction from the $\kappa$ portion of the path.
Practically the convergence of the expectation value has to be monitored with respect to the back propagation time $\kappa$.
It should be emphasized that in ph-AFQMC the walker wavefunction is a single determinant wavefunction.

It was found in Ref.\citenum{motta_back_prop} that the standard back-propagation algorithm described in \cref{eq:bp_obs} can yield poor results in ph-AFQMC when applied to {\it ab-initio} systems. 
The authors devised a number of additional steps to reduce the phaseless error, the most accurate of which was to partially restore the phase and cosine factors along the back propagation portion of the path.
In this work we restore phases along the back propagated path as well as along the forward direction.
Practically this amounts to storing the phases and cosine factors between $[\tau-\kappa, \tau+\kappa]$ and multiplying these by the weights appearing in \cref{eq:bp_obs}.
This additional restoration of paths along the forward direction was not described in Ref.\citenum{motta_back_prop} but was used in practice\cite{MottaPrivate,ChenSolidsBP2020} and we found it necessary to obtain more accurate results for the systems studied here.

In EKT1-AFQMC, we directly sample $\mathbf F_\pm$ using the back-propagated estimator form.
This boils down to the evaluation of the 2-RDM appearing in \cref{eq:2-RDM} using the back propagated 1-RDM via Wick's theorem:
\begin{equation}
\Gamma_{pt}^{rs} 
%= 
%\frac{\langle 
%\Psi (\kappa)
%|
%\hat{a}_p^\dagger
%\hat{a}_t^\dagger
%\hat{a}_s
%\hat{a}_r
%| \Psi(\tau)\rangle}{
%\langle 
%\Psi (\kappa)
%| \Psi(\tau)\rangle
%}
= P_{pr}P_{ts} - P_{ps}P_{tr} .
\label{eq:2-RDMback}
\end{equation}
%where $\mathbf P$ here is $\mathbf P(\kappa,\tau)$.
With these ingredients, we can evaluate $\mathbf F_{\pm}$ by contracting the one- and two-body matrix elements with the back propagated 1- and 2-RDMs.

While efficient implementations are not the focus of our efforts in this work,
we mention the computational cost of producing one back-propagated sample of $\mathbf F^\pm$ and $\mathbf S^\pm$. 
A sample of $\mathbf S^\pm$ has the same overhead as computing a back-propagated 1-RDM sample which scales as $\mathcal O(NM^2)$ where $M$ is the number of orbitals and $N$ is the number of occupied orbitals. The cost for producing a sample of $\mathbf F^\pm$ is more involved and depends on the integral factorization that one chooses to use. Using the most common integral factorization, i.e., the Cholesky factorization, it can be shown that the cost scales as $\mathcal O(M^3 X)$ where $X$ is the number of Cholesky vectors (also note that the 2-RDM is never explicitly formed).
With tensor hypercontraction,\cite{thc1,thc2,thc3,malone_isdf,lee2019systematically} the cost can be brought down to overall cubic. 
If one were to just implement a matrix-vector product for iterative eigensolvers, the Cholesky factorization can achieve cubic scaling per matrix-vector product as well. The cost is increased in EKT3-AFQMC where each matrix-vector product sample costs $\mathcal O(M^5)$.
It is potentially possible to reduce this cost further by also factorizing EKT amplitudes in a THC format, as is done in Ref. \citenum{thc3}.

We leave the exploration of EKT3-AFQMC for the future study and focus on EKT1-AFQMC in this work.

\subsection{Uniform Electron Gas}
Aside from small molecular benchmarks, we also study the spectral properties of the uniform electron gas (UEG) model. The UEG model is usually defined in the plane-wave basis set, which gives
the one-body operator
\begin{equation}
\hat{\mathcal H}_1 = \sum_{\mathbf K} \frac{|\mathbf K|^2}{2} a_{\mathbf K}^\dagger a_{\mathbf K}
\end{equation}
and the electron-electron interaction operator is (in a spin-orbital basis)
\begin{equation}
\hat{\mathcal H}_2 = \frac{1}{2\Omega	}
\sum_{\mathbf{K}\ne\mathbf 0,\mathbf{K}_1,\mathbf{K}_2}
\frac{4\pi}{|\mathbf{K}|^2}
a_{\mathbf{K}_1+\mathbf{K}}^\dagger
a_{\mathbf{K}_2-\mathbf{K}}^\dagger
a_{\mathbf{K}_2}
a_{\mathbf{K}_1},
\label{eq:ueg2}
\end{equation}
where $\mathbf K$ here is a planewave vector and $\Omega$ is the volume of the unit cell.
In addition to $\hat{\mathcal H}_1$ and $\hat{\mathcal H}_2$, there is a constant term that arises due to a
finite-size effect.
Specifically, the Madelung energy $E_M$ should be included to account for self-interactions associated with the Ewald sum under periodic boundary conditions\cite{schoof_prl} via
\begin{equation}
%E_M = -2.837297 \times \left(\frac{3}{4\pi}\right)^{1/3}N^{2/3}r_s^{-1},
E_M = \frac{N}2 \xi ,
\end{equation}
with 
\begin{equation}
\xi = -2 \times 2.837297 \times \left(\frac{3}{4\pi}\right)^{1/3}N^{-1/3}r_s^{-1} ,
\end{equation}
where $N$ is the number of electrons in the unit cell and $r_s$ is the Wigner-Seitz radius.
We define the UEG Hamiltonian as a sum of these three terms,
\begin{equation}
\hat{H}_\text{UEG} = \hat{\mathcal H}_1 + \hat{\mathcal H}_2+ E_M .
\label{eq:uegham}
\end{equation}

The Madelung constant can be either included in the Hamiltonian as written in \cref{eq:uegham}
or it can be included as an {\it a posteriori} correction to the simulation done without it.
When the latter choice is made, the spectral functions have to be shifted accordingly 
in order to compare the results obtained from the former approach. The corresponding shift can be
derived from a shift in the poles
\begin{align}
\text{IP} &= E(N-1) - E(N) \rightarrow -\frac{\xi}2 ,\\
\text{EA} &= E(N+1) - E(N) \rightarrow \frac{\xi}2 .
\end{align}
Regardless of whether we included the Madelung constant in the Hamiltonian, there is an additional correction of $-\frac{\xi}2$ for both IP and EA to remove spurious image interactions coming from
the excess charge created.\cite{Yang2020Feb} Therefore, overall electron removal poles are shifted by $-\xi$ and electron addition poles remain the same.

For many molecular quantum chemistry methods, often the two-electron integral tensor is assumed to be 8-fold symmetric.
Practical implementations utilize this symmetry to simplify equations as well. As such, without the 8-fold symmetry, molecular quantum chemistry methods would not produce correct answers even though the UEG Hamiltonian only contains real-valued matrix elements.
This complication of the UEG Hamiltonian becomes more obvious once we write it in the form of \cref{eq:h2chem}
with
\begin{align}\nonumber
\langle pq | rs \rangle 
%=(pr|qs)
= &
%\begin{cases}
%0 ~~~~\text{if} ~\mathbf K_p - \mathbf K_r = \mathbf 0
%\\
\frac1\Omega
\frac{4\pi}{|\mathbf K_p - \mathbf K_r|^2}
\delta_{\mathbf K_p - \mathbf K_r, \mathbf K_q - \mathbf K_s}\\
%~~ \text{otherwise}
&\times(1-\delta_{\mathbf K_p,\mathbf K_r}) .
%\end{cases}
\label{eq:chemeri}
\end{align}
The permutation between $p$ and $r$ or between $q$ and $s$ alters
the value of the integral tensor because of the Kronecker delta term.
%The integral tensor is therefore 
This is a direct consequence of using a planewave basis which is complex, unlike the usual Gaussian orbitals.
To circumvent any complications due to this, we perform a unitary transformation that rotates the planewave basis into
a real-valued basis. 
Namely, for given $\mathbf K$ and $-\mathbf K$ (assuming $\mathbf K \ne \mathbf 0$), we use
\begin{equation}
\mathbf U = 
\frac{1}{\sqrt{2}}
\begin{pmatrix}
1 & -i \\
1 & i 
\end{pmatrix} .
\end{equation}
We apply this transformation to every pair of $\mathbf K$ and $-\mathbf K$ in the two-electron integral tensor in \cref{eq:chemeri}.
The resulting transformed integral tensor now recovers the full 8-fold symmetry.
One can also transform observables such as spectral functions back to the original basis using $\mathbf U^\dagger$ when necessary.

\section{Computational Details}
%\subsection{Heat-Bath Configuration Interaction}
All quantum chemistry calculations are performed with \texttt{PySCF}\cite{Sun2018Jan} which include mean-field (HF) calculations, coupled-cluster with singles and doubles (CCSD) and CCSD with perturbative triples (CCSD(T)).
All one- and two-electron integrals needed for ph-AFQMC were also generated with \texttt{PySCF}.
ph-AFQMC calculations were mostly performed with \texttt{QMCPACK}\cite{Kent2020May} and \texttt{PAUXY}\cite{pauxy} was used to crosscheck some results.

We use a selected configuration interaction (CI) method called heat-bath CI (HCI)\cite{Holmes2016Aug,Sharma2017Apr,Smith2017Nov} to produce
numerically exact IPs within a basis whenever possible.
Furthermore, we compute the HCI IPs within EKT1 (i.e., EKT1-HCI) using the 1- and 2-RDM from variational HCI wavefunctions along with \cref{eq:fmcomp}.
%EKT1-HCI is used to quantify the back-propagation error in ph-AFQMC 
Since there can be an inherent bias of EKT1 itself, we provide EKT1-HCI as an ``exact'' result for IPs within the limits of the EKT1 approach.
This can be used to quantify the phaseless and back-propagation errors in EKT1-AFQMC.
In HCI, there is a single tunable parameter, $\epsilon_1$ that controls the variational energy which is used to select determinants
to be included in the variational expansion. 
We also use the 3-RDM of the variational wavefunction to compute the 4-RDM via the cumulant construction.\cite{Zgid2009May,Saitow2013Jul}
These 3- and 4-RDMs are further used to construct the EKT3 Fock matrix in \cref{eq:fmekt3}. 
This approach is referred to as EKT3-HCI.
The eigenvalues and eigenvectors of the EKT3 Fock matrix (with its pertinent metric) can then be used to produce IPs and spectral functions of EKT3.
%This then produces spectral functions and IPs of EKT3-HCI.
We tuned $\epsilon_1$ to be such that the resulting second-order Epstein-Nesbet perturbation energy is no greater than 1 m$E_h$ for every system except the UEG model. In the UEG model, we observed a PT2 correction of 3 m$E_h$ at $r_s=4$. 
This was found to be sufficient to produce accurate EKT IPs for systems studied here.
All calculations are performed with a locally modified version of \texttt{Dice}.\cite{Holmes2016Aug,Sharma2017Apr,Smith2017Nov}
%\subsection{Dynamical Lanczos}

We used a timestep of 0.01 au and the pair branch population control method\cite{wagner_qwalk} used the hybrid propagation scheme\citep{Purwanto2004Nov} for all ph-AFQMC simulations.
For the small atoms and molecules and 14 electron UEG examples, we used 2880 walkers while for the 54 electron UEG we used 1152 walkers.
All calculations used restricted Hartree--Fock (RHF) trial wavefunctions except for the charged species where we instead used unrestricted Hartree--Fock wavefunctions (UHF).
The exception to this was \ce{CH4} where we used the same RHF orbitals for the charged species as we found that the UHF solution broke spatial symmetry and led to a large phaseless constraint bias. 
All AFQMC results are performed with the phaseless approximation so we simply refer ph-AFQMC to as AFQMC in the following sections.

We adapted the standard dynamical Lanczos algorithm\cite{Haydock1975Aug,DagottoLanczos1994} to obtain spectral functions of \cref{eq:uegham}.
Even by exploiting symmetry and using a distributed sparse Hamiltonian, dynamical Lanczos results could only be obtained for the smallest UEG system of 14 electrons in 19 plane waves, corresponding to an $N$-electron Hilbert space size of $2.5\times10^9$ determinants.
In the dynamical Lanczos algorithm, one first obtains the $N$-particle ground state, $|\Psi_0^N\rangle$ iterating within the Lanczos Krylov subspace.
Ultimately, our goal is to compute \cref{eq:ImG} which then requires another run of the Lanczos algorithm.
For an electron removal problem, we pick an orbital index $i$ and generate an initial vector in the $(N-1)$-electron sector, $|f_0\rangle$ = $\hat{a}_i |\Psi_0^N\rangle / \langle \Psi_0^N |\hat{a}_i^\dagger\hat{a}_i|\Psi_0^N\rangle$.
Each Lanczos iteration then generates coefficients, $\{a_k\}$ and $\{b_k\}$, for the following continued fraction expression:\cite{Haydock1975Aug,DagottoLanczos1994}
\begin{equation}
A_{ii}(\omega, \eta)
=
-\frac{1}\pi \text{Im}\frac{\langle \Psi_0^N |\hat{a}_i^\dagger\hat{a}_i|\Psi_0^N\rangle}
{z - a_0 - \frac{b_1^2}{z-a_1 - \frac{b_2^2}{z-a_2 \cdots}}} ,
\end{equation}
where $z = E_0^{(N)} - \omega + i \eta$ with some spectral broadening constant $\eta$.
We take a total of 50 Lanczos iterations to generate the continued fraction coefficients and
this was enough to converge the low-energy spectrum within the energy scale that is relevant in this work.

\section{Results and Discussion}
While the EKT approach is valid for both electron removal and electron addition, 
for numerical results we focus on electron removal processes (IP energies and electron removal spectral functions) for simplicity.
We benchmark the IP energies from the proposed EKT1-AFQMC approach over
several small chemical systems and 
the UEG model.
Furthermore, we also show
promising improvements over EKT1 using EKT3-HCI when satellite peaks are important.
We use a ``$\Delta$ method'' to denote a scheme where we run the pertinent method for both $N$- and $(N-1)$-electron systems and 
obtain the IP as an energy difference.

\subsection{Small chemical systems in the aug-cc-pVDZ basis}
\begin{table}
\begin{tabular}{|c|c|c|c|}
\hline
 & Experiment & $\Delta$HCI \\ \hline
He & 24.59 & -0.23 \\ \hline
Be & 9.32 & -0.03 \\ \hline
Ne & 21.56 & -0.13 \\ \hline
FH & 16.19 & -0.12 \\ \hline
\ce{N2} & 15.6 & -0.34 \\ \hline
\ce{CH4} & 14.35 & -0.06 \\ \hline
\ce{H2O} & 12.62 & -0.07 \\ \hline
\end{tabular}
\caption{
Experimental first ionization potentials (eV) and deviation (eV) of the numerically exact $\Delta$HCI from these results for chemical species considered.
$\Delta$HCI employed the aug-cc-pVDZ basis.
}
\label{tab:basis}
\end{table}

In this section, we study seven small chemical systems (He, Be, Ne, FH, \ce{N2}, \ce{CH4}, and \ce{H2O})
that have well-documented experimental IPs.\cite{vanSetten2015Dec}
We use the nomenclature FH for the hydrogen fluoride molecule to distinguish it from the abbreviation for Hartree-Fock (HF).
All geometries were taken from ref. \citenum{vanSetten2015Dec}.
We used a relatively small basis set, namely aug-cc-pVDZ,\cite{Dunning1989} to obtain good statistics in 
back-propagated estimators.
We have used more than 3000 back-propagated estimator samples in all cases considered here,
each of which requires a back-propagation time of greater than 4 a.u.
This results in a total propagation time longer than 12000 a.u., which is unusually long for standard AFQMC calculations.
The use of this basis set also allows for a direct comparison between AFQMC and
numerically exact HCI within this basis set.

The goal of this numerical section is to quantify the three sources of error in addition to the basis set incompleteness error in EKT1-AFQMC
based on simple examples where exact simulations are possible.
These three sources of error are:
\begin{enumerate}
\item {\it Phaseless constraint errors.}
As mentioned, the phaseless constraint is necessary to remove the phase problem that arises in the imaginary-time propagation.
However, due to this constraint, the resulting ground state energies and properties (e.g., RDMs) are biased.
\item {\it Back-propagation errors.}
The back-propagation algorithm incurs additional errors.
This was noted and studied in detail in ref. \citenum{motta_back_prop}. For instance, in ref. \citenum{motta_back_prop}, it was shown that for neon the phaseless error with a simple trial wavefunction is negligible (below 1 m$E_h$) but the error in the one-body energy from the back-propagated 1-RDM was about 5 m$E_h$.
\item {\it EKT1 errors.}
While systematically improvable with higher-order excitations, EKT1 is not an exact approach to quasiparticle spectra unless all orders of excitations are included. 
Nonetheless, for the first IP, it has been numerically and analytically suggested that 
EKT1 approaches the exact IP in the basis set limit
if the exact 1- and 2-RDMs are used.\cite{Morrison1992a,Sundholm1993,Ernzerhof2009,Vanfleteren2009}
Beyond the first IP, we will show that EKT1 qualitatively fails to capture satellite peaks that arise in the case of the core spectrum of the UEG model.
\end{enumerate}

In \cref{tab:basis}, we present numerically exact first IPs of molecules within this basis set using $\Delta$HCI.
The basis set incompleteness error can be as large as 0.3 eV in these molecules and
therefore we will only compare AFQMC results to these numerically exact results in the same basis set
as opposed to comparing to the experimental data.
We do not expect the qualitative conclusions of our study to change with larger basis sets.

\begin{table}
\begin{tabular}{|c|c|c|c|}
\hline
 & Ground state & IP \\ \hline
He & 0.00 & 0.00 \\ \hline
Be & 0.01 & -0.01 \\ \hline
Ne & -0.03 & 0.07 \\ \hline
FH & -0.03 & 0.07 \\ \hline
\ce{N2} & -0.04 & 0.06 \\ \hline
\ce{CH4} & -0.03 & 0.06 \\ \hline
\ce{H2O} & -0.03 & 0.04 \\ \hline
\end{tabular}
\caption{
Error (eV) in the AFQMC $N$-electron system ground state energy and in the first ionization potential with respect to the corresponding HCI results. 
The statistical error bar of AFQMC is less than 0.01 eV and therefore we do not present them here.
}
\label{tab:phaseless}
\end{table}
Next, we assess the phaseless bias in the $N$-electron 
system ground state energy as well as the error in the $\Delta$AFQMC IP energies compared to the $\Delta$HCI IPs.
In \cref{tab:phaseless}, we present numerical data that detail the phaseless bias in these quantities. 
The ground state energy error is less than 0.04 eV which is in the neighborhood of the usual standard of accuracy, 1 m$E_h$. 
Unfortunately, in many cases $(N-1)$-electron systems incur as large a phaseless bias as do the $N$-electron systems, albeit with an opposite sign of the error.
Thus, AFQMC does not benefit from a cancellation of errors for the IP energy, which results in IP errors that are larger than those for the ground state energy. The largest IP error we find is around 0.07 eV. 
%Nonetheless, since EKT theory takes RDMs of the $N$-particle ground state, the small phaseless error in the ground state %energy is very encouraging.

\begin{table}
\begin{tabular}{|c|c|c|c|}
\hline
 & EKT1-HCI & KT & EKT1-AFQMC \\ \hline
He & 24.36 & 0.60 & 0.02 \\ \hline
Be & 9.29 & -0.87 & 0.06 \\ \hline
Ne & 21.48 & 1.73 & -0.04 \\ \hline
FH & 16.13 & 1.58 & 0.01 \\ \hline
\ce{N2} & 15.34 & 1.92 & 0.15 \\ \hline
\ce{CH4} & 14.13 & 0.68 & 0.19 \\ \hline
\ce{H2O} & 12.60 & 1.27 & -0.04 \\ \hline\end{tabular}
\caption{
Error (eV) in the first IP obtained from EKT1-AFQMC and the Koopmans' theorem (KT) relative to EKT1-HCI.
The statistical error bar of EKT1-AFQMC cannot be estimated without bias (see main text for discussion).\cite{Ceperley1988Nov,Blunt2018Aug}
}
\label{tab:ekt1}
\end{table}
We present the EKT1 results in \cref{tab:ekt1}. We refer readers to \cref{sec:numdet}
where a detailed description of the EKT1 calculation is given.
Theoretically ``exact'' EKT1 results can be obtained by using exact RDMs from HCI.
To quantify the back-propagation error of AFQMC (given that the phaseless error is very small for these systems), we shall compare EKT1-AFQMC to EKT1-HCI.
We also computed simple Koopmans' theorem (KT) IPs using HF and report these results in \cref{tab:ekt1}.
The error of EKT1-AFQMC is small for most chemical species, but it becomes as large as 0.19 eV for \ce{CH4}.
Even though the phaseless bias in the ground state energy was found to be very small (0.04 eV or less), EKT1-AFQMC errors from using the back-propagated 1-RDM and the EKT Fock matrix can be five times larger. 
Therefore, we attribute this error mainly to the back-propagation error.
Nonetheless, the comparison against the simpler KT suggests that EKT1-AFQMC
can readily recover the correlation contribution to quasiparticle energies with errors less than 0.2 eV in these examples.

We note that EKT1-AFQMC IPs do not have any statistical error bars.
This is not because these numbers are deterministic but arises because EKT eigenvalues are not unbiased estimators.
This was also observed in a similar approach called Krylov-projected full configuration interaction QMC.
Interested readers are referred to Ref. \citenum{Blunt2018Aug} for details.
In essence, eigenvalues of a noisy matrix where each element is normally distributed are not normally distributed. 
Therefore, statistical error bars are difficult to estimate and are not simply associated with variances of Gaussian distributions.
Given the small statistical error bars (on the order of $3\times 10^{-4}$ or less) on the diagonal elements of the EKT1 Fock matrix and 1-RDM, we expect that these results are reproducible up to 0.01 eV if one follows exactly the same numerical protocol given in \cref{sec:numdet}.

\begin{table}
\begin{tabular}{|c|c|c|c|}
\hline
 & EKT1-HCI & EKT1-AFQMC & EOM-IP-CCSD \\ \hline
He & 0.00 & 0.02 & 0.00 \\ \hline
Be & 0.00 & 0.06 & 0.00 \\ \hline
Ne & 0.05 & 0.02 & -0.28 \\ \hline
FH & 0.06 & 0.07 & -0.22 \\ \hline
\ce{N2} & 0.05 & 0.20 & 0.14 \\ \hline
\ce{CH4} & -0.15 & 0.04 & -0.02 \\ \hline
\ce{H2O} & 0.05 & 0.01 & -0.15 \\ \hline
\end{tabular}
\caption{
Error (eV) in the first IP obtained from EKT1-HCI, EKT1-AFQMC, and EOM-IP-CCSD relative to $\Delta$HCI (given in \cref{tab:basis}) in the aug-cc-pVDZ basis.
The statistical error bar of EKT1-AFQMC cannot be estimated without biases (see main text for discussion).\cite{Ceperley1988Nov,Blunt2018Aug}
}
\label{tab:ektvsccsd}
\end{table}
Finally, we discuss the inherent error of EKT1 by comparing EKT1-HCI and EKT1-AFQMC against $\Delta$HCI as in \cref{tab:ektvsccsd}.
We also compare a more widely used approach called equation-of-motion coupled cluster ionization potential with singles and doubles (EOM-IP-CCSD)
to gauge the magnitude of EKT1 errors.
Since this is a benchmark on the first IP of molecules, EKT1-HCI is expected to be quite accurate. 
This expectation is due to the general belief that EKT1 with exact RDMs yields exact first IP.\cite{Morrison1992a,Sundholm1993,Ernzerhof2009,Vanfleteren2009} 
Given this, the small errors of EKT1-HCI in \cref{tab:ektvsccsd} are not surprising.
The only exception is \ce{CH4} with an error of -0.15 eV, which we believe will still approach the exact IP in the complete basis set limit.
EKT1-AFQMC appears to be as good as EKT1-HCI, with an outlier for the case of \ce{N2}. The error of EKT1-AFQMC relative to EKT1-HCI is comparable to the error of EKT1-HCI relative to $\Delta$HCI in this basis set. %This is a quite encouraging result for EKT1-AFQMC.
EOM-IP-CCSD generally does not work as well as the EKT1 approaches with a maximum error of -0.28 eV on the neon atom.
While these finite basis set comparisons are informative, we emphasize that more fair comparisons should be conducted in the complete basis set limit and we hope to carry these out in the future. Regardless, the EKT1-AFQMC results are encouraging.

\begin{figure*}
    \centering
    \includegraphics[scale=0.65]{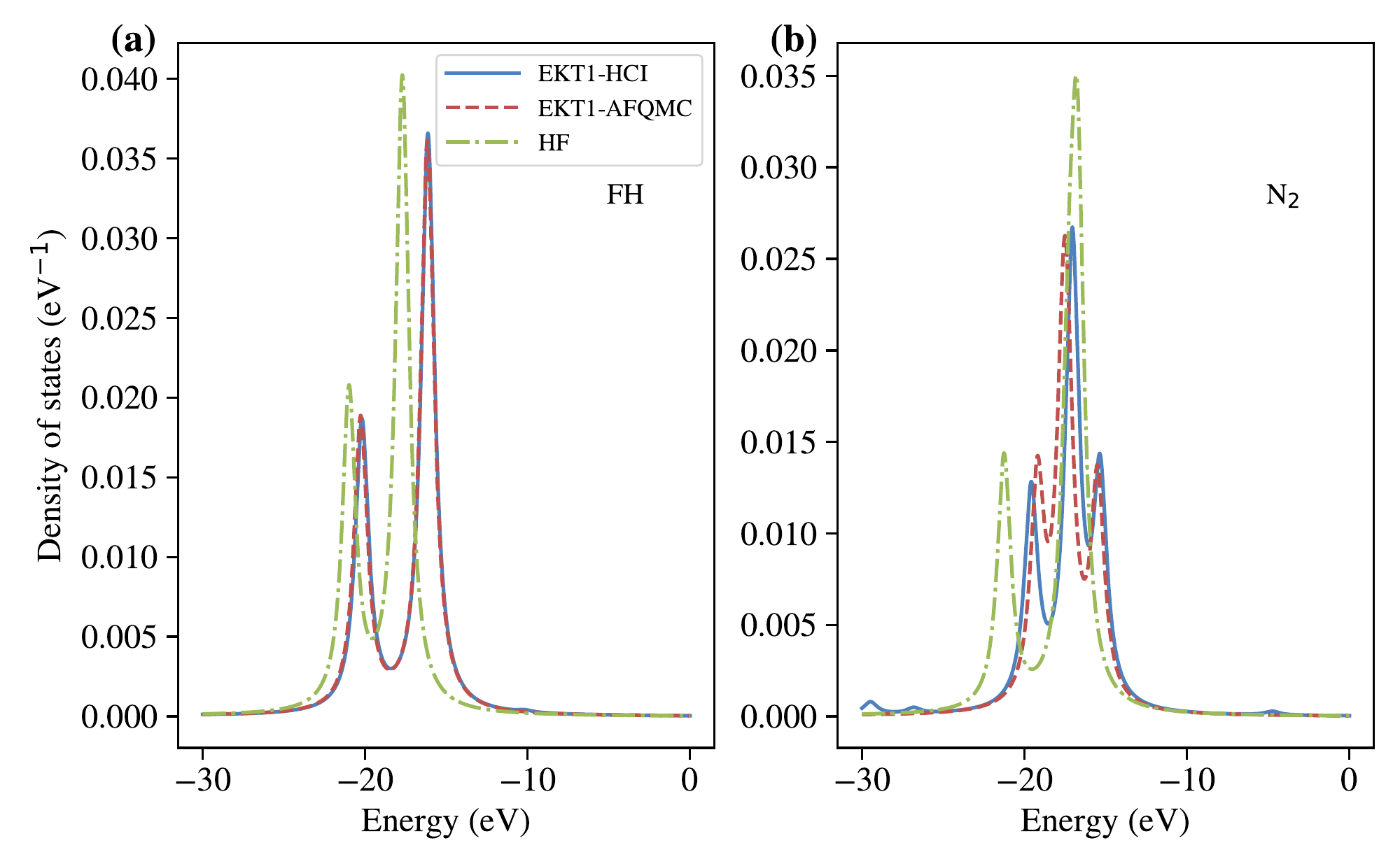}
    \caption{
Electron removal spectral functions from various methods (EKT1-HCI, EKT1-AFQMC, and HF) of (a) FH and (b) \ce{N2} in the aug-cc-pVDZ basis set. 
Note that in (a) EKT1-AFQMC is right on top of EKT1-HCI on the plotted scale.
A broadening parameter $\eta = 0.5$ eV was used.
     }
    \label{fig:specchem}
\end{figure*}

The main motivation for performing EKT1 within AFQMC was to obtain spectral functions.
Poles alone can be obtained using the ground state AFQMC algorithm (i.e., $\Delta$AFQMC) by imposing a proper constraint for excited state descriptions. While the choice of proper trials is a challenge for this purpose, such an approach avoids the complications
due to back-propagation.
However, spectral weights cannot be obtained from $\Delta$AFQMC.
We show EKT1-AFQMC spectral functions for FH and \ce{N2} in \cref{fig:specchem}.
Based on the results shown in \cref{tab:ekt1}, EKT1-AFQMC is in good agreement with EKT1-HCI for FH, but not for \ce{N2}. 
Therefore, comparing these two cases is useful for understanding how back-propagation errors are reflected in spectral functions.
In the case of FH, we do not see any visible differences between EKT1-HCI and EKT1-AFQMC on the plotted energy scale.
However, for \ce{N2} we can clearly see some deviation between EKT1-AFQMC and EKT1-HCI. Nonetheless, the main features of the spectral function are reproduced, namely three large quasiparticle peaks with the middle peak being the largest. We note that there are peaks with very small spectral weights in EKT1-HCI close to -30 eV, -27 eV, and -5 eV and that these features these do not appear in EKT1-AFQMC. We attribute this to a relatively large linear dependency cutoff ($10^{-4}$) needed in EKT1-AFQMC to stabilize the generalized eigenvalue problem as explained in \cref{sec:numdet}.
In both molecules, there are insignificant differences between the EKT1 and HF spectra in terms of the peak heights, locations and the number of peaks with a brodening parameter of $\eta = 0.5$ eV. %The fact that EKT1-AFQMC closely approximates EKT1-HCI is again encouraging.

\subsection{The uniform electron gas (UEG) model}
AFQMC has emerged as a unique tool for simulating correlated solids.\cite{lee_2019_UEG,zhang_nio,malone2020gpu,Malone2020Oct,Lee2020Dec} 
A model solid that describes the basic physics of metallic systems
is the UEG model.
The accuracy and scope of AFQMC in studying the UEG model has been
well documented
at zero temperature and finite temperature.\cite{lee_2019_UEG,Lee2020Dec}
Motivated by these studies, we investigate the spectral properties of the UEG model 
within the EKT approaches (EKT1 and EKT3).

\subsubsection{14-electrons/19-planewaves}
\begin{figure*}
    \centering
    \includegraphics[scale=0.65]{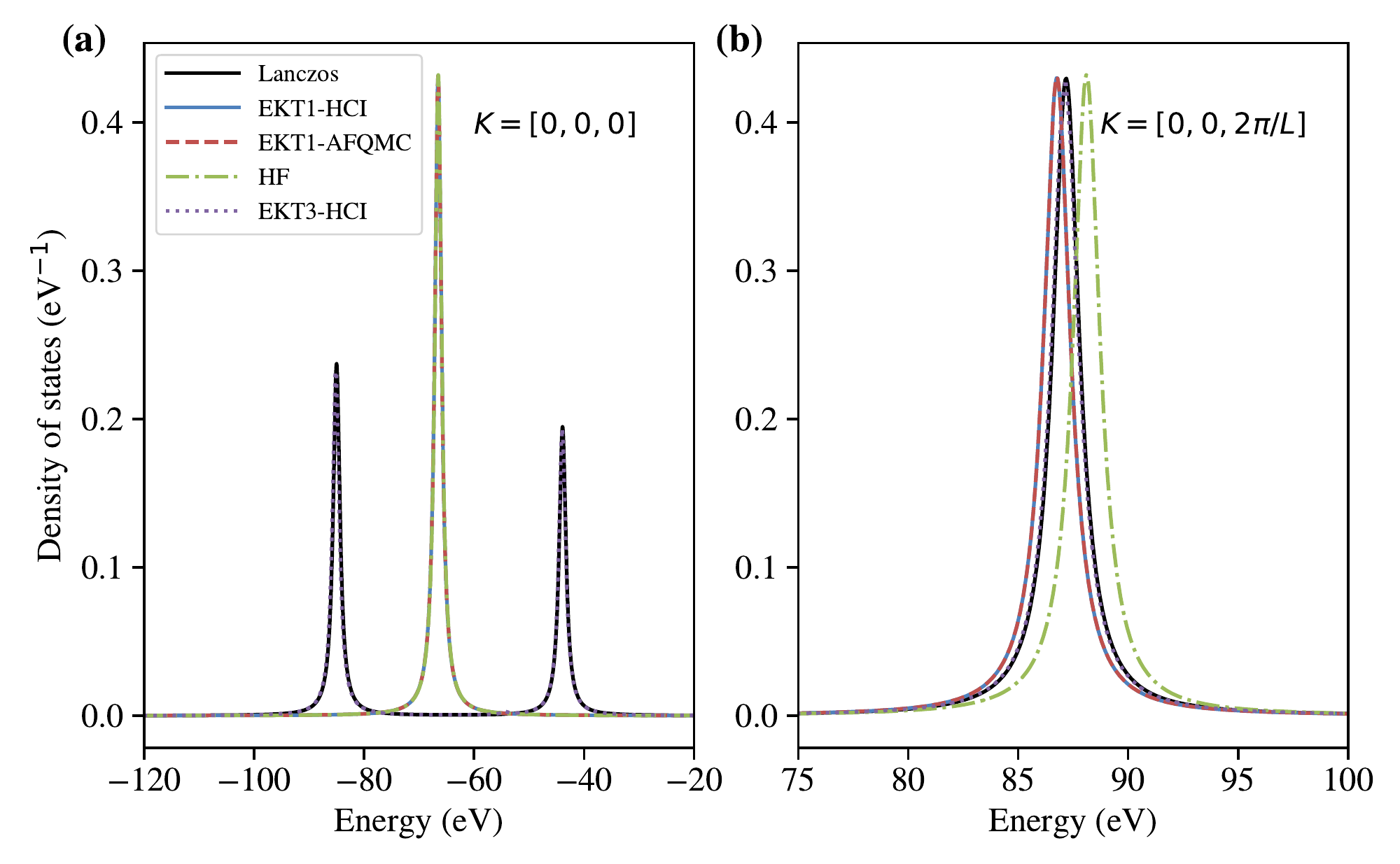}
    \caption{
Electron removal spectral functions of the 14-electron in 19-planewave UEG model from
various methods at
$r_s = 0.5$: (a) $\mathbf K = [0,0,0]$ and (b) $\mathbf K = [0,0, 2\pi/L]$.
In (a), note that
EKT1-AFQMC, EKT1-HCI, and HF are right on top of each other.
In (b), EKT1-AFQMC and EKT1-HCI are right on top of each other. 
In both (a) an (b), Lanczos and EKT3-HCI are right on top of each other.
A broadening parameter of 0.2 eV was used for all plots.
     }
\label{fig:rs05}
\end{figure*}
\begin{figure*}
    \centering
    \includegraphics[scale=0.65]{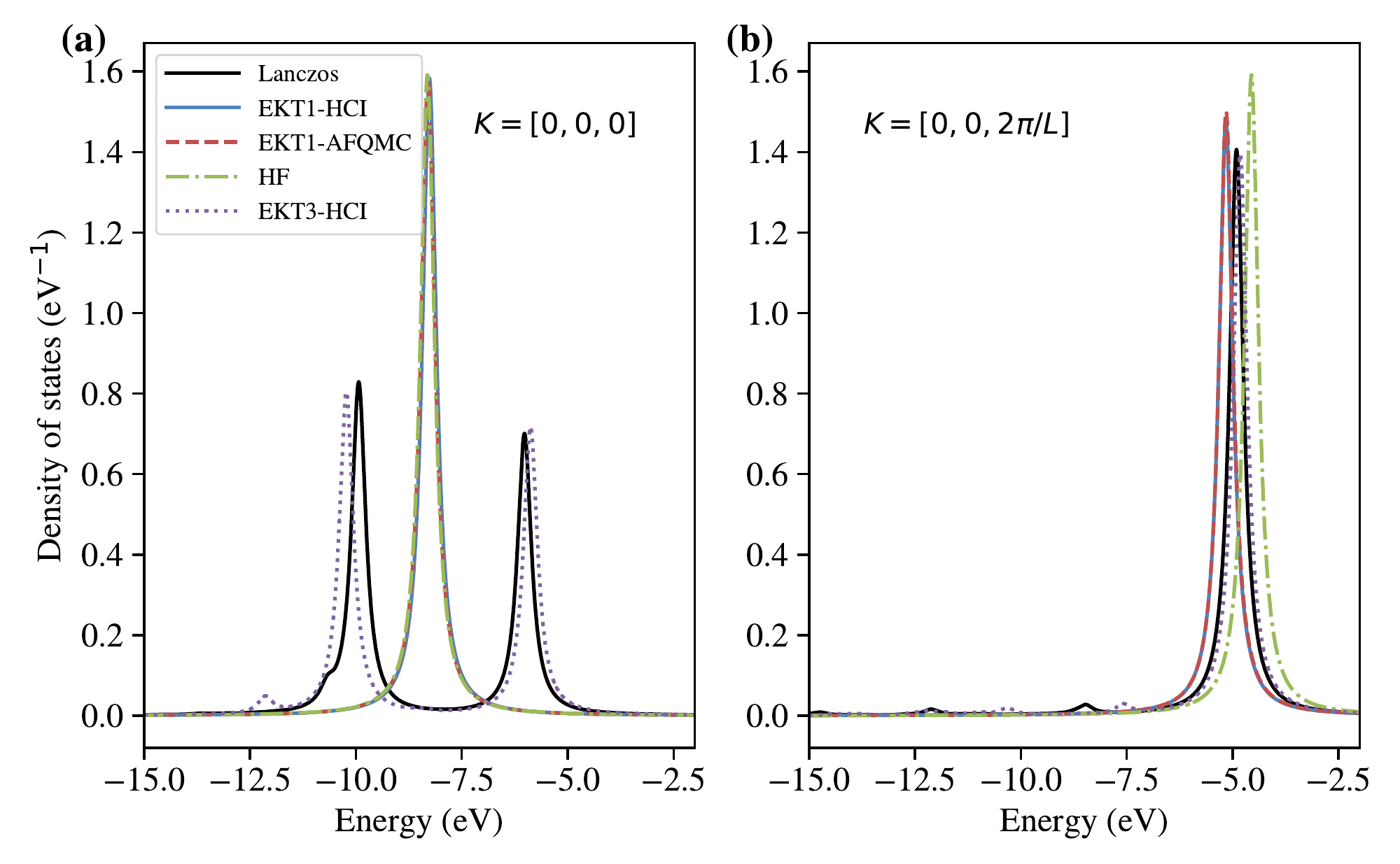}
    \caption{
Electron removal spectral functions of the 14-electron in 19-planewave UEG model from
various methods at
$r_s = 4.0$: (a) $\mathbf K = [0,0,0]$ and (b) $\mathbf K = [0,0, 2\pi/L]$.
In (a), note that
EKT1-AFQMC, EKT1-HCI, and HF are right on top of each other.
In (b), EKT1-AFQMC and EKT1-HCI are nearly on top of each other. 
A broadening parameter of 0.2 eV was used for all plots.
     }
\label{fig:rs4}
\end{figure*}
The first example that we consider is
a relatively small UEG supercell with only 19 planewaves.
It is far from the basis set limit as well as from the thermodynamic limit.
However, it is small enough for one to produce unbiased 
EKT results using HCI and numerically exact
dynamical Lanczos results.
We note that
the spectral function of this benchmark UEG model at $r_s = 4$
was first presented in Ref. \citenum{McClain2016Jun}.
We produced around 3000 back-propagation samples,
which yielded the largest statistical error in the Fock matrix and 1-RDM
on the order of $5\times10^{-4}$.

We consider two Wigner-Seitz radii, $r_s = 0.5$ and $r_s = 4$. 
Based on our previous benchmark study of AFQMC on this system, 
we expect that the phaseless error in the ground state at $r_s = 0.5$ is negligible while
the error is relatively more noticeable at $r_s= 4$.\cite{lee_2019_UEG}
Compared to the numerically exact energies, 
it was found that the constraint bias in AFQMC
is only -0.0118(6) eV at $r_s = 0.5$ and 0.185(2) eV at $r_s = 4$.
Given this small ground state bias, we expect the EKT1-AFQMC approach would be
as accurate as EKT1-HCI if good statistics and accuracy in the back-propagated estimators can be achieved.

In \cref{fig:rs05}, we present spectral functions of this model at $r_s =0.5$
for two momenta. The first is at $\mathbf K = [0,0,0]$ which represents removal of an electron from the 
core shell of the UEG model. 
Core spectra have been shown to have rich satellite features
where different many body methods do not agree in terms of the precise satellite structure.\cite{McClain2016Jun}
However, such features are very simple in a small supercell like this one.
As can be seen in \cref{fig:rs05}(a), 
we only have two peaks from the Lanczos method.
Neither of these peaks corresponds to a single Koopmans-like excitation.
Namely, they cannot be found from a simple single ionization process that either HF (i.e., Koopmans theorem) or EKT1 describes well.
As a result of this, HF, EKT1-AFQMC, and EKT-HCI all fail to capture the peak split and only yield a single peak.
The correlation effect is very marginal in the sense that nearly no improvement was observed with the EKT1 methods compared to HF.

A significant improvement can be made by incorporating higher-order terms such as 2h1p excitations. 
In other words, core ionization satellite states (up to leading order) require excitations such as
\begin{equation}
(\sum_{\mathbf{K}' \ne \mathbf 0}C_{\mathbf{K}'}\hat{a}^\dagger_{\mathbf{K}'} \hat{a}_{\mathbf {K}'}) \hat{a}_{\mathbf K = \mathbf 0}
\end{equation}
where $C_{\mathbf{K}'}$ are the excitation amplitudes.
All of these excitations are included in EKT3.
EKT3-HCI can nearly completely reproduce the exact
spectral function despite the use of the cumulant approximation for the 4-RDM. The cumulant approximation error is small especially in weakly correlated cases such as $r_{s}=0.5$ where the connected component of 4-RDM is expected to be small.
In \cref{fig:rs05}(b), we emphasize that we observe meaningful improvements even from the EKT1 methods compared to HF. 
$\mathbf K = [0,0,2\pi/L]$ corresponds to the top of the valence band corresponding to the first IP of the UEG model.
There is no satellite peak visible at this momentum and the single peak found from the EKT1 methods is reasonable.
The peak location of HF is displaced by about +0.9 eV from the correct location while the EKT1 methods yield a peak shifted by about -0.6 eV. We note that EKT1-AFQMC is practically indistinguishable from EKT1-HCI for both momenta which 
indicates a small phaseless bias, a small back-propagation error, and good statistics for the estimators.

A similar conclusion can be drawn for $r_s = 4$ as shown in \cref{fig:rs4}.
For the core excitation spectrum in \cref{fig:rs4}(a), we observe the same split peak structure observed at $r_s = 0.5$.
We see another smaller peak emerging on the left shoulder of the peak near -12 eV. 
While EKT3-HCI is no longer exact, 
it reproduces most of the features in the exact spectral function including the emergence of the third peak.
While EKT1-HCI and EKT1-AFQMC agree well, there is no visible improvement over HF.
The EKT1 methods all yield a single peak which is qualitatively wrong.
The valence excitation structure illustrated in \cref{fig:rs4}(b) is relatively featureless, but there are small peaks emerging
in the high energy (more negative) region of the spectrum. 
EKT3-HCI shows good agreement with Lanczos for the main quasiparticle peak and also
produces satellite features. 
There is a visible improvement of EKT1 approaches (with a deviation of the peak energy of approximately -0.25 eV) compared to HF (with an approximate deviation of +0.34 eV).
A slight deviation of EKT1-AFQMC from EKT1-HCI is observed, but the difference in the main quasiparticle peak location is only about 0.01 eV.

Overall, in this small benchmark study, the EKT1 approaches provide some improvement over HF 
for valence excitations and qualitatively fails for the core region. 
The agreement between the EKT1 valence peaks and the Lanczos peaks is not perfect, with 
an error of  about -0.6 eV for $r_s = 0.5$ and -0.25 eV for $r_s = 4$. 
However, we emphasize again that we expect EKT1 to become exact for the first IP as the complete basis set limit is approached. 
Finally, EKT1-AFQMC is able to reproduce EKT1-HCI nearly perfectly even for $r_s = 4$, where the phaseless error in the ground state energy
is about 0.185(2) eV. 
\subsubsection{54-electrons/287-planewaves}
Next, we consider a
larger UEG supercell (54 electrons in 287 planewaves)
where
obtaining many back-propagation samples
is difficult.
We study $r_s = 2$, where AFQMC can be reliably extrapolated to the basis set limit.\cite{lee_2019_UEG}
We produced 600 back-propagation samples with a back propagation time of 8 a.u. 
This amounts to a total of 4800 a.u. propagation time, which is a long propagation for this system size.
Our approach yielded a maximum statistical error in 1-RDM and Fock matrices of $4\times 10^{-3}$.
While this error is not small, the procedure described in \cref{sec:numdet}
was enough to stabilize the final results.
Unlike previous cases, we take the upper triangular part of the Fock matrix and explicitly symmetrize the Fock matrix. A linear dependency cutoff of $10^{-3}$ was used in EKT1-AFQMC.
It is difficult to generate highly accurate 1- and 2-RDMs from HCI for this system size,
so for this system we do not have an exact benchmark reference to compare to our EKT1-AFQMC results.
Similarly, EKT3-HCI is also intractable for this system size.
Instead, we have performed EOM-IP-CCSD and 
$\Delta$AFQMC to compare with and to gauge the magnitude of the errors of EKT1-AFQMC.

\begin{figure}
    \centering
    \includegraphics[scale=0.4]{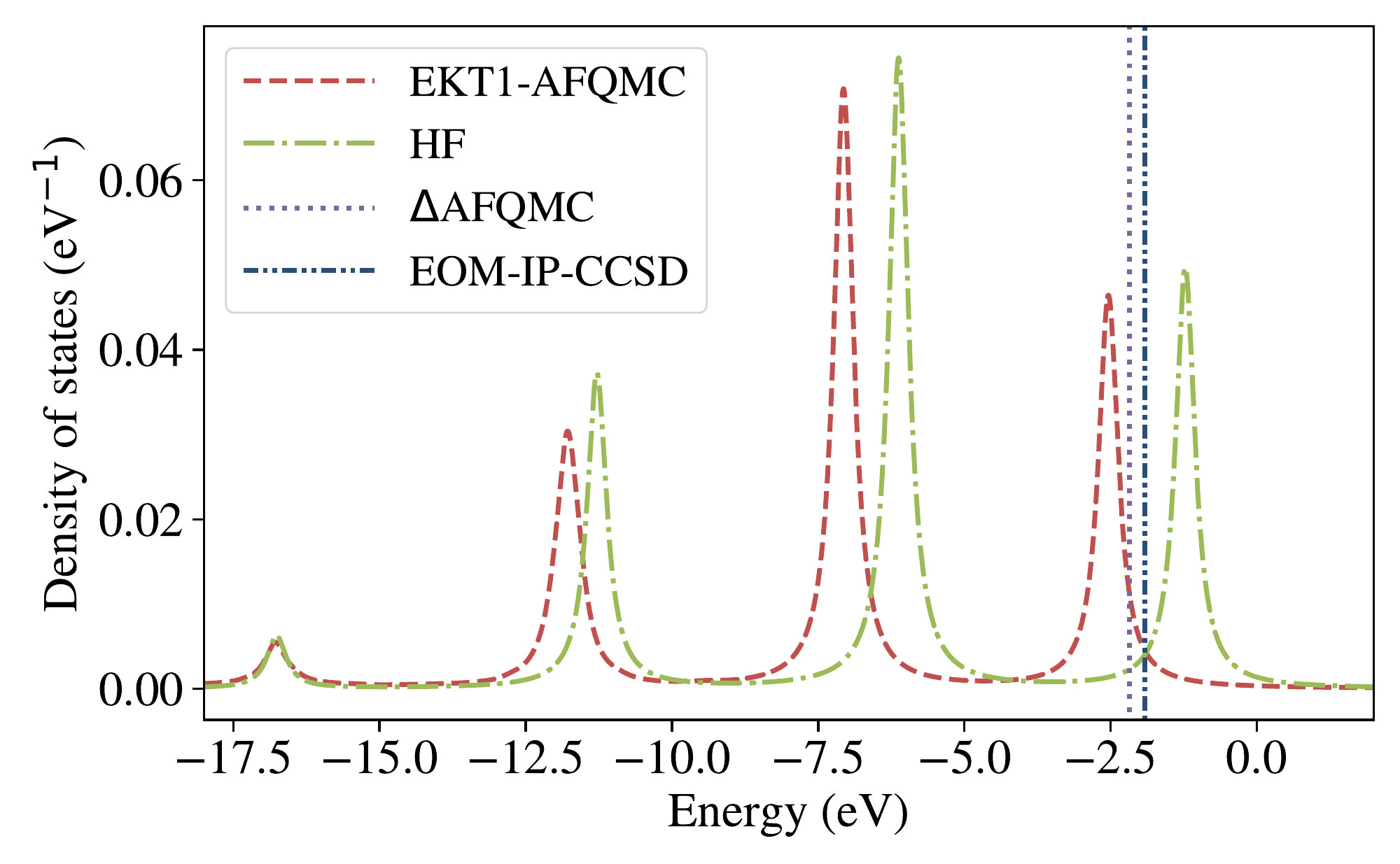}
    \caption{
Electron removal spectral functions of the 54-electron in 287-planewave UEG model at $r_s = 2$ from
EKT1-AFQMC and HF.
The first IPs from $\Delta$AFQMC and EOM-IP-CCSD are shown for comparisons.
A broadening parameter of 0.2 eV was used for all plots.
     }
\label{fig:54e}
\end{figure}
In \cref{fig:54e}, 
EKT1-AFQMC and HF spectral functions are shown.
As expected, EKT1-AFQMC does not show any satellite peaks at all
and EKT1-AFQMC only introduces a shift to the HF spectrum.
Correlation effects in the peak height do not appear large, but
the peak location changes by about 1 eV going from HF to EKT1-AFQMC. 
We also produced $\Delta$AFQMC and EOM-IP-CCSD for comparison.
The $\Delta$AFQMC IP is 2.18(1) eV, EOM-IP-CCSD yields 1.91 eV, and
EKT1-AFQMC gives 2.51 eV in this basis set. 
The deviation of EKT1-AFQMC from $\Delta$AFQMC is 0.33(1) eV whereas EOM-IP-CCSD deviates by 0.27(1) eV.
These deviations are similar in magnitude but with opposite signs.
The accuracy of EOM-IP-CCSD is unclear because
the ground state energy found from CCSD is higher than that of AFQMC by 0.0291(1) eV per electron.
In the basis set limit, AFQMC was found to be 
as accurate as state-of-the-art diffusion Monte Carlo for the ground state energy, differing  by 0.0088(9) eV per electron or less.\cite{lee_2019_UEG} 
This suggests that the ground state correlation energy of CCSD may be on the order of 0.03 eV per electron.
How much of this error is propagated to the EOM-IP-CCSD calculation remains unclear.
In the complete basis set limit,
we believe that the first IP from EKT1-AFQMC will become more accurate and closer to that of $\Delta$AFQMC.
A more complete comparison should be conducted in this limit, but such calculations are very difficult due to the need to procure many back-propagation samples in EKT1-AFQMC.

\subsection{Towards {\it ab-initio} solids: diamond at the $\Gamma$-point}
\begin{figure}
    \centering
    \includegraphics[scale=0.4]{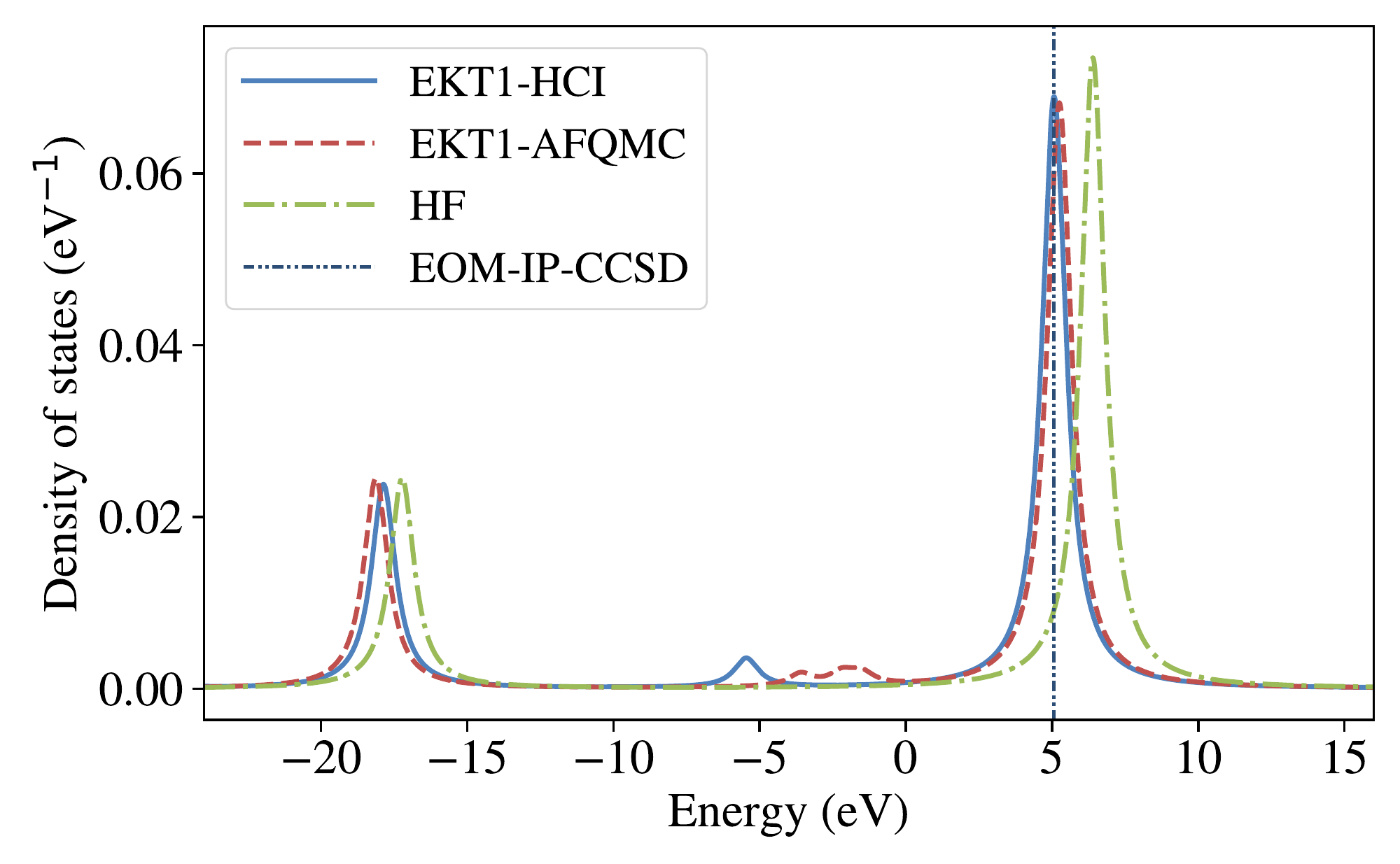}
    \caption{
EKT1-HCI and EKT1-AFQMC electron removal spectral functions of diamond at the $\Gamma$-point.
A broadening parameter $\eta = 0.5$ eV was used.
     }
\label{fig:diamond}
\end{figure}

With recent advances in
open-source software such as \texttt{PySCF},\cite{Sun2018Jan}
performing calculations on {\it ab-initio} solids
is relatively straightforward.
While an implementation of AFQMC with $\mathbf k$-points 
has been previously presented,\cite{Motta2019Jul,malone2020gpu}
we only present a $\Gamma$-point result in this work.
This is mainly because our current EKT1 implementation  
does not explicitly consider $\mathbf k$-points.
We chose to study diamond because it is one of the simplest solids just with two carbon atoms in its unit cell.
We used the GTH-PADE pseudopotential\cite{Goedecker1996Jul} and the GTH-DZVP basis set.\cite{VandeVondele2007Sep}

This system is overall as small as the smaller systems considered in this work.
Therefore, we could obtain over 6000 back-propagation samples
with a 4 a.u. back-propagation time. 
Even with these many samples, 
the largest error bar in the Fock and 1-RDM matrices was $6\times 10^{-3}$.
Due to this large statistical error in the matrix elements,
we used a linear dependency cutoff of 0.01
and symmetrized the Fock matrix by taking only the upper triangular part of it (see \cref{sec:numdet} for details).
We also included the shift by the Madelung constant in the spectral functions.

As shown in \cref{fig:diamond},
EKT1-AFQMC successfully reproduces 
EKT1-HCI.
A small quasiparticle peak near -5 eV
is not accurately captured by EKT1-AFQMC
largely due to statistical noise. 
Small peaks are difficult to resolve in EKT1-AFQMC without reducing the statistical errors on each matrix element further.
Nonetheless, two other larger quasiparticle peaks are well represented.
Both peaks are well reproduced within 0.25 eV from EKT1-HCI.
The improvement over HF is about 1 eV or so and
the first IP from EOM-IP-CCSD is within 0.02 eV of the EKT1-HCI result.
It will be interesting to revisit the assessment of EKT1-AFQMC in the basis set and thermodynamic limits
via direct comparisons to experiments.

\section{Conclusions}
In this work,
we have explored
the extended Koopmans' theorem (EKT) approach to the computation of
spectral functions via phaseless auxiliary-field quantum Monte Carlo (AFQMC).
Previous attempts in AFQMC to obtaining spectral functions have resorted to analytic continuation\cite{Motta2015Oct,VitaliDynamical2016} which has well-documented drawbacks.\cite{,Goulko2017Jan,DornheimDynamicalStructure2018}
The EKT approach is attractive because it requires neither an explicit representation of the ground state wavefunction nor analytic continuation to compute spectral functions. 
Instead, its only inputs are N-particle reduced density matrices (N-RDMs) which can be computed in AFQMC via the back-propagation algorithm.
The motivation of our work was thus to use the EKT approach with the aim of avoiding numerical problems arising in analytic continuation for the accurate assessment of real-frequency spectral information.
While many studies have so far focused on the simplest level of the EKT, the EKT approach is systematically improvable with increasing order of excitations: 1h, 1p2h, etc. for electron removal and 1p, 1h2p, etc. for electron addition. We presented the implementation of EKT1 (1h or 1p) and EKT3 (1p2h or 1h2p). For EKT3, we proposed the use of a cumulant approximation to the 4-RDM to avoid the steep storage requirements.

We produced preliminary results using EKT1 within AFQMC (EKT1-AFQMC) for small molecular systems, uniform electron gas (UEG) 14-electron and 54-electron supercells, and
diamond at the $\Gamma$-point. 
We focused on studying the first ionization potential (IP) and electron removal spectral functions of these systems.
By comparing numerically exact EKT1 results based on heat-bath configuration interaction (i.e., EKT1-HCI), we showed that
despite statistical noise, EKT1-AFQMC can capture most qualitative features of EKT1-HCI.
We provide a more detailed summary on our findings as follows:
\begin{enumerate}
\item In small molecular benchmarks within the aug-cc-pVDZ basis, we found the maximum deviation of EKT1-AFQMC from EKT1-HCI in the first IP to be 0.19 eV. These molecules have quite small phaseless biases in the ground state energy ($\le0.04$ eV) so we attributed additional biases to back-propagation. Electron removal spectral functions from EKT1-AFQMC look qualitatively similar to that of EKT1-HCI even in the least accurate case (\ce{N2}). 
\item For the 14-electron UEG supercell (19-planewave) benchmark, we observed a qualitative failure of EKT1 due to its inability to describe
satellite states at $\mathbf K = \mathbf 0$. We showed that EKT3 (within HCI) significantly improves this. 
Despite these failures of EKT1, we found EKT1-AFQMC to have peak locations that are nearly identical (within 0.01 eV) to EKT1-HCI for both $r_s = 0.5$ and $r_s = 4$. 
Given the noticeable phaseless bias at $r_s=4$, this result is quite encouraging. Lastly, for the valence region of the electron removal spectral function, we observed reasonable accuracy of EKT1 compared to the exact spectral function. The location of the first IP was off by 0.4 eV for $r_s=0.5$ and 0.25 eV for $r_s = 4.0$, which we expect to improve in larger bases.
\item For the 54-electron UEG supercell (257-planewave) benchmark, we could not obtain EKT1-HCI due computational expense. Therefore, we attempted to assess the accuracy of EKT1-AFQMC by comparing the first IP of EKT1-AFQMC with that of equation-of-motion IP coupled-cluster with singles and doubles (EOM-IP-CCSD) and $\Delta$-AFQMC. However, all three methods differ from each other by more than 0.25 eV and a more thorough benchmark in the basis set limit is highly desirable.
\item For diamond at the $\Gamma$-point, EKT1-AFQMC produced a qualitatively correct electron removal spectral function which agrees well with EKT1-HCI.
However, EKT1-AFQMC peak locations were off by 0.2 eV from those of EKT1-HCI. We also noted that EKT1-HCI first IP agrees with that of EOM-IP-CCSD within 0.01 eV. 
\end{enumerate}

While a more extensive benchmark study is highly desirable, we cautiously
conclude that EKT1-AFQMC is useful for
charge excitations that are heavily dominated by Koopmans-like excitations.
EKT1-AFQMC errors in peak locations can be as large as 0.25 eV compared to EKT1-HCI, but the line shapes of EKT1-AFQMC closely follow those of EKT1-HCI in all systems considered in this work.

The greatest challenge of EKT1-AFQMC is currently the statistical inefficiency in obtaining relevant back-propagated quantities with error bars small enough to enable the construction of stable EKT1-AFQMC spectral functions. 
Future work must first be dedicated to improving the statistical efficiency of back-propagation.
Furthermore, better back-propagation algorithms are needed to reduce the
back-propagation error further.
A practical implementation of EKT3-AFQMC using an iterative eigenvalue solver will be an interesting topic to explore in the future.
Several interesting extensions are immediately possible.
First, extending the EKT framework to neutral excitations\cite{Pavlyukh2018,Pavlyukh2019} is relatively straightforward and could be interesting to explore. Next, the extension of the EKT framework for finite-temperature coupled electron-phonon problems would provide a way to compute temperature-dependent vibronic spectra directly from AFQMC.\cite{Lee2020Dec,Lee2020Dec2}
We also leave the comparison of these EKT-based spectral functions to analytically continued spectral functions for a future study.

\section{Acknowledgements}
We thank Sandeep Sharma for providing access to a version of \texttt{Dice} with the 3-RDM and 4-RDM capability which was used in testing and producing EKT3-HCI results presented in this work. We also thank Garnet Chan for discussion about finite-size corrections to ionization energies. DRR acknowledges support of NSF CHE- 1954791.
The work of FDM and MAM was performed under the auspices of the U.S. Department of Energy
(DOE) by LLNL under Contract No. DE-AC52-07NA27344 and was supported by 
the U.S. DOE, Office of Science, Basic Energy Sciences, Materials Sciences and
Engineering Division, as part of the Computational Materials Sciences Program
and Center for Predictive Simulation of Functional Materials (CPSFM).  Computing support for this work came from the LLNL Institutional Computing Grand Challenge program.

\appendix{}
\section{Numerical Details of EKT}\label{sec:numdet}
Determining the eigenvalues and eigenvectors of \cref{eq:gen_eig} from noisy QMC density matrices is non-trivial.
We first diagonalise the metric matrices $S_{\pm}$,
\begin{equation}
S_{\pm} = \mathbf U
\Lambda_\pm \mathbf U^\dagger,
\end{equation}
 and discard eigenvalues below a given threshold.
 In this work, unless noted otherwise, in the main text we used $10^{-4}$ for all EKT1-AFQMC results, which is on the order of the largest statistical error in the EKT1 Fock matrix. 
We next construct the transformation matrix
\begin{equation}
	\mathbf{X}_{\pm} = \mathbf{U}\mathbf{\Lambda_\pm^{-1/2}}.
\end{equation}
This procedure is often referred to as {\it canonical} orthogonalization in quantum chemistry.
Then, we solve
\begin{equation}
	\tilde{\mathbf{F}}_{\pm} \tilde{\mathbf{c}}^\nu_\pm = \epsilon_{\pm}^{\nu} \tilde{\mathbf{c}}^{\nu}_\pm ,
\end{equation}
where
\begin{equation}
\tilde{\mathbf{F}}_{\pm} = \mathbf{X}_{\pm}{^\dagger} \mathbf{F}_{\pm} \mathbf{X}_{\pm}.
\end{equation}
Finally the eigenvectors in the original basis can be determined from
\begin{equation}
	(\mathbf{c}^{\nu}_\pm)_p = \sum_{I} (\mathbf{X}_{\pm})_{pI} (\tilde{\mathbf{c}^{\nu}}_\pm)_I.
\end{equation}
Following Kent {\em et al.},\cite{Kent1998} 
we explicitly zero out all matrix elements whose magnitude is smaller than two times the corresponding statistical error bar.
We explicitly symmetrize RDMs, but leave the Fock matrix asymmetric as required for approximate wavefunctions.
However, for the more difficult problems considered in this work, such as the 54-electron UEG electron
and diamond, we found that symmetrizing the Fock matrix is useful so we choose to symmetrize the Fock matrix in such cases (by taking only the upper triangular part of the Fock matrix).
These steps improved the numerical stability of the eigenvalue problem. 

We also note that there are generic numerical issues arising in EKT even without any statistical sampling error.
This was observed in both EKT1-HCI and EKT3-HCI, where spurious solutions with large negative IPs appear. 
These states stem from the fact that the metric matrix in EKT problems are generally low-rank, as they are related to RDMs. 
For instance, in the EKT1 formulation, the metric we diagonalize for the IP problem is a 1-RDM whose rank is not so much larger than the number of electrons in the system.
These spurious states can be removed with larger cutoffs while often affecting peak locations of quasiparticle states.
Interestingly, these spurious states all carry negligible spectral weights and do not appear in spectra.
Motivated by this observation, the most satisfying solution we found was to use as a small threshold as possible and to measure the overlap between Koopmans states and eigenvectors to identify quasiparticle-like eigenvectors.
This was enough to identify physical IP excitations that are quasi-particle-like. The same principle is applicable to EKT3-HCI and also the EA calculations.

Spectral functions are plotted by approximating the $\delta$-function in the spectral function expression as
a Lorentzian function
\begin{equation}
\delta(\omega) \simeq \frac{1}\pi \left(\frac{\eta}{\omega^2+\eta^2}\right)
\end{equation}
for some small constant $\eta$. 
To ensure reproducibility, we specified the value of $\eta$ in all relevant figures.
%
%Discuss numerical issues, discarded eigenvalues, error estimation etc.
%
%Dirac deltas are evaluated numerically as Lorentzians with $\eta=0.05$ Ha.
%\begin{equation}
%
%\end{equation}

\bibliography{refs}

%merlin.mbs apsrev4-1.bst 2010-07-25 4.21a (PWD, AO, DPC) hacked
%Control: key (0)
%Control: author (8) initials jnrlst
%Control: editor formatted (1) identically to author
%Control: production of article title (-1) disabled
%Control: page (0) single
%Control: year (1) truncated
%Control: production of eprint (0) enabled
\begin{thebibliography}{148}%
\makeatletter
\providecommand \@ifxundefined [1]{%
 \@ifx{#1\undefined}
}%
\providecommand \@ifnum [1]{%
 \ifnum #1\expandafter \@firstoftwo
 \else \expandafter \@secondoftwo
 \fi
}%
\providecommand \@ifx [1]{%
 \ifx #1\expandafter \@firstoftwo
 \else \expandafter \@secondoftwo
 \fi
}%
\providecommand \natexlab [1]{#1}%
\providecommand \enquote  [1]{``#1''}%
\providecommand \bibnamefont  [1]{#1}%
\providecommand \bibfnamefont [1]{#1}%
\providecommand \citenamefont [1]{#1}%
\providecommand \href@noop [0]{\@secondoftwo}%
\providecommand \href [0]{\begingroup \@sanitize@url \@href}%
\providecommand \@href[1]{\@@startlink{#1}\@@href}%
\providecommand \@@href[1]{\endgroup#1\@@endlink}%
\providecommand \@sanitize@url [0]{\catcode `\\12\catcode `\$12\catcode
  `\&12\catcode `\#12\catcode `\^12\catcode `\_12\catcode `\%12\relax}%
\providecommand \@@startlink[1]{}%
\providecommand \@@endlink[0]{}%
\providecommand \url  [0]{\begingroup\@sanitize@url \@url }%
\providecommand \@url [1]{\endgroup\@href {#1}{\urlprefix }}%
\providecommand \urlprefix  [0]{URL }%
\providecommand \Eprint [0]{\href }%
\providecommand \doibase [0]{http://dx.doi.org/}%
\providecommand \selectlanguage [0]{\@gobble}%
\providecommand \bibinfo  [0]{\@secondoftwo}%
\providecommand \bibfield  [0]{\@secondoftwo}%
\providecommand \translation [1]{[#1]}%
\providecommand \BibitemOpen [0]{}%
\providecommand \bibitemStop [0]{}%
\providecommand \bibitemNoStop [0]{.\EOS\space}%
\providecommand \EOS [0]{\spacefactor3000\relax}%
\providecommand \BibitemShut  [1]{\csname bibitem#1\endcsname}%
\let\auto@bib@innerbib\@empty
%</preamble>
\bibitem [{\citenamefont {Lu}\ \emph {et~al.}(2012)\citenamefont {Lu},
  \citenamefont {Vishik}, \citenamefont {Yi}, \citenamefont {Chen},
  \citenamefont {Moore},\ and\ \citenamefont {Shen}}]{lu2012angle}%
  \BibitemOpen
  \bibfield  {author} {\bibinfo {author} {\bibfnamefont {D.}~\bibnamefont
  {Lu}}, \bibinfo {author} {\bibfnamefont {I.~M.}\ \bibnamefont {Vishik}},
  \bibinfo {author} {\bibfnamefont {M.}~\bibnamefont {Yi}}, \bibinfo {author}
  {\bibfnamefont {Y.}~\bibnamefont {Chen}}, \bibinfo {author} {\bibfnamefont
  {R.~G.}\ \bibnamefont {Moore}}, \ and\ \bibinfo {author} {\bibfnamefont
  {Z.-X.}\ \bibnamefont {Shen}},\ }\href@noop {} {\bibfield  {journal}
  {\bibinfo  {journal} {Annu. Rev. Condens. Matter Phys.}\ }\textbf {\bibinfo
  {volume} {3}},\ \bibinfo {pages} {129} (\bibinfo {year} {2012})}\BibitemShut
  {NoStop}%
\bibitem [{\citenamefont {Egerton}(2008)}]{egerton2008electron}%
  \BibitemOpen
  \bibfield  {author} {\bibinfo {author} {\bibfnamefont {R.~F.}\ \bibnamefont
  {Egerton}},\ }\href@noop {} {\bibfield  {journal} {\bibinfo  {journal} {Rep.
  Prog. Phys.}\ }\textbf {\bibinfo {volume} {72}},\ \bibinfo {pages} {016502}
  (\bibinfo {year} {2008})}\BibitemShut {NoStop}%
\bibitem [{\citenamefont {Andreani}\ \emph {et~al.}(2005)\citenamefont
  {Andreani}, \citenamefont {Colognesi}, \citenamefont {Mayers}, \citenamefont
  {Reiter},\ and\ \citenamefont {Senesi}}]{andreani2005measurement}%
  \BibitemOpen
  \bibfield  {author} {\bibinfo {author} {\bibfnamefont {C.}~\bibnamefont
  {Andreani}}, \bibinfo {author} {\bibfnamefont {D.}~\bibnamefont {Colognesi}},
  \bibinfo {author} {\bibfnamefont {J.}~\bibnamefont {Mayers}}, \bibinfo
  {author} {\bibfnamefont {G.}~\bibnamefont {Reiter}}, \ and\ \bibinfo {author}
  {\bibfnamefont {R.}~\bibnamefont {Senesi}},\ }\href@noop {} {\bibfield
  {journal} {\bibinfo  {journal} {Adv. Phys.}\ }\textbf {\bibinfo {volume}
  {54}},\ \bibinfo {pages} {377} (\bibinfo {year} {2005})}\BibitemShut
  {NoStop}%
\bibitem [{\citenamefont {Fetter}\ and\ \citenamefont
  {Walecka}(2003)}]{alexanderfetter2003}%
  \BibitemOpen
  \bibfield  {author} {\bibinfo {author} {\bibfnamefont {A.~L.}\ \bibnamefont
  {Fetter}}\ and\ \bibinfo {author} {\bibfnamefont {J.~D.}\ \bibnamefont
  {Walecka}},\ }\href {https://www.xarg.org/ref/a/0486428273/} {\emph {\bibinfo
  {title} {Quantum Theory of Many-Particle Systems (Dover Books on Physics)}}}\
  (\bibinfo  {publisher} {Dover Publications},\ \bibinfo {year}
  {2003})\BibitemShut {NoStop}%
\bibitem [{\citenamefont {Onida}\ \emph {et~al.}(2002)\citenamefont {Onida},
  \citenamefont {Reining},\ and\ \citenamefont {Rubio}}]{onida2002electronic}%
  \BibitemOpen
  \bibfield  {author} {\bibinfo {author} {\bibfnamefont {G.}~\bibnamefont
  {Onida}}, \bibinfo {author} {\bibfnamefont {L.}~\bibnamefont {Reining}}, \
  and\ \bibinfo {author} {\bibfnamefont {A.}~\bibnamefont {Rubio}},\
  }\href@noop {} {\bibfield  {journal} {\bibinfo  {journal} {Rev. Mod. Phys.}\
  }\textbf {\bibinfo {volume} {74}},\ \bibinfo {pages} {601} (\bibinfo {year}
  {2002})}\BibitemShut {NoStop}%
\bibitem [{\citenamefont {Cederbaum}\ and\ \citenamefont
  {Domcke}(1977)}]{cederbaum1977theoretical}%
  \BibitemOpen
  \bibfield  {author} {\bibinfo {author} {\bibfnamefont {L.}~\bibnamefont
  {Cederbaum}}\ and\ \bibinfo {author} {\bibfnamefont {W.}~\bibnamefont
  {Domcke}},\ }\href@noop {} {\bibfield  {journal} {\bibinfo  {journal} {Adv.
  Chem. Phys}\ }\textbf {\bibinfo {volume} {36}},\ \bibinfo {pages} {205}
  (\bibinfo {year} {1977})}\BibitemShut {NoStop}%
\bibitem [{\citenamefont {Hedin}\ \emph {et~al.}(1998)\citenamefont {Hedin},
  \citenamefont {Michiels},\ and\ \citenamefont
  {Inglesfield}}]{hedin1998transition}%
  \BibitemOpen
  \bibfield  {author} {\bibinfo {author} {\bibfnamefont {L.}~\bibnamefont
  {Hedin}}, \bibinfo {author} {\bibfnamefont {J.}~\bibnamefont {Michiels}}, \
  and\ \bibinfo {author} {\bibfnamefont {J.}~\bibnamefont {Inglesfield}},\
  }\href@noop {} {\bibfield  {journal} {\bibinfo  {journal} {Phys. Rev. B}\
  }\textbf {\bibinfo {volume} {58}},\ \bibinfo {pages} {15565} (\bibinfo {year}
  {1998})}\BibitemShut {NoStop}%
\bibitem [{\citenamefont {Duchemin}\ and\ \citenamefont
  {Blase}(2020)}]{duchemin2020robust}%
  \BibitemOpen
  \bibfield  {author} {\bibinfo {author} {\bibfnamefont {I.}~\bibnamefont
  {Duchemin}}\ and\ \bibinfo {author} {\bibfnamefont {X.}~\bibnamefont
  {Blase}},\ }\href@noop {} {\bibfield  {journal} {\bibinfo  {journal} {J.
  Chem. Theory Comput.}\ }\textbf {\bibinfo {volume} {16}},\ \bibinfo {pages}
  {1742} (\bibinfo {year} {2020})}\BibitemShut {NoStop}%
\bibitem [{\citenamefont {Hybertsen}\ and\ \citenamefont
  {Louie}(1986)}]{hybertsen1986electron}%
  \BibitemOpen
  \bibfield  {author} {\bibinfo {author} {\bibfnamefont {M.~S.}\ \bibnamefont
  {Hybertsen}}\ and\ \bibinfo {author} {\bibfnamefont {S.~G.}\ \bibnamefont
  {Louie}},\ }\href@noop {} {\bibfield  {journal} {\bibinfo  {journal} {Phys.
  Rev. B}\ }\textbf {\bibinfo {volume} {34}},\ \bibinfo {pages} {5390}
  (\bibinfo {year} {1986})}\BibitemShut {NoStop}%
\bibitem [{\citenamefont {Hedin}(1965)}]{hedin1965new}%
  \BibitemOpen
  \bibfield  {author} {\bibinfo {author} {\bibfnamefont {L.}~\bibnamefont
  {Hedin}},\ }\href@noop {} {\bibfield  {journal} {\bibinfo  {journal} {Phys.
  Rev.}\ }\textbf {\bibinfo {volume} {139}},\ \bibinfo {pages} {A796} (\bibinfo
  {year} {1965})}\BibitemShut {NoStop}%
\bibitem [{\citenamefont {Reining}(2018)}]{reining2018gw}%
  \BibitemOpen
  \bibfield  {author} {\bibinfo {author} {\bibfnamefont {L.}~\bibnamefont
  {Reining}},\ }\href@noop {} {\bibfield  {journal} {\bibinfo  {journal} {Wiley
  Interdiscip. Rev. Comput. Mol. Sci.}\ }\textbf {\bibinfo {volume} {8}},\
  \bibinfo {pages} {e1344} (\bibinfo {year} {2018})}\BibitemShut {NoStop}%
\bibitem [{\citenamefont {Rinke}\ \emph {et~al.}(2005)\citenamefont {Rinke},
  \citenamefont {Qteish}, \citenamefont {Neugebauer}, \citenamefont
  {Freysoldt},\ and\ \citenamefont {Scheffler}}]{rinke2005combining}%
  \BibitemOpen
  \bibfield  {author} {\bibinfo {author} {\bibfnamefont {P.}~\bibnamefont
  {Rinke}}, \bibinfo {author} {\bibfnamefont {A.}~\bibnamefont {Qteish}},
  \bibinfo {author} {\bibfnamefont {J.}~\bibnamefont {Neugebauer}}, \bibinfo
  {author} {\bibfnamefont {C.}~\bibnamefont {Freysoldt}}, \ and\ \bibinfo
  {author} {\bibfnamefont {M.}~\bibnamefont {Scheffler}},\ }\href@noop {}
  {\bibfield  {journal} {\bibinfo  {journal} {New J. Phys.}\ }\textbf {\bibinfo
  {volume} {7}},\ \bibinfo {pages} {126} (\bibinfo {year} {2005})}\BibitemShut
  {NoStop}%
\bibitem [{\citenamefont {Fuchs}\ \emph {et~al.}(2007)\citenamefont {Fuchs},
  \citenamefont {Furthm{\"u}ller}, \citenamefont {Bechstedt}, \citenamefont
  {Shishkin},\ and\ \citenamefont {Kresse}}]{fuchs2007quasiparticle}%
  \BibitemOpen
  \bibfield  {author} {\bibinfo {author} {\bibfnamefont {F.}~\bibnamefont
  {Fuchs}}, \bibinfo {author} {\bibfnamefont {J.}~\bibnamefont
  {Furthm{\"u}ller}}, \bibinfo {author} {\bibfnamefont {F.}~\bibnamefont
  {Bechstedt}}, \bibinfo {author} {\bibfnamefont {M.}~\bibnamefont {Shishkin}},
  \ and\ \bibinfo {author} {\bibfnamefont {G.}~\bibnamefont {Kresse}},\
  }\href@noop {} {\bibfield  {journal} {\bibinfo  {journal} {Phys. Rev. B}\
  }\textbf {\bibinfo {volume} {76}},\ \bibinfo {pages} {115109} (\bibinfo
  {year} {2007})}\BibitemShut {NoStop}%
\bibitem [{\citenamefont {Marom}\ \emph {et~al.}(2012)\citenamefont {Marom},
  \citenamefont {Caruso}, \citenamefont {Ren}, \citenamefont {Hofmann},
  \citenamefont {K{\"o}rzd{\"o}rfer}, \citenamefont {Chelikowsky},
  \citenamefont {Rubio}, \citenamefont {Scheffler},\ and\ \citenamefont
  {Rinke}}]{marom2012benchmark}%
  \BibitemOpen
  \bibfield  {author} {\bibinfo {author} {\bibfnamefont {N.}~\bibnamefont
  {Marom}}, \bibinfo {author} {\bibfnamefont {F.}~\bibnamefont {Caruso}},
  \bibinfo {author} {\bibfnamefont {X.}~\bibnamefont {Ren}}, \bibinfo {author}
  {\bibfnamefont {O.~T.}\ \bibnamefont {Hofmann}}, \bibinfo {author}
  {\bibfnamefont {T.}~\bibnamefont {K{\"o}rzd{\"o}rfer}}, \bibinfo {author}
  {\bibfnamefont {J.~R.}\ \bibnamefont {Chelikowsky}}, \bibinfo {author}
  {\bibfnamefont {A.}~\bibnamefont {Rubio}}, \bibinfo {author} {\bibfnamefont
  {M.}~\bibnamefont {Scheffler}}, \ and\ \bibinfo {author} {\bibfnamefont
  {P.}~\bibnamefont {Rinke}},\ }\href@noop {} {\bibfield  {journal} {\bibinfo
  {journal} {Phys. Rev. B}\ }\textbf {\bibinfo {volume} {86}},\ \bibinfo
  {pages} {245127} (\bibinfo {year} {2012})}\BibitemShut {NoStop}%
\bibitem [{\citenamefont {Bruneval}(2012)}]{bruneval2012ionization}%
  \BibitemOpen
  \bibfield  {author} {\bibinfo {author} {\bibfnamefont {F.}~\bibnamefont
  {Bruneval}},\ }\href@noop {} {\bibfield  {journal} {\bibinfo  {journal} {J.
  Chem. Phys.}\ }\textbf {\bibinfo {volume} {136}},\ \bibinfo {pages} {194107}
  (\bibinfo {year} {2012})}\BibitemShut {NoStop}%
\bibitem [{\citenamefont {Langreth}(1970)}]{langreth1970singularities}%
  \BibitemOpen
  \bibfield  {author} {\bibinfo {author} {\bibfnamefont {D.~C.}\ \bibnamefont
  {Langreth}},\ }\href@noop {} {\bibfield  {journal} {\bibinfo  {journal}
  {Phys. Rev. B}\ }\textbf {\bibinfo {volume} {1}},\ \bibinfo {pages} {471}
  (\bibinfo {year} {1970})}\BibitemShut {NoStop}%
\bibitem [{\citenamefont {Aryasetiawan}\ \emph {et~al.}(1996)\citenamefont
  {Aryasetiawan}, \citenamefont {Hedin},\ and\ \citenamefont
  {Karlsson}}]{aryasetiawan1996multiple}%
  \BibitemOpen
  \bibfield  {author} {\bibinfo {author} {\bibfnamefont {F.}~\bibnamefont
  {Aryasetiawan}}, \bibinfo {author} {\bibfnamefont {L.}~\bibnamefont {Hedin}},
  \ and\ \bibinfo {author} {\bibfnamefont {K.}~\bibnamefont {Karlsson}},\
  }\href@noop {} {\bibfield  {journal} {\bibinfo  {journal} {Phys. Rev. Lett.}\
  }\textbf {\bibinfo {volume} {77}},\ \bibinfo {pages} {2268} (\bibinfo {year}
  {1996})}\BibitemShut {NoStop}%
\bibitem [{\citenamefont {Guzzo}\ \emph {et~al.}(2011)\citenamefont {Guzzo},
  \citenamefont {Lani}, \citenamefont {Sottile}, \citenamefont {Romaniello},
  \citenamefont {Gatti}, \citenamefont {Kas}, \citenamefont {Rehr},
  \citenamefont {Silly}, \citenamefont {Sirotti},\ and\ \citenamefont
  {Reining}}]{guzzo2011valence}%
  \BibitemOpen
  \bibfield  {author} {\bibinfo {author} {\bibfnamefont {M.}~\bibnamefont
  {Guzzo}}, \bibinfo {author} {\bibfnamefont {G.}~\bibnamefont {Lani}},
  \bibinfo {author} {\bibfnamefont {F.}~\bibnamefont {Sottile}}, \bibinfo
  {author} {\bibfnamefont {P.}~\bibnamefont {Romaniello}}, \bibinfo {author}
  {\bibfnamefont {M.}~\bibnamefont {Gatti}}, \bibinfo {author} {\bibfnamefont
  {J.~J.}\ \bibnamefont {Kas}}, \bibinfo {author} {\bibfnamefont {J.~J.}\
  \bibnamefont {Rehr}}, \bibinfo {author} {\bibfnamefont {M.~G.}\ \bibnamefont
  {Silly}}, \bibinfo {author} {\bibfnamefont {F.}~\bibnamefont {Sirotti}}, \
  and\ \bibinfo {author} {\bibfnamefont {L.}~\bibnamefont {Reining}},\ }\href
  {\doibase 10.1103/PhysRevLett.107.166401} {\bibfield  {journal} {\bibinfo
  {journal} {Phys. Rev. Lett.}\ }\textbf {\bibinfo {volume} {107}},\ \bibinfo
  {pages} {166401} (\bibinfo {year} {2011})}\BibitemShut {NoStop}%
\bibitem [{\citenamefont {Lischner}\ \emph {et~al.}(2013)\citenamefont
  {Lischner}, \citenamefont {Vigil-Fowler},\ and\ \citenamefont
  {Louie}}]{lischner2013physical}%
  \BibitemOpen
  \bibfield  {author} {\bibinfo {author} {\bibfnamefont {J.}~\bibnamefont
  {Lischner}}, \bibinfo {author} {\bibfnamefont {D.}~\bibnamefont
  {Vigil-Fowler}}, \ and\ \bibinfo {author} {\bibfnamefont {S.~G.}\
  \bibnamefont {Louie}},\ }\href {\doibase 10.1103/PhysRevLett.110.146801}
  {\bibfield  {journal} {\bibinfo  {journal} {Phys. Rev. Lett.}\ }\textbf
  {\bibinfo {volume} {110}},\ \bibinfo {pages} {146801} (\bibinfo {year}
  {2013})}\BibitemShut {NoStop}%
\bibitem [{\citenamefont {Stan}\ \emph {et~al.}(2009)\citenamefont {Stan},
  \citenamefont {Dahlen},\ and\ \citenamefont {Van~Leeuwen}}]{stan2009levels}%
  \BibitemOpen
  \bibfield  {author} {\bibinfo {author} {\bibfnamefont {A.}~\bibnamefont
  {Stan}}, \bibinfo {author} {\bibfnamefont {N.~E.}\ \bibnamefont {Dahlen}}, \
  and\ \bibinfo {author} {\bibfnamefont {R.}~\bibnamefont {Van~Leeuwen}},\
  }\href@noop {} {\bibfield  {journal} {\bibinfo  {journal} {J. Chem. Phys.}\
  }\textbf {\bibinfo {volume} {130}},\ \bibinfo {pages} {114105} (\bibinfo
  {year} {2009})}\BibitemShut {NoStop}%
\bibitem [{\citenamefont {von Barth}\ and\ \citenamefont
  {Holm}(1996)}]{von1996self}%
  \BibitemOpen
  \bibfield  {author} {\bibinfo {author} {\bibfnamefont {U.}~\bibnamefont {von
  Barth}}\ and\ \bibinfo {author} {\bibfnamefont {B.}~\bibnamefont {Holm}},\
  }\href@noop {} {\bibfield  {journal} {\bibinfo  {journal} {Phys. Rev. B}\
  }\textbf {\bibinfo {volume} {54}},\ \bibinfo {pages} {8411} (\bibinfo {year}
  {1996})}\BibitemShut {NoStop}%
\bibitem [{\citenamefont {Holm}\ and\ \citenamefont {von
  Barth}(1998)}]{holm1998fully}%
  \BibitemOpen
  \bibfield  {author} {\bibinfo {author} {\bibfnamefont {B.}~\bibnamefont
  {Holm}}\ and\ \bibinfo {author} {\bibfnamefont {U.}~\bibnamefont {von
  Barth}},\ }\href@noop {} {\bibfield  {journal} {\bibinfo  {journal} {Phys.
  Rev. B}\ }\textbf {\bibinfo {volume} {57}},\ \bibinfo {pages} {2108}
  (\bibinfo {year} {1998})}\BibitemShut {NoStop}%
\bibitem [{\citenamefont {Faleev}\ \emph {et~al.}(2004)\citenamefont {Faleev},
  \citenamefont {Van~Schilfgaarde},\ and\ \citenamefont
  {Kotani}}]{faleev2004all}%
  \BibitemOpen
  \bibfield  {author} {\bibinfo {author} {\bibfnamefont {S.~V.}\ \bibnamefont
  {Faleev}}, \bibinfo {author} {\bibfnamefont {M.}~\bibnamefont
  {Van~Schilfgaarde}}, \ and\ \bibinfo {author} {\bibfnamefont
  {T.}~\bibnamefont {Kotani}},\ }\href@noop {} {\bibfield  {journal} {\bibinfo
  {journal} {Phys. Rev. Lett.}\ }\textbf {\bibinfo {volume} {93}},\ \bibinfo
  {pages} {126406} (\bibinfo {year} {2004})}\BibitemShut {NoStop}%
\bibitem [{\citenamefont {van Schilfgaarde}\ \emph {et~al.}(2006)\citenamefont
  {van Schilfgaarde}, \citenamefont {Kotani},\ and\ \citenamefont
  {Faleev}}]{van2006quasiparticle}%
  \BibitemOpen
  \bibfield  {author} {\bibinfo {author} {\bibfnamefont {M.}~\bibnamefont {van
  Schilfgaarde}}, \bibinfo {author} {\bibfnamefont {T.}~\bibnamefont {Kotani}},
  \ and\ \bibinfo {author} {\bibfnamefont {S.}~\bibnamefont {Faleev}},\
  }\href@noop {} {\bibfield  {journal} {\bibinfo  {journal} {Phys. Rev. Lett.}\
  }\textbf {\bibinfo {volume} {96}},\ \bibinfo {pages} {226402} (\bibinfo
  {year} {2006})}\BibitemShut {NoStop}%
\bibitem [{\citenamefont {Shishkin}\ and\ \citenamefont
  {Kresse}(2007)}]{shishkin2007self}%
  \BibitemOpen
  \bibfield  {author} {\bibinfo {author} {\bibfnamefont {M.}~\bibnamefont
  {Shishkin}}\ and\ \bibinfo {author} {\bibfnamefont {G.}~\bibnamefont
  {Kresse}},\ }\href@noop {} {\bibfield  {journal} {\bibinfo  {journal} {Phys.
  Rev. B}\ }\textbf {\bibinfo {volume} {75}},\ \bibinfo {pages} {235102}
  (\bibinfo {year} {2007})}\BibitemShut {NoStop}%
\bibitem [{\citenamefont {Rostgaard}\ \emph {et~al.}(2010)\citenamefont
  {Rostgaard}, \citenamefont {Jacobsen},\ and\ \citenamefont
  {Thygesen}}]{rostgaard2010fully}%
  \BibitemOpen
  \bibfield  {author} {\bibinfo {author} {\bibfnamefont {C.}~\bibnamefont
  {Rostgaard}}, \bibinfo {author} {\bibfnamefont {K.~W.}\ \bibnamefont
  {Jacobsen}}, \ and\ \bibinfo {author} {\bibfnamefont {K.~S.}\ \bibnamefont
  {Thygesen}},\ }\href@noop {} {\bibfield  {journal} {\bibinfo  {journal}
  {Phys. Rev. B}\ }\textbf {\bibinfo {volume} {81}},\ \bibinfo {pages} {085103}
  (\bibinfo {year} {2010})}\BibitemShut {NoStop}%
\bibitem [{\citenamefont {Koval}\ \emph {et~al.}(2014)\citenamefont {Koval},
  \citenamefont {Foerster},\ and\ \citenamefont
  {S{\'a}nchez-Portal}}]{koval2014fully}%
  \BibitemOpen
  \bibfield  {author} {\bibinfo {author} {\bibfnamefont {P.}~\bibnamefont
  {Koval}}, \bibinfo {author} {\bibfnamefont {D.}~\bibnamefont {Foerster}}, \
  and\ \bibinfo {author} {\bibfnamefont {D.}~\bibnamefont
  {S{\'a}nchez-Portal}},\ }\href@noop {} {\bibfield  {journal} {\bibinfo
  {journal} {Phys. Rev. B}\ }\textbf {\bibinfo {volume} {89}},\ \bibinfo
  {pages} {155417} (\bibinfo {year} {2014})}\BibitemShut {NoStop}%
\bibitem [{\citenamefont {Bobbert}\ and\ \citenamefont
  {Van~Haeringen}(1994)}]{bobbert1994lowest}%
  \BibitemOpen
  \bibfield  {author} {\bibinfo {author} {\bibfnamefont {P.}~\bibnamefont
  {Bobbert}}\ and\ \bibinfo {author} {\bibfnamefont {W.}~\bibnamefont
  {Van~Haeringen}},\ }\href@noop {} {\bibfield  {journal} {\bibinfo  {journal}
  {Phys. Rev. B}\ }\textbf {\bibinfo {volume} {49}},\ \bibinfo {pages} {10326}
  (\bibinfo {year} {1994})}\BibitemShut {NoStop}%
\bibitem [{\citenamefont {Schindlmayr}\ and\ \citenamefont
  {Godby}(1998)}]{schindlmayr1998systematic}%
  \BibitemOpen
  \bibfield  {author} {\bibinfo {author} {\bibfnamefont {A.}~\bibnamefont
  {Schindlmayr}}\ and\ \bibinfo {author} {\bibfnamefont {R.~W.}\ \bibnamefont
  {Godby}},\ }\href@noop {} {\bibfield  {journal} {\bibinfo  {journal} {Phys.
  Rev. Lett.}\ }\textbf {\bibinfo {volume} {80}},\ \bibinfo {pages} {1702}
  (\bibinfo {year} {1998})}\BibitemShut {NoStop}%
\bibitem [{\citenamefont {Romaniello}\ \emph {et~al.}(2009)\citenamefont
  {Romaniello}, \citenamefont {Guyot},\ and\ \citenamefont
  {Reining}}]{romaniello2009self}%
  \BibitemOpen
  \bibfield  {author} {\bibinfo {author} {\bibfnamefont {P.}~\bibnamefont
  {Romaniello}}, \bibinfo {author} {\bibfnamefont {S.}~\bibnamefont {Guyot}}, \
  and\ \bibinfo {author} {\bibfnamefont {L.}~\bibnamefont {Reining}},\
  }\href@noop {} {\bibfield  {journal} {\bibinfo  {journal} {J. Chem. Phys.}\
  }\textbf {\bibinfo {volume} {131}},\ \bibinfo {pages} {154111} (\bibinfo
  {year} {2009})}\BibitemShut {NoStop}%
\bibitem [{\citenamefont {Chen}\ and\ \citenamefont
  {Pasquarello}(2015)}]{chen2015accurate}%
  \BibitemOpen
  \bibfield  {author} {\bibinfo {author} {\bibfnamefont {W.}~\bibnamefont
  {Chen}}\ and\ \bibinfo {author} {\bibfnamefont {A.}~\bibnamefont
  {Pasquarello}},\ }\href@noop {} {\bibfield  {journal} {\bibinfo  {journal}
  {Phys. Rev. B}\ }\textbf {\bibinfo {volume} {92}},\ \bibinfo {pages} {041115}
  (\bibinfo {year} {2015})}\BibitemShut {NoStop}%
\bibitem [{\citenamefont {Maggio}\ and\ \citenamefont
  {Kresse}(2017)}]{maggio2017gw}%
  \BibitemOpen
  \bibfield  {author} {\bibinfo {author} {\bibfnamefont {E.}~\bibnamefont
  {Maggio}}\ and\ \bibinfo {author} {\bibfnamefont {G.}~\bibnamefont
  {Kresse}},\ }\href@noop {} {\bibfield  {journal} {\bibinfo  {journal} {J.
  Chem. Theory Comput.}\ }\textbf {\bibinfo {volume} {13}},\ \bibinfo {pages}
  {4765} (\bibinfo {year} {2017})}\BibitemShut {NoStop}%
\bibitem [{\citenamefont {Holm}\ and\ \citenamefont
  {Aryasetiawan}(1997)}]{holm1997self}%
  \BibitemOpen
  \bibfield  {author} {\bibinfo {author} {\bibfnamefont {B.}~\bibnamefont
  {Holm}}\ and\ \bibinfo {author} {\bibfnamefont {F.}~\bibnamefont
  {Aryasetiawan}},\ }\href {\doibase 10.1103/PhysRevB.56.12825} {\bibfield
  {journal} {\bibinfo  {journal} {Phys. Rev. B}\ }\textbf {\bibinfo {volume}
  {56}},\ \bibinfo {pages} {12825} (\bibinfo {year} {1997})}\BibitemShut
  {NoStop}%
\bibitem [{\citenamefont {Kas}\ \emph {et~al.}(2014)\citenamefont {Kas},
  \citenamefont {Rehr},\ and\ \citenamefont {Reining}}]{kas2014cumulant}%
  \BibitemOpen
  \bibfield  {author} {\bibinfo {author} {\bibfnamefont {J.~J.}\ \bibnamefont
  {Kas}}, \bibinfo {author} {\bibfnamefont {J.~J.}\ \bibnamefont {Rehr}}, \
  and\ \bibinfo {author} {\bibfnamefont {L.}~\bibnamefont {Reining}},\ }\href
  {\doibase 10.1103/PhysRevB.90.085112} {\bibfield  {journal} {\bibinfo
  {journal} {Phys. Rev. B}\ }\textbf {\bibinfo {volume} {90}},\ \bibinfo
  {pages} {085112} (\bibinfo {year} {2014})}\BibitemShut {NoStop}%
\bibitem [{\citenamefont {Lischner}\ \emph {et~al.}(2014)\citenamefont
  {Lischner}, \citenamefont {Vigil-Fowler},\ and\ \citenamefont
  {Louie}}]{lischner2014satellite}%
  \BibitemOpen
  \bibfield  {author} {\bibinfo {author} {\bibfnamefont {J.}~\bibnamefont
  {Lischner}}, \bibinfo {author} {\bibfnamefont {D.}~\bibnamefont
  {Vigil-Fowler}}, \ and\ \bibinfo {author} {\bibfnamefont {S.~G.}\
  \bibnamefont {Louie}},\ }\href {\doibase 10.1103/PhysRevB.89.125430}
  {\bibfield  {journal} {\bibinfo  {journal} {Phys. Rev. B}\ }\textbf {\bibinfo
  {volume} {89}},\ \bibinfo {pages} {125430} (\bibinfo {year}
  {2014})}\BibitemShut {NoStop}%
\bibitem [{\citenamefont {Caruso}\ and\ \citenamefont
  {Giustino}(2016)}]{caruso2016gw}%
  \BibitemOpen
  \bibfield  {author} {\bibinfo {author} {\bibfnamefont {F.}~\bibnamefont
  {Caruso}}\ and\ \bibinfo {author} {\bibfnamefont {F.}~\bibnamefont
  {Giustino}},\ }\href {https://doi.org/10.1140/epjb/e2016-70028-4} {\bibfield
  {journal} {\bibinfo  {journal} {Eur. Phys. J. B}\ }\textbf {\bibinfo {volume}
  {89}} (\bibinfo {year} {2016})}\BibitemShut {NoStop}%
\bibitem [{\citenamefont {Mayers}\ \emph {et~al.}(2016)\citenamefont {Mayers},
  \citenamefont {Hybertsen},\ and\ \citenamefont {Reichman}}]{mayers2016descr}%
  \BibitemOpen
  \bibfield  {author} {\bibinfo {author} {\bibfnamefont {M.~Z.}\ \bibnamefont
  {Mayers}}, \bibinfo {author} {\bibfnamefont {M.~S.}\ \bibnamefont
  {Hybertsen}}, \ and\ \bibinfo {author} {\bibfnamefont {D.~R.}\ \bibnamefont
  {Reichman}},\ }\href {\doibase 10.1103/PhysRevB.94.081109} {\bibfield
  {journal} {\bibinfo  {journal} {Phys. Rev. B}\ }\textbf {\bibinfo {volume}
  {94}},\ \bibinfo {pages} {081109} (\bibinfo {year} {2016})}\BibitemShut
  {NoStop}%
\bibitem [{\citenamefont {Jeckelmann}(2002)}]{Jeckelmann2002Jul}%
  \BibitemOpen
  \bibfield  {author} {\bibinfo {author} {\bibfnamefont {E.}~\bibnamefont
  {Jeckelmann}},\ }\href {\doibase 10.1103/PhysRevB.66.045114} {\bibfield
  {journal} {\bibinfo  {journal} {Phys. Rev. B}\ }\textbf {\bibinfo {volume}
  {66}},\ \bibinfo {pages} {045114} (\bibinfo {year} {2002})}\BibitemShut
  {NoStop}%
\bibitem [{\citenamefont {Schollw{\ifmmode\ddot{o}\else\"{o}\fi}ck}\ and\
  \citenamefont {White}(2006)}]{Schollwock2006Feb}%
  \BibitemOpen
  \bibfield  {author} {\bibinfo {author} {\bibfnamefont {U.}~\bibnamefont
  {Schollw{\ifmmode\ddot{o}\else\"{o}\fi}ck}}\ and\ \bibinfo {author}
  {\bibfnamefont {S.~R.}\ \bibnamefont {White}},\ }\href {\doibase
  10.1063/1.2178041} {\bibfield  {journal} {\bibinfo  {journal} {AIP Conf.
  Proc.}\ }\textbf {\bibinfo {volume} {816}},\ \bibinfo {pages} {155} (\bibinfo
  {year} {2006})}\BibitemShut {NoStop}%
\bibitem [{\citenamefont {Schirmer}(1982)}]{Schirmer1982Nov}%
  \BibitemOpen
  \bibfield  {author} {\bibinfo {author} {\bibfnamefont {J.}~\bibnamefont
  {Schirmer}},\ }\href {\doibase 10.1103/PhysRevA.26.2395} {\bibfield
  {journal} {\bibinfo  {journal} {Phys. Rev. A}\ }\textbf {\bibinfo {volume}
  {26}},\ \bibinfo {pages} {2395} (\bibinfo {year} {1982})}\BibitemShut
  {NoStop}%
\bibitem [{\citenamefont {Schirmer}\ \emph {et~al.}(1983)\citenamefont
  {Schirmer}, \citenamefont {Cederbaum},\ and\ \citenamefont
  {Walter}}]{Schirmer1983Sep}%
  \BibitemOpen
  \bibfield  {author} {\bibinfo {author} {\bibfnamefont {J.}~\bibnamefont
  {Schirmer}}, \bibinfo {author} {\bibfnamefont {L.~S.}\ \bibnamefont
  {Cederbaum}}, \ and\ \bibinfo {author} {\bibfnamefont {O.}~\bibnamefont
  {Walter}},\ }\href {\doibase 10.1103/PhysRevA.28.1237} {\bibfield  {journal}
  {\bibinfo  {journal} {Phys. Rev. A}\ }\textbf {\bibinfo {volume} {28}},\
  \bibinfo {pages} {1237} (\bibinfo {year} {1983})}\BibitemShut {NoStop}%
\bibitem [{\citenamefont {Tarantelli}\ and\ \citenamefont
  {Cederbaum}(1989)}]{Tarantelli1989Feb}%
  \BibitemOpen
  \bibfield  {author} {\bibinfo {author} {\bibfnamefont {A.}~\bibnamefont
  {Tarantelli}}\ and\ \bibinfo {author} {\bibfnamefont {L.~S.}\ \bibnamefont
  {Cederbaum}},\ }\href {\doibase 10.1103/PhysRevA.39.1656} {\bibfield
  {journal} {\bibinfo  {journal} {Phys. Rev. A}\ }\textbf {\bibinfo {volume}
  {39}},\ \bibinfo {pages} {1656} (\bibinfo {year} {1989})}\BibitemShut
  {NoStop}%
\bibitem [{\citenamefont {Wenzel}\ \emph {et~al.}(2014)\citenamefont {Wenzel},
  \citenamefont {Wormit},\ and\ \citenamefont {Dreuw}}]{Wenzel2014Oct}%
  \BibitemOpen
  \bibfield  {author} {\bibinfo {author} {\bibfnamefont {J.}~\bibnamefont
  {Wenzel}}, \bibinfo {author} {\bibfnamefont {M.}~\bibnamefont {Wormit}}, \
  and\ \bibinfo {author} {\bibfnamefont {A.}~\bibnamefont {Dreuw}},\ }\href
  {\doibase 10.1021/ct5006888} {\bibfield  {journal} {\bibinfo  {journal} {J.
  Chem. Theory Comput.}\ }\textbf {\bibinfo {volume} {10}},\ \bibinfo {pages}
  {4583} (\bibinfo {year} {2014})}\BibitemShut {NoStop}%
\bibitem [{\citenamefont {Dreuw}\ and\ \citenamefont
  {Wormit}(2015)}]{Dreuw2015Jan}%
  \BibitemOpen
  \bibfield  {author} {\bibinfo {author} {\bibfnamefont {A.}~\bibnamefont
  {Dreuw}}\ and\ \bibinfo {author} {\bibfnamefont {M.}~\bibnamefont {Wormit}},\
  }\href {\doibase 10.1002/wcms.1206} {\bibfield  {journal} {\bibinfo
  {journal} {WIREs Comput. Mol. Sci.}\ }\textbf {\bibinfo {volume} {5}},\
  \bibinfo {pages} {82} (\bibinfo {year} {2015})}\BibitemShut {NoStop}%
\bibitem [{\citenamefont {Sokolov}(2018)}]{Sokolov2018Nov}%
  \BibitemOpen
  \bibfield  {author} {\bibinfo {author} {\bibfnamefont {A.~{\relax Yu}.}\
  \bibnamefont {Sokolov}},\ }\href {\doibase 10.1063/1.5055380} {\bibfield
  {journal} {\bibinfo  {journal} {J. Chem. Phys.}\ }\textbf {\bibinfo {volume}
  {149}},\ \bibinfo {pages} {204113} (\bibinfo {year} {2018})}\BibitemShut
  {NoStop}%
\bibitem [{\citenamefont {Monkhorst}(1977)}]{Monkhorst1977Jan}%
  \BibitemOpen
  \bibfield  {author} {\bibinfo {author} {\bibfnamefont {H.~J.}\ \bibnamefont
  {Monkhorst}},\ }\href {\doibase 10.1002/qua.560120850} {\bibfield  {journal}
  {\bibinfo  {journal} {Int. J. Quantum Chem.}\ }\textbf {\bibinfo {volume}
  {12}},\ \bibinfo {pages} {421} (\bibinfo {year} {1977})}\BibitemShut
  {NoStop}%
\bibitem [{\citenamefont {Nooijen}\ and\ \citenamefont
  {Snijders}(1992)}]{Nooijen1992Mar}%
  \BibitemOpen
  \bibfield  {author} {\bibinfo {author} {\bibfnamefont {M.}~\bibnamefont
  {Nooijen}}\ and\ \bibinfo {author} {\bibfnamefont {J.~G.}\ \bibnamefont
  {Snijders}},\ }\href {\doibase 10.1002/qua.560440808} {\bibfield  {journal}
  {\bibinfo  {journal} {Int. J. Quantum Chem.}\ }\textbf {\bibinfo {volume}
  {44}},\ \bibinfo {pages} {55} (\bibinfo {year} {1992})}\BibitemShut {NoStop}%
\bibitem [{\citenamefont {Nooijen}\ and\ \citenamefont
  {Snijders}(1993)}]{Nooijen1993Oct}%
  \BibitemOpen
  \bibfield  {author} {\bibinfo {author} {\bibfnamefont {M.}~\bibnamefont
  {Nooijen}}\ and\ \bibinfo {author} {\bibfnamefont {J.~G.}\ \bibnamefont
  {Snijders}},\ }\href {\doibase 10.1002/qua.560480103} {\bibfield  {journal}
  {\bibinfo  {journal} {Int. J. Quantum Chem.}\ }\textbf {\bibinfo {volume}
  {48}},\ \bibinfo {pages} {15} (\bibinfo {year} {1993})}\BibitemShut {NoStop}%
\bibitem [{\citenamefont {Peng}\ and\ \citenamefont
  {Kowalski}(2016)}]{Peng2016Dec}%
  \BibitemOpen
  \bibfield  {author} {\bibinfo {author} {\bibfnamefont {B.}~\bibnamefont
  {Peng}}\ and\ \bibinfo {author} {\bibfnamefont {K.}~\bibnamefont
  {Kowalski}},\ }\href {\doibase 10.1103/PhysRevA.94.062512} {\bibfield
  {journal} {\bibinfo  {journal} {Phys. Rev. A}\ }\textbf {\bibinfo {volume}
  {94}},\ \bibinfo {pages} {062512} (\bibinfo {year} {2016})}\BibitemShut
  {NoStop}%
\bibitem [{\citenamefont {McClain}\ \emph {et~al.}(2016)\citenamefont
  {McClain}, \citenamefont {Lischner}, \citenamefont {Watson}, \citenamefont
  {Matthews}, \citenamefont {Ronca}, \citenamefont {Louie}, \citenamefont
  {Berkelbach},\ and\ \citenamefont {Chan}}]{McClain2016Jun}%
  \BibitemOpen
  \bibfield  {author} {\bibinfo {author} {\bibfnamefont {J.}~\bibnamefont
  {McClain}}, \bibinfo {author} {\bibfnamefont {J.}~\bibnamefont {Lischner}},
  \bibinfo {author} {\bibfnamefont {T.}~\bibnamefont {Watson}}, \bibinfo
  {author} {\bibfnamefont {D.~A.}\ \bibnamefont {Matthews}}, \bibinfo {author}
  {\bibfnamefont {E.}~\bibnamefont {Ronca}}, \bibinfo {author} {\bibfnamefont
  {S.~G.}\ \bibnamefont {Louie}}, \bibinfo {author} {\bibfnamefont {T.~C.}\
  \bibnamefont {Berkelbach}}, \ and\ \bibinfo {author} {\bibfnamefont
  {G.~K.-L.}\ \bibnamefont {Chan}},\ }\href {\doibase
  10.1103/PhysRevB.93.235139} {\bibfield  {journal} {\bibinfo  {journal} {Phys.
  Rev. B}\ }\textbf {\bibinfo {volume} {93}},\ \bibinfo {pages} {235139}
  (\bibinfo {year} {2016})}\BibitemShut {NoStop}%
\bibitem [{\citenamefont {McClain}\ \emph {et~al.}(2017)\citenamefont
  {McClain}, \citenamefont {Sun}, \citenamefont {Chan},\ and\ \citenamefont
  {Berkelbach}}]{McClain2017Mar}%
  \BibitemOpen
  \bibfield  {author} {\bibinfo {author} {\bibfnamefont {J.}~\bibnamefont
  {McClain}}, \bibinfo {author} {\bibfnamefont {Q.}~\bibnamefont {Sun}},
  \bibinfo {author} {\bibfnamefont {G.~K.-L.}\ \bibnamefont {Chan}}, \ and\
  \bibinfo {author} {\bibfnamefont {T.~C.}\ \bibnamefont {Berkelbach}},\ }\href
  {\doibase 10.1021/acs.jctc.7b00049} {\bibfield  {journal} {\bibinfo
  {journal} {J. Chem. Theory Comput.}\ }\textbf {\bibinfo {volume} {13}},\
  \bibinfo {pages} {1209} (\bibinfo {year} {2017})}\BibitemShut {NoStop}%
\bibitem [{\citenamefont {Furukawa}\ \emph {et~al.}(2018)\citenamefont
  {Furukawa}, \citenamefont {Kosugi}, \citenamefont {Nishi},\ and\
  \citenamefont {Matsushita}}]{Furukawa2018May}%
  \BibitemOpen
  \bibfield  {author} {\bibinfo {author} {\bibfnamefont {Y.}~\bibnamefont
  {Furukawa}}, \bibinfo {author} {\bibfnamefont {T.}~\bibnamefont {Kosugi}},
  \bibinfo {author} {\bibfnamefont {H.}~\bibnamefont {Nishi}}, \ and\ \bibinfo
  {author} {\bibfnamefont {Y.-i.}\ \bibnamefont {Matsushita}},\ }\href
  {\doibase 10.1063/1.5029537} {\bibfield  {journal} {\bibinfo  {journal} {J.
  Chem. Phys.}\ }\textbf {\bibinfo {volume} {148}},\ \bibinfo {pages} {204109}
  (\bibinfo {year} {2018})}\BibitemShut {NoStop}%
\bibitem [{\citenamefont {Peng}\ and\ \citenamefont
  {Kowalski}(2018)}]{Peng2018Mar}%
  \BibitemOpen
  \bibfield  {author} {\bibinfo {author} {\bibfnamefont {B.}~\bibnamefont
  {Peng}}\ and\ \bibinfo {author} {\bibfnamefont {K.}~\bibnamefont
  {Kowalski}},\ }\href {\doibase 10.1080/00268976.2017.1351630} {\bibfield
  {journal} {\bibinfo  {journal} {Mol. Phys.}\ }\textbf {\bibinfo {volume}
  {116}},\ \bibinfo {pages} {561} (\bibinfo {year} {2018})}\BibitemShut
  {NoStop}%
\bibitem [{\citenamefont {Shee}\ and\ \citenamefont
  {Zgid}(2019)}]{Shee2019Nov}%
  \BibitemOpen
  \bibfield  {author} {\bibinfo {author} {\bibfnamefont {A.}~\bibnamefont
  {Shee}}\ and\ \bibinfo {author} {\bibfnamefont {D.}~\bibnamefont {Zgid}},\
  }\href {\doibase 10.1021/acs.jctc.9b00603} {\bibfield  {journal} {\bibinfo
  {journal} {J. Chem. Theory Comput.}\ }\textbf {\bibinfo {volume} {15}},\
  \bibinfo {pages} {6010} (\bibinfo {year} {2019})}\BibitemShut {NoStop}%
\bibitem [{\citenamefont {Blankenbecler}\ and\ \citenamefont
  {Sugar}(1983)}]{Blankenbecler1983Mar}%
  \BibitemOpen
  \bibfield  {author} {\bibinfo {author} {\bibfnamefont {R.}~\bibnamefont
  {Blankenbecler}}\ and\ \bibinfo {author} {\bibfnamefont {R.~L.}\ \bibnamefont
  {Sugar}},\ }\href {\doibase 10.1103/PhysRevD.27.1304} {\bibfield  {journal}
  {\bibinfo  {journal} {Phys. Rev. D}\ }\textbf {\bibinfo {volume} {27}},\
  \bibinfo {pages} {1304} (\bibinfo {year} {1983})}\BibitemShut {NoStop}%
\bibitem [{\citenamefont {Becca}\ and\ \citenamefont
  {Sorella}(2017)}]{Becca2017Nov}%
  \BibitemOpen
  \bibfield  {author} {\bibinfo {author} {\bibfnamefont {F.}~\bibnamefont
  {Becca}}\ and\ \bibinfo {author} {\bibfnamefont {S.}~\bibnamefont
  {Sorella}},\ }\href {\doibase 10.1017/9781316417041} {\emph {\bibinfo {title}
  {{Quantum Monte Carlo Approaches for Correlated Systems}}}}\ (\bibinfo
  {publisher} {Cambridge University Press},\ \bibinfo {address} {Cambridge,
  England, UK},\ \bibinfo {year} {2017})\BibitemShut {NoStop}%
\bibitem [{\citenamefont {Foulkes}\ \emph {et~al.}(2001)\citenamefont
  {Foulkes}, \citenamefont {Mitas}, \citenamefont {Needs},\ and\ \citenamefont
  {Rajagopal}}]{foulkes_rmp}%
  \BibitemOpen
  \bibfield  {author} {\bibinfo {author} {\bibfnamefont {W.~M.~C.}\
  \bibnamefont {Foulkes}}, \bibinfo {author} {\bibfnamefont {L.}~\bibnamefont
  {Mitas}}, \bibinfo {author} {\bibfnamefont {R.~J.}\ \bibnamefont {Needs}}, \
  and\ \bibinfo {author} {\bibfnamefont {G.}~\bibnamefont {Rajagopal}},\ }\href
  {\doibase 10.1103/RevModPhys.73.33} {\bibfield  {journal} {\bibinfo
  {journal} {Rev. Mod. Phys.}\ }\textbf {\bibinfo {volume} {73}},\ \bibinfo
  {pages} {33} (\bibinfo {year} {2001})}\BibitemShut {NoStop}%
\bibitem [{\citenamefont {Silver}\ \emph {et~al.}(1990)\citenamefont {Silver},
  \citenamefont {Sivia},\ and\ \citenamefont {Gubernatis}}]{SilverMaxEnt1990}%
  \BibitemOpen
  \bibfield  {author} {\bibinfo {author} {\bibfnamefont {R.~N.}\ \bibnamefont
  {Silver}}, \bibinfo {author} {\bibfnamefont {D.~S.}\ \bibnamefont {Sivia}}, \
  and\ \bibinfo {author} {\bibfnamefont {J.~E.}\ \bibnamefont {Gubernatis}},\
  }\href {\doibase 10.1103/PhysRevB.41.2380} {\bibfield  {journal} {\bibinfo
  {journal} {Phys. Rev. B}\ }\textbf {\bibinfo {volume} {41}},\ \bibinfo
  {pages} {2380} (\bibinfo {year} {1990})}\BibitemShut {NoStop}%
\bibitem [{\citenamefont {Gubernatis}\ \emph {et~al.}(1991)\citenamefont
  {Gubernatis}, \citenamefont {Jarrell}, \citenamefont {Silver},\ and\
  \citenamefont {Sivia}}]{GubernatisMaxEnt1991}%
  \BibitemOpen
  \bibfield  {author} {\bibinfo {author} {\bibfnamefont {J.~E.}\ \bibnamefont
  {Gubernatis}}, \bibinfo {author} {\bibfnamefont {M.}~\bibnamefont {Jarrell}},
  \bibinfo {author} {\bibfnamefont {R.~N.}\ \bibnamefont {Silver}}, \ and\
  \bibinfo {author} {\bibfnamefont {D.~S.}\ \bibnamefont {Sivia}},\ }\href
  {\doibase 10.1103/PhysRevB.44.6011} {\bibfield  {journal} {\bibinfo
  {journal} {Phys. Rev. B}\ }\textbf {\bibinfo {volume} {44}},\ \bibinfo
  {pages} {6011} (\bibinfo {year} {1991})}\BibitemShut {NoStop}%
\bibitem [{\citenamefont {Jarrell}\ and\ \citenamefont
  {Gubernatis}(1996)}]{Jarrell1996May}%
  \BibitemOpen
  \bibfield  {author} {\bibinfo {author} {\bibfnamefont {M.}~\bibnamefont
  {Jarrell}}\ and\ \bibinfo {author} {\bibfnamefont {J.~E.}\ \bibnamefont
  {Gubernatis}},\ }\href {\doibase 10.1016/0370-1573(95)00074-7} {\bibfield
  {journal} {\bibinfo  {journal} {Phys. Rep.}\ }\textbf {\bibinfo {volume}
  {269}},\ \bibinfo {pages} {133} (\bibinfo {year} {1996})}\BibitemShut
  {NoStop}%
\bibitem [{\citenamefont {Motta}\ \emph {et~al.}(2014)\citenamefont {Motta},
  \citenamefont {Galli}, \citenamefont {Moroni},\ and\ \citenamefont
  {Vitali}}]{Motta2014Jan}%
  \BibitemOpen
  \bibfield  {author} {\bibinfo {author} {\bibfnamefont {M.}~\bibnamefont
  {Motta}}, \bibinfo {author} {\bibfnamefont {D.~E.}\ \bibnamefont {Galli}},
  \bibinfo {author} {\bibfnamefont {S.}~\bibnamefont {Moroni}}, \ and\ \bibinfo
  {author} {\bibfnamefont {E.}~\bibnamefont {Vitali}},\ }\href {\doibase
  10.1063/1.4861227} {\bibfield  {journal} {\bibinfo  {journal} {J. Chem.
  Phys.}\ }\textbf {\bibinfo {volume} {140}},\ \bibinfo {pages} {024107}
  (\bibinfo {year} {2014})}\BibitemShut {NoStop}%
\bibitem [{\citenamefont {Motta}\ \emph {et~al.}(2015)\citenamefont {Motta},
  \citenamefont {Galli}, \citenamefont {Moroni},\ and\ \citenamefont
  {Vitali}}]{Motta2015Oct}%
  \BibitemOpen
  \bibfield  {author} {\bibinfo {author} {\bibfnamefont {M.}~\bibnamefont
  {Motta}}, \bibinfo {author} {\bibfnamefont {D.~E.}\ \bibnamefont {Galli}},
  \bibinfo {author} {\bibfnamefont {S.}~\bibnamefont {Moroni}}, \ and\ \bibinfo
  {author} {\bibfnamefont {E.}~\bibnamefont {Vitali}},\ }\href {\doibase
  10.1063/1.4934666} {\bibfield  {journal} {\bibinfo  {journal} {J. Chem.
  Phys.}\ }\textbf {\bibinfo {volume} {143}},\ \bibinfo {pages} {164108}
  (\bibinfo {year} {2015})}\BibitemShut {NoStop}%
\bibitem [{\citenamefont {Otsuki}\ \emph {et~al.}(2017)\citenamefont {Otsuki},
  \citenamefont {Ohzeki}, \citenamefont {Shinaoka},\ and\ \citenamefont
  {Yoshimi}}]{Otsuki2017Jun}%
  \BibitemOpen
  \bibfield  {author} {\bibinfo {author} {\bibfnamefont {J.}~\bibnamefont
  {Otsuki}}, \bibinfo {author} {\bibfnamefont {M.}~\bibnamefont {Ohzeki}},
  \bibinfo {author} {\bibfnamefont {H.}~\bibnamefont {Shinaoka}}, \ and\
  \bibinfo {author} {\bibfnamefont {K.}~\bibnamefont {Yoshimi}},\ }\href
  {\doibase 10.1103/PhysRevE.95.061302} {\bibfield  {journal} {\bibinfo
  {journal} {Phys. Rev. E}\ }\textbf {\bibinfo {volume} {95}},\ \bibinfo
  {pages} {061302} (\bibinfo {year} {2017})}\BibitemShut {NoStop}%
\bibitem [{\citenamefont {Bertaina}\ \emph {et~al.}(2017)\citenamefont
  {Bertaina}, \citenamefont {Galli},\ and\ \citenamefont
  {Vitali}}]{Bertaina2017Mar}%
  \BibitemOpen
  \bibfield  {author} {\bibinfo {author} {\bibfnamefont {G.}~\bibnamefont
  {Bertaina}}, \bibinfo {author} {\bibfnamefont {D.~E.}\ \bibnamefont {Galli}},
  \ and\ \bibinfo {author} {\bibfnamefont {E.}~\bibnamefont {Vitali}},\ }\href
  {\doibase 10.1080/23746149.2017.1288585} {\bibfield  {journal} {\bibinfo
  {journal} {Adv. Phys-X}\ }\textbf {\bibinfo {volume} {2}},\ \bibinfo {pages}
  {302} (\bibinfo {year} {2017})}\BibitemShut {NoStop}%
\bibitem [{\citenamefont {Reichman}\ and\ \citenamefont
  {Rabani}(2009)}]{Reichman2009Aug}%
  \BibitemOpen
  \bibfield  {author} {\bibinfo {author} {\bibfnamefont {D.~R.}\ \bibnamefont
  {Reichman}}\ and\ \bibinfo {author} {\bibfnamefont {E.}~\bibnamefont
  {Rabani}},\ }\href {\doibase 10.1063/1.3185728} {\bibfield  {journal}
  {\bibinfo  {journal} {J. Chem. Phys.}\ }\textbf {\bibinfo {volume} {131}},\
  \bibinfo {pages} {054502} (\bibinfo {year} {2009})}\BibitemShut {NoStop}%
\bibitem [{\citenamefont {Goulko}\ \emph {et~al.}(2017)\citenamefont {Goulko},
  \citenamefont {Mishchenko}, \citenamefont {Pollet}, \citenamefont
  {Prokof'ev},\ and\ \citenamefont {Svistunov}}]{Goulko2017Jan}%
  \BibitemOpen
  \bibfield  {author} {\bibinfo {author} {\bibfnamefont {O.}~\bibnamefont
  {Goulko}}, \bibinfo {author} {\bibfnamefont {A.~S.}\ \bibnamefont
  {Mishchenko}}, \bibinfo {author} {\bibfnamefont {L.}~\bibnamefont {Pollet}},
  \bibinfo {author} {\bibfnamefont {N.}~\bibnamefont {Prokof'ev}}, \ and\
  \bibinfo {author} {\bibfnamefont {B.}~\bibnamefont {Svistunov}},\ }\href
  {\doibase 10.1103/PhysRevB.95.014102} {\bibfield  {journal} {\bibinfo
  {journal} {Phys. Rev. B}\ }\textbf {\bibinfo {volume} {95}},\ \bibinfo
  {pages} {014102} (\bibinfo {year} {2017})}\BibitemShut {NoStop}%
\bibitem [{\citenamefont {Dornheim}\ \emph {et~al.}(2018)\citenamefont
  {Dornheim}, \citenamefont {Groth}, \citenamefont {Vorberger},\ and\
  \citenamefont {Bonitz}}]{DornheimDynamicalStructure2018}%
  \BibitemOpen
  \bibfield  {author} {\bibinfo {author} {\bibfnamefont {T.}~\bibnamefont
  {Dornheim}}, \bibinfo {author} {\bibfnamefont {S.}~\bibnamefont {Groth}},
  \bibinfo {author} {\bibfnamefont {J.}~\bibnamefont {Vorberger}}, \ and\
  \bibinfo {author} {\bibfnamefont {M.}~\bibnamefont {Bonitz}},\ }\href
  {\doibase 10.1103/PhysRevLett.121.255001} {\bibfield  {journal} {\bibinfo
  {journal} {Phys. Rev. Lett.}\ }\textbf {\bibinfo {volume} {121}},\ \bibinfo
  {pages} {255001} (\bibinfo {year} {2018})}\BibitemShut {NoStop}%
\bibitem [{\citenamefont {Ceperley}\ and\ \citenamefont
  {Bernu}(1988)}]{Ceperley1988Nov}%
  \BibitemOpen
  \bibfield  {author} {\bibinfo {author} {\bibfnamefont {D.~M.}\ \bibnamefont
  {Ceperley}}\ and\ \bibinfo {author} {\bibfnamefont {B.}~\bibnamefont
  {Bernu}},\ }\href {\doibase 10.1063/1.455398} {\bibfield  {journal} {\bibinfo
   {journal} {J. Chem. Phys.}\ }\textbf {\bibinfo {volume} {89}},\ \bibinfo
  {pages} {6316} (\bibinfo {year} {1988})}\BibitemShut {NoStop}%
\bibitem [{\citenamefont {Blunt}\ \emph {et~al.}(2015)\citenamefont {Blunt},
  \citenamefont {Alavi},\ and\ \citenamefont {Booth}}]{Blunt2015Jul}%
  \BibitemOpen
  \bibfield  {author} {\bibinfo {author} {\bibfnamefont {N.~S.}\ \bibnamefont
  {Blunt}}, \bibinfo {author} {\bibfnamefont {A.}~\bibnamefont {Alavi}}, \ and\
  \bibinfo {author} {\bibfnamefont {G.~H.}\ \bibnamefont {Booth}},\ }\href
  {\doibase 10.1103/PhysRevLett.115.050603} {\bibfield  {journal} {\bibinfo
  {journal} {Phys. Rev. Lett.}\ }\textbf {\bibinfo {volume} {115}},\ \bibinfo
  {pages} {050603} (\bibinfo {year} {2015})}\BibitemShut {NoStop}%
\bibitem [{\citenamefont {Blunt}\ \emph {et~al.}(2018)\citenamefont {Blunt},
  \citenamefont {Alavi},\ and\ \citenamefont {Booth}}]{Blunt2018Aug}%
  \BibitemOpen
  \bibfield  {author} {\bibinfo {author} {\bibfnamefont {N.~S.}\ \bibnamefont
  {Blunt}}, \bibinfo {author} {\bibfnamefont {A.}~\bibnamefont {Alavi}}, \ and\
  \bibinfo {author} {\bibfnamefont {G.~H.}\ \bibnamefont {Booth}},\ }\href
  {\doibase 10.1103/PhysRevB.98.085118} {\bibfield  {journal} {\bibinfo
  {journal} {Phys. Rev. B}\ }\textbf {\bibinfo {volume} {98}},\ \bibinfo
  {pages} {085118} (\bibinfo {year} {2018})}\BibitemShut {NoStop}%
\bibitem [{\citenamefont {Booth}\ \emph {et~al.}(2009)\citenamefont {Booth},
  \citenamefont {Thom},\ and\ \citenamefont {Alavi}}]{Booth2009Aug}%
  \BibitemOpen
  \bibfield  {author} {\bibinfo {author} {\bibfnamefont {G.~H.}\ \bibnamefont
  {Booth}}, \bibinfo {author} {\bibfnamefont {A.~J.~W.}\ \bibnamefont {Thom}},
  \ and\ \bibinfo {author} {\bibfnamefont {A.}~\bibnamefont {Alavi}},\ }\href
  {\doibase 10.1063/1.3193710} {\bibfield  {journal} {\bibinfo  {journal} {J.
  Chem. Phys.}\ }\textbf {\bibinfo {volume} {131}},\ \bibinfo {pages} {054106}
  (\bibinfo {year} {2009})}\BibitemShut {NoStop}%
\bibitem [{\citenamefont {Day}\ \emph {et~al.}(1975)\citenamefont {Day},
  \citenamefont {Smith},\ and\ \citenamefont {Morrison}}]{Day1975}%
  \BibitemOpen
  \bibfield  {author} {\bibinfo {author} {\bibfnamefont {O.~W.}\ \bibnamefont
  {Day}}, \bibinfo {author} {\bibfnamefont {D.~W.}\ \bibnamefont {Smith}}, \
  and\ \bibinfo {author} {\bibfnamefont {R.~C.}\ \bibnamefont {Morrison}},\
  }\href {\doibase 10.1063/1.430254} {\bibfield  {journal} {\bibinfo  {journal}
  {J. Chem. Phys.}\ }\textbf {\bibinfo {volume} {62}},\ \bibinfo {pages} {115}
  (\bibinfo {year} {1975})}\BibitemShut {NoStop}%
\bibitem [{\citenamefont {Smith}\ and\ \citenamefont {Day}(1975)}]{Smith1975}%
  \BibitemOpen
  \bibfield  {author} {\bibinfo {author} {\bibfnamefont {D.~W.}\ \bibnamefont
  {Smith}}\ and\ \bibinfo {author} {\bibfnamefont {O.~W.}\ \bibnamefont
  {Day}},\ }\href {\doibase 10.1063/1.430253} {\bibfield  {journal} {\bibinfo
  {journal} {J. Chem. Phys.}\ }\textbf {\bibinfo {volume} {62}},\ \bibinfo
  {pages} {113} (\bibinfo {year} {1975})}\BibitemShut {NoStop}%
\bibitem [{\citenamefont {Pickup}(1975)}]{Pickup1975}%
  \BibitemOpen
  \bibfield  {author} {\bibinfo {author} {\bibfnamefont {B.~T.}\ \bibnamefont
  {Pickup}},\ }\href {\doibase 10.1016/0009-2614(75)85744-7} {\bibfield
  {journal} {\bibinfo  {journal} {Chem. Phys. Lett.}\ }\textbf {\bibinfo
  {volume} {33}},\ \bibinfo {pages} {422} (\bibinfo {year} {1975})}\BibitemShut
  {NoStop}%
\bibitem [{\citenamefont {Morrison}\ \emph {et~al.}(1975)\citenamefont
  {Morrison}, \citenamefont {Day},\ and\ \citenamefont
  {Smith}}]{Morrison1975Jan}%
  \BibitemOpen
  \bibfield  {author} {\bibinfo {author} {\bibfnamefont {R.~C.}\ \bibnamefont
  {Morrison}}, \bibinfo {author} {\bibfnamefont {O.~W.}\ \bibnamefont {Day}}, \
  and\ \bibinfo {author} {\bibfnamefont {D.~W.}\ \bibnamefont {Smith}},\ }\href
  {\doibase 10.1002/qua.560090829} {\bibfield  {journal} {\bibinfo  {journal}
  {Int. J. Quantum Chem.}\ }\textbf {\bibinfo {volume} {9}},\ \bibinfo {pages}
  {229} (\bibinfo {year} {1975})}\BibitemShut {NoStop}%
\bibitem [{\citenamefont {Ellenbogen}\ \emph {et~al.}(1977)\citenamefont
  {Ellenbogen}, \citenamefont {Day}, \citenamefont {Smith},\ and\ \citenamefont
  {Morrison}}]{Ellenbogen1977Jun}%
  \BibitemOpen
  \bibfield  {author} {\bibinfo {author} {\bibfnamefont {J.~C.}\ \bibnamefont
  {Ellenbogen}}, \bibinfo {author} {\bibfnamefont {O.~W.}\ \bibnamefont {Day}},
  \bibinfo {author} {\bibfnamefont {D.~W.}\ \bibnamefont {Smith}}, \ and\
  \bibinfo {author} {\bibfnamefont {R.~C.}\ \bibnamefont {Morrison}},\ }\href
  {\doibase 10.1063/1.433842} {\bibfield  {journal} {\bibinfo  {journal} {J.
  Chem. Phys.}\ }\textbf {\bibinfo {volume} {66}},\ \bibinfo {pages} {4795}
  (\bibinfo {year} {1977})}\BibitemShut {NoStop}%
\bibitem [{\citenamefont {Morrison}\ and\ \citenamefont
  {Liu}(1992)}]{Morrison1992}%
  \BibitemOpen
  \bibfield  {author} {\bibinfo {author} {\bibfnamefont {R.~C.}\ \bibnamefont
  {Morrison}}\ and\ \bibinfo {author} {\bibfnamefont {G.}~\bibnamefont {Liu}},\
  }\href {\doibase 10.1002/jcc.540130811} {\bibfield  {journal} {\bibinfo
  {journal} {J. Comp. Chem.}\ }\textbf {\bibinfo {volume} {13}},\ \bibinfo
  {pages} {1004} (\bibinfo {year} {1992})}\BibitemShut {NoStop}%
\bibitem [{\citenamefont {Morrison}(1992)}]{Morrison1992a}%
  \BibitemOpen
  \bibfield  {author} {\bibinfo {author} {\bibfnamefont {R.~C.}\ \bibnamefont
  {Morrison}},\ }\href {\doibase 10.1063/1.461875} {\bibfield  {journal}
  {\bibinfo  {journal} {J. Chem. Phys.}\ }\textbf {\bibinfo {volume} {96}},\
  \bibinfo {pages} {3718} (\bibinfo {year} {1992})}\BibitemShut {NoStop}%
\bibitem [{\citenamefont {Sundholm}\ and\ \citenamefont
  {Olsen}(1993)}]{Sundholm1993}%
  \BibitemOpen
  \bibfield  {author} {\bibinfo {author} {\bibfnamefont {D.}~\bibnamefont
  {Sundholm}}\ and\ \bibinfo {author} {\bibfnamefont {J.}~\bibnamefont
  {Olsen}},\ }\href {\doibase 10.1063/1.464028} {\bibfield  {journal} {\bibinfo
   {journal} {J. Chem. Phys.}\ }\textbf {\bibinfo {volume} {98}},\ \bibinfo
  {pages} {3999} (\bibinfo {year} {1993})}\BibitemShut {NoStop}%
\bibitem [{\citenamefont {Cioslowski}\ \emph {et~al.}(1997)\citenamefont
  {Cioslowski}, \citenamefont {Piskorz},\ and\ \citenamefont
  {Liu}}]{Cioslowski1997}%
  \BibitemOpen
  \bibfield  {author} {\bibinfo {author} {\bibfnamefont {J.}~\bibnamefont
  {Cioslowski}}, \bibinfo {author} {\bibfnamefont {P.}~\bibnamefont {Piskorz}},
  \ and\ \bibinfo {author} {\bibfnamefont {G.}~\bibnamefont {Liu}},\ }\href
  {\doibase 10.1063/1.474921} {\bibfield  {journal} {\bibinfo  {journal} {J.
  Chem. Phys.}\ }\textbf {\bibinfo {volume} {107}},\ \bibinfo {pages} {6804}
  (\bibinfo {year} {1997})}\BibitemShut {NoStop}%
\bibitem [{\citenamefont {Kent}\ \emph {et~al.}(1998)\citenamefont {Kent},
  \citenamefont {Hood}, \citenamefont {Towler}, \citenamefont {Needs},\ and\
  \citenamefont {Rajagopal}}]{Kent1998}%
  \BibitemOpen
  \bibfield  {author} {\bibinfo {author} {\bibfnamefont {P.}~\bibnamefont
  {Kent}}, \bibinfo {author} {\bibfnamefont {R.~Q.}\ \bibnamefont {Hood}},
  \bibinfo {author} {\bibfnamefont {M.}~\bibnamefont {Towler}}, \bibinfo
  {author} {\bibfnamefont {R.}~\bibnamefont {Needs}}, \ and\ \bibinfo {author}
  {\bibfnamefont {G.}~\bibnamefont {Rajagopal}},\ }\href@noop {} {\bibfield
  {journal} {\bibinfo  {journal} {Phys. Rev. B - Condensed Matter and Materials
  Physics}\ }\textbf {\bibinfo {volume} {57}},\ \bibinfo {pages} {15293}
  (\bibinfo {year} {1998})}\BibitemShut {NoStop}%
\bibitem [{\citenamefont {Olsen}\ and\ \citenamefont
  {Sundholm}(1998)}]{Olsen1998May}%
  \BibitemOpen
  \bibfield  {author} {\bibinfo {author} {\bibfnamefont {J.}~\bibnamefont
  {Olsen}}\ and\ \bibinfo {author} {\bibfnamefont {D.}~\bibnamefont
  {Sundholm}},\ }\href {\doibase 10.1016/S0009-2614(98)00302-9} {\bibfield
  {journal} {\bibinfo  {journal} {Chem. Phys. Lett.}\ }\textbf {\bibinfo
  {volume} {288}},\ \bibinfo {pages} {282} (\bibinfo {year}
  {1998})}\BibitemShut {NoStop}%
\bibitem [{\citenamefont {Pernal}\ and\ \citenamefont
  {Cioslowski}(2001)}]{Pernal2001Mar}%
  \BibitemOpen
  \bibfield  {author} {\bibinfo {author} {\bibfnamefont {K.}~\bibnamefont
  {Pernal}}\ and\ \bibinfo {author} {\bibfnamefont {J.}~\bibnamefont
  {Cioslowski}},\ }\href {\doibase 10.1063/1.1336543} {\bibfield  {journal}
  {\bibinfo  {journal} {J. Chem. Phys.}\ }\textbf {\bibinfo {volume} {114}},\
  \bibinfo {pages} {4359} (\bibinfo {year} {2001})}\BibitemShut {NoStop}%
\bibitem [{\citenamefont {Farnum}\ and\ \citenamefont
  {Mazziotti}(2004)}]{Farnum2004}%
  \BibitemOpen
  \bibfield  {author} {\bibinfo {author} {\bibfnamefont {J.~D.}\ \bibnamefont
  {Farnum}}\ and\ \bibinfo {author} {\bibfnamefont {D.~A.}\ \bibnamefont
  {Mazziotti}},\ }\href {\doibase 10.1016/j.cplett.2004.10.075} {\bibfield
  {journal} {\bibinfo  {journal} {Chem. Phys. Lett.}\ }\textbf {\bibinfo
  {volume} {400}},\ \bibinfo {pages} {90} (\bibinfo {year} {2004})}\BibitemShut
  {NoStop}%
\bibitem [{\citenamefont {Pernal}\ and\ \citenamefont
  {Cioslowski}(2005)}]{Pernal2005Aug}%
  \BibitemOpen
  \bibfield  {author} {\bibinfo {author} {\bibfnamefont {K.}~\bibnamefont
  {Pernal}}\ and\ \bibinfo {author} {\bibfnamefont {J.}~\bibnamefont
  {Cioslowski}},\ }\href {\doibase 10.1016/j.cplett.2005.06.103} {\bibfield
  {journal} {\bibinfo  {journal} {Chem. Phys. Lett.}\ }\textbf {\bibinfo
  {volume} {412}},\ \bibinfo {pages} {71} (\bibinfo {year} {2005})}\BibitemShut
  {NoStop}%
\bibitem [{\citenamefont {Ernzerhof}(2009)}]{Ernzerhof2009}%
  \BibitemOpen
  \bibfield  {author} {\bibinfo {author} {\bibfnamefont {M.}~\bibnamefont
  {Ernzerhof}},\ }\href {\doibase 10.1021/ct800552k} {\bibfield  {journal}
  {\bibinfo  {journal} {J. Chem. Theory Comput.}\ }\textbf {\bibinfo {volume}
  {5}},\ \bibinfo {pages} {793} (\bibinfo {year} {2009})}\BibitemShut {NoStop}%
\bibitem [{\citenamefont {Vanfleteren}\ \emph {et~al.}(2009)\citenamefont
  {Vanfleteren}, \citenamefont {{Van Neck}}, \citenamefont {Ayers},
  \citenamefont {Morrison},\ and\ \citenamefont {Bultinck}}]{Vanfleteren2009}%
  \BibitemOpen
  \bibfield  {author} {\bibinfo {author} {\bibfnamefont {D.}~\bibnamefont
  {Vanfleteren}}, \bibinfo {author} {\bibfnamefont {D.}~\bibnamefont {{Van
  Neck}}}, \bibinfo {author} {\bibfnamefont {P.~W.}\ \bibnamefont {Ayers}},
  \bibinfo {author} {\bibfnamefont {R.~C.}\ \bibnamefont {Morrison}}, \ and\
  \bibinfo {author} {\bibfnamefont {P.}~\bibnamefont {Bultinck}},\ }\href
  {\doibase 10.1063/1.3130044} {\bibfield  {journal} {\bibinfo  {journal} {J.
  Chem. Phys.}\ }\textbf {\bibinfo {volume} {130}} (\bibinfo {year} {2009}),\
  10.1063/1.3130044}\BibitemShut {NoStop}%
\bibitem [{\citenamefont {Piris}\ \emph {et~al.}(2012)\citenamefont {Piris},
  \citenamefont {Matxain}, \citenamefont {Lopez},\ and\ \citenamefont
  {Ugalde}}]{Piris2012May}%
  \BibitemOpen
  \bibfield  {author} {\bibinfo {author} {\bibfnamefont {M.}~\bibnamefont
  {Piris}}, \bibinfo {author} {\bibfnamefont {J.~M.}\ \bibnamefont {Matxain}},
  \bibinfo {author} {\bibfnamefont {X.}~\bibnamefont {Lopez}}, \ and\ \bibinfo
  {author} {\bibfnamefont {J.~M.}\ \bibnamefont {Ugalde}},\ }\href {\doibase
  10.1063/1.4709769} {\bibfield  {journal} {\bibinfo  {journal} {J. Chem.
  Phys.}\ }\textbf {\bibinfo {volume} {136}},\ \bibinfo {pages} {174116}
  (\bibinfo {year} {2012})}\BibitemShut {NoStop}%
\bibitem [{\citenamefont {Piris}\ \emph {et~al.}(2013)\citenamefont {Piris},
  \citenamefont {Matxain}, \citenamefont {Lopez},\ and\ \citenamefont
  {Ugalde}}]{Piris2013Jan}%
  \BibitemOpen
  \bibfield  {author} {\bibinfo {author} {\bibfnamefont {M.}~\bibnamefont
  {Piris}}, \bibinfo {author} {\bibfnamefont {J.~M.}\ \bibnamefont {Matxain}},
  \bibinfo {author} {\bibfnamefont {X.}~\bibnamefont {Lopez}}, \ and\ \bibinfo
  {author} {\bibfnamefont {J.~M.}\ \bibnamefont {Ugalde}},\ }in\ \href
  {\doibase 10.1007/978-3-642-41272-1_2} {\emph {\bibinfo {booktitle} {{8th
  Congress on Electronic Structure: Principles and Applications (ESPA 2012): A
  Conference Selection from Theoretical Chemistry Accounts}}}}\ (\bibinfo
  {publisher} {Springer},\ \bibinfo {address} {Berlin, Germany},\ \bibinfo
  {year} {2013})\ pp.\ \bibinfo {pages} {5--15}\BibitemShut {NoStop}%
\bibitem [{\citenamefont {Bozkaya}(2013)}]{Bozkaya2013}%
  \BibitemOpen
  \bibfield  {author} {\bibinfo {author} {\bibfnamefont {U.}~\bibnamefont
  {Bozkaya}},\ }\href {\doibase 10.1063/1.4825041} {\bibfield  {journal}
  {\bibinfo  {journal} {J. Chem. Phys.}\ }\textbf {\bibinfo {volume} {139}}
  (\bibinfo {year} {2013}),\ 10.1063/1.4825041}\BibitemShut {NoStop}%
\bibitem [{\citenamefont {Welden}\ \emph {et~al.}(2015)\citenamefont {Welden},
  \citenamefont {Phillips},\ and\ \citenamefont {Zgid}}]{Welden2015}%
  \BibitemOpen
  \bibfield  {author} {\bibinfo {author} {\bibfnamefont {A.~R.}\ \bibnamefont
  {Welden}}, \bibinfo {author} {\bibfnamefont {J.~J.}\ \bibnamefont
  {Phillips}}, \ and\ \bibinfo {author} {\bibfnamefont {D.}~\bibnamefont
  {Zgid}},\ }\href {http://arxiv.org/abs/1505.05575} {\ ,\ \bibinfo {pages} {1}
  (\bibinfo {year} {2015})},\ \Eprint {http://arxiv.org/abs/1505.05575}
  {arXiv:1505.05575} \BibitemShut {NoStop}%
\bibitem [{\citenamefont {Bozkaya}\ and\ \citenamefont
  {{\"{U}}nal}(2018)}]{Bozkaya2018}%
  \BibitemOpen
  \bibfield  {author} {\bibinfo {author} {\bibfnamefont {U.}~\bibnamefont
  {Bozkaya}}\ and\ \bibinfo {author} {\bibfnamefont {A.}~\bibnamefont
  {{\"{U}}nal}},\ }\href {\doibase 10.1021/acs.jpca.8b01851} {\bibfield
  {journal} {\bibinfo  {journal} {J. Phys. Chem. A}\ }\textbf {\bibinfo
  {volume} {122}},\ \bibinfo {pages} {4375} (\bibinfo {year}
  {2018})}\BibitemShut {NoStop}%
\bibitem [{\citenamefont {Pavlyukh}(2018)}]{Pavlyukh2018}%
  \BibitemOpen
  \bibfield  {author} {\bibinfo {author} {\bibfnamefont {Y.}~\bibnamefont
  {Pavlyukh}},\ }\href {\doibase 10.1103/PhysRevA.98.052508} {\bibfield
  {journal} {\bibinfo  {journal} {Phys. Rev. A}\ }\textbf {\bibinfo {volume}
  {98}},\ \bibinfo {pages} {1} (\bibinfo {year} {2018})}\BibitemShut {NoStop}%
\bibitem [{\citenamefont {Pavlyukh}(2019)}]{Pavlyukh2019}%
  \BibitemOpen
  \bibfield  {author} {\bibinfo {author} {\bibfnamefont {Y.}~\bibnamefont
  {Pavlyukh}},\ }\href {\doibase 10.1002/pssb.201800591} {\bibfield  {journal}
  {\bibinfo  {journal} {Phys. Status Solidi B Basic Res.}\ }\textbf {\bibinfo
  {volume} {256}},\ \bibinfo {pages} {1} (\bibinfo {year} {2019})}\BibitemShut
  {NoStop}%
\bibitem [{\citenamefont {Zheng}(2016)}]{Zheng2016Jul}%
  \BibitemOpen
  \bibfield  {author} {\bibinfo {author} {\bibfnamefont {H.}~\bibnamefont
  {Zheng}},\ }\href {https://www.ideals.illinois.edu/handle/2142/92794}
  {\enquote {\bibinfo {title} {{First principles quantum Monte Carlo study of
  correlated electronic systems}},}\ } (\bibinfo {year} {2016}),\ \bibinfo
  {note} {[Online; accessed 12. Jan. 2021]}\BibitemShut {NoStop}%
\bibitem [{\citenamefont {Zhang}\ \emph {et~al.}(1995)\citenamefont {Zhang},
  \citenamefont {Carlson},\ and\ \citenamefont {Gubernatis}}]{Zhang1995May}%
  \BibitemOpen
  \bibfield  {author} {\bibinfo {author} {\bibfnamefont {S.}~\bibnamefont
  {Zhang}}, \bibinfo {author} {\bibfnamefont {J.}~\bibnamefont {Carlson}}, \
  and\ \bibinfo {author} {\bibfnamefont {J.~E.}\ \bibnamefont {Gubernatis}},\
  }\href {\doibase 10.1103/PhysRevLett.74.3652} {\bibfield  {journal} {\bibinfo
   {journal} {Phys. Rev. Lett.}\ }\textbf {\bibinfo {volume} {74}},\ \bibinfo
  {pages} {3652} (\bibinfo {year} {1995})}\BibitemShut {NoStop}%
\bibitem [{\citenamefont {Zhang}\ \emph {et~al.}(1997)\citenamefont {Zhang},
  \citenamefont {Carlson},\ and\ \citenamefont {Gubernatis}}]{Zhang1997Mar}%
  \BibitemOpen
  \bibfield  {author} {\bibinfo {author} {\bibfnamefont {S.}~\bibnamefont
  {Zhang}}, \bibinfo {author} {\bibfnamefont {J.}~\bibnamefont {Carlson}}, \
  and\ \bibinfo {author} {\bibfnamefont {J.~E.}\ \bibnamefont {Gubernatis}},\
  }\href {\doibase 10.1103/PhysRevB.55.7464} {\bibfield  {journal} {\bibinfo
  {journal} {Phys. Rev. B}\ }\textbf {\bibinfo {volume} {55}},\ \bibinfo
  {pages} {7464} (\bibinfo {year} {1997})}\BibitemShut {NoStop}%
\bibitem [{\citenamefont {Zhang}\ and\ \citenamefont
  {Krakauer}(2003)}]{zhang2003quantum}%
  \BibitemOpen
  \bibfield  {author} {\bibinfo {author} {\bibfnamefont {S.}~\bibnamefont
  {Zhang}}\ and\ \bibinfo {author} {\bibfnamefont {H.}~\bibnamefont
  {Krakauer}},\ }\href@noop {} {\bibfield  {journal} {\bibinfo  {journal}
  {Phys. Rev. Lett.}\ }\textbf {\bibinfo {volume} {90}},\ \bibinfo {pages}
  {136401} (\bibinfo {year} {2003})}\BibitemShut {NoStop}%
\bibitem [{\citenamefont {Carlson}\ \emph {et~al.}(1999)\citenamefont
  {Carlson}, \citenamefont {Gubernatis}, \citenamefont {Ortiz},\ and\
  \citenamefont {Zhang}}]{carlson1999issues}%
  \BibitemOpen
  \bibfield  {author} {\bibinfo {author} {\bibfnamefont {J.}~\bibnamefont
  {Carlson}}, \bibinfo {author} {\bibfnamefont {J.}~\bibnamefont {Gubernatis}},
  \bibinfo {author} {\bibfnamefont {G.}~\bibnamefont {Ortiz}}, \ and\ \bibinfo
  {author} {\bibfnamefont {S.}~\bibnamefont {Zhang}},\ }\href@noop {}
  {\bibfield  {journal} {\bibinfo  {journal} {Phys. Rev. B}\ }\textbf {\bibinfo
  {volume} {59}},\ \bibinfo {pages} {12788} (\bibinfo {year}
  {1999})}\BibitemShut {NoStop}%
\bibitem [{\citenamefont {LeBlanc}\ \emph {et~al.}(2015)\citenamefont
  {LeBlanc}, \citenamefont {Antipov}, \citenamefont {Becca}, \citenamefont
  {Bulik}, \citenamefont {Chan}, \citenamefont {Chung}, \citenamefont {Deng},
  \citenamefont {Ferrero}, \citenamefont {Henderson}, \citenamefont
  {Jim{\'e}nez-Hoyos} \emph {et~al.}}]{leblanc2015solutions}%
  \BibitemOpen
  \bibfield  {author} {\bibinfo {author} {\bibfnamefont {J.~P.}\ \bibnamefont
  {LeBlanc}}, \bibinfo {author} {\bibfnamefont {A.~E.}\ \bibnamefont
  {Antipov}}, \bibinfo {author} {\bibfnamefont {F.}~\bibnamefont {Becca}},
  \bibinfo {author} {\bibfnamefont {I.~W.}\ \bibnamefont {Bulik}}, \bibinfo
  {author} {\bibfnamefont {G.~K.-L.}\ \bibnamefont {Chan}}, \bibinfo {author}
  {\bibfnamefont {C.-M.}\ \bibnamefont {Chung}}, \bibinfo {author}
  {\bibfnamefont {Y.}~\bibnamefont {Deng}}, \bibinfo {author} {\bibfnamefont
  {M.}~\bibnamefont {Ferrero}}, \bibinfo {author} {\bibfnamefont {T.~M.}\
  \bibnamefont {Henderson}}, \bibinfo {author} {\bibfnamefont {C.~A.}\
  \bibnamefont {Jim{\'e}nez-Hoyos}},  \emph {et~al.},\ }\href@noop {}
  {\bibfield  {journal} {\bibinfo  {journal} {Phys. Rev. X}\ }\textbf {\bibinfo
  {volume} {5}},\ \bibinfo {pages} {041041} (\bibinfo {year}
  {2015})}\BibitemShut {NoStop}%
\bibitem [{\citenamefont {Zheng}\ \emph {et~al.}(2017)\citenamefont {Zheng},
  \citenamefont {Chung}, \citenamefont {Corboz}, \citenamefont {Ehlers},
  \citenamefont {Qin}, \citenamefont {Noack}, \citenamefont {Shi},
  \citenamefont {White}, \citenamefont {Zhang},\ and\ \citenamefont
  {Chan}}]{zheng2017stripe}%
  \BibitemOpen
  \bibfield  {author} {\bibinfo {author} {\bibfnamefont {B.-X.}\ \bibnamefont
  {Zheng}}, \bibinfo {author} {\bibfnamefont {C.-M.}\ \bibnamefont {Chung}},
  \bibinfo {author} {\bibfnamefont {P.}~\bibnamefont {Corboz}}, \bibinfo
  {author} {\bibfnamefont {G.}~\bibnamefont {Ehlers}}, \bibinfo {author}
  {\bibfnamefont {M.-P.}\ \bibnamefont {Qin}}, \bibinfo {author} {\bibfnamefont
  {R.~M.}\ \bibnamefont {Noack}}, \bibinfo {author} {\bibfnamefont
  {H.}~\bibnamefont {Shi}}, \bibinfo {author} {\bibfnamefont {S.~R.}\
  \bibnamefont {White}}, \bibinfo {author} {\bibfnamefont {S.}~\bibnamefont
  {Zhang}}, \ and\ \bibinfo {author} {\bibfnamefont {G.~K.-L.}\ \bibnamefont
  {Chan}},\ }\href@noop {} {\bibfield  {journal} {\bibinfo  {journal}
  {Science}\ }\textbf {\bibinfo {volume} {358}},\ \bibinfo {pages} {1155}
  (\bibinfo {year} {2017})}\BibitemShut {NoStop}%
\bibitem [{\citenamefont {Motta}\ \emph {et~al.}(2017)\citenamefont {Motta},
  \citenamefont {Ceperley}, \citenamefont {Chan}, \citenamefont {Gomez},
  \citenamefont {Gull}, \citenamefont {Guo}, \citenamefont {Jim{\'e}nez-Hoyos},
  \citenamefont {Lan}, \citenamefont {Li}, \citenamefont {Ma} \emph
  {et~al.}}]{motta2017towards}%
  \BibitemOpen
  \bibfield  {author} {\bibinfo {author} {\bibfnamefont {M.}~\bibnamefont
  {Motta}}, \bibinfo {author} {\bibfnamefont {D.~M.}\ \bibnamefont {Ceperley}},
  \bibinfo {author} {\bibfnamefont {G.~K.-L.}\ \bibnamefont {Chan}}, \bibinfo
  {author} {\bibfnamefont {J.~A.}\ \bibnamefont {Gomez}}, \bibinfo {author}
  {\bibfnamefont {E.}~\bibnamefont {Gull}}, \bibinfo {author} {\bibfnamefont
  {S.}~\bibnamefont {Guo}}, \bibinfo {author} {\bibfnamefont {C.~A.}\
  \bibnamefont {Jim{\'e}nez-Hoyos}}, \bibinfo {author} {\bibfnamefont {T.~N.}\
  \bibnamefont {Lan}}, \bibinfo {author} {\bibfnamefont {J.}~\bibnamefont
  {Li}}, \bibinfo {author} {\bibfnamefont {F.}~\bibnamefont {Ma}},  \emph
  {et~al.},\ }\href@noop {} {\bibfield  {journal} {\bibinfo  {journal} {Phys.
  Rev. X}\ }\textbf {\bibinfo {volume} {7}},\ \bibinfo {pages} {031059}
  (\bibinfo {year} {2017})}\BibitemShut {NoStop}%
\bibitem [{\citenamefont {Zhang}\ \emph {et~al.}(2018)\citenamefont {Zhang},
  \citenamefont {Malone},\ and\ \citenamefont {Morales}}]{zhang_nio}%
  \BibitemOpen
  \bibfield  {author} {\bibinfo {author} {\bibfnamefont {S.}~\bibnamefont
  {Zhang}}, \bibinfo {author} {\bibfnamefont {F.~D.}\ \bibnamefont {Malone}}, \
  and\ \bibinfo {author} {\bibfnamefont {M.~A.}\ \bibnamefont {Morales}},\
  }\href {\doibase 10.1063/1.5040900} {\bibfield  {journal} {\bibinfo
  {journal} {J. Chem. Phys.}\ }\textbf {\bibinfo {volume} {149}},\ \bibinfo
  {pages} {164102} (\bibinfo {year} {2018})}\BibitemShut {NoStop}%
\bibitem [{\citenamefont {Motta}\ \emph {et~al.}(2020)\citenamefont {Motta},
  \citenamefont {Genovese}, \citenamefont {Ma}, \citenamefont {Cui},
  \citenamefont {Sawaya}, \citenamefont {Chan}, \citenamefont {Chepiga},
  \citenamefont {Helms}, \citenamefont {Jim{\'e}nez-Hoyos}, \citenamefont
  {Millis} \emph {et~al.}}]{motta2019ground}%
  \BibitemOpen
  \bibfield  {author} {\bibinfo {author} {\bibfnamefont {M.}~\bibnamefont
  {Motta}}, \bibinfo {author} {\bibfnamefont {C.}~\bibnamefont {Genovese}},
  \bibinfo {author} {\bibfnamefont {F.}~\bibnamefont {Ma}}, \bibinfo {author}
  {\bibfnamefont {Z.-H.}\ \bibnamefont {Cui}}, \bibinfo {author} {\bibfnamefont
  {R.}~\bibnamefont {Sawaya}}, \bibinfo {author} {\bibfnamefont {G.~K.-L.}\
  \bibnamefont {Chan}}, \bibinfo {author} {\bibfnamefont {N.}~\bibnamefont
  {Chepiga}}, \bibinfo {author} {\bibfnamefont {P.}~\bibnamefont {Helms}},
  \bibinfo {author} {\bibfnamefont {C.}~\bibnamefont {Jim{\'e}nez-Hoyos}},
  \bibinfo {author} {\bibfnamefont {A.~J.}\ \bibnamefont {Millis}},  \emph
  {et~al.},\ }\href@noop {} {\bibfield  {journal} {\bibinfo  {journal} {Phys.
  Rev. X}\ }\textbf {\bibinfo {volume} {10}},\ \bibinfo {pages} {031058}
  (\bibinfo {year} {2020})}\BibitemShut {NoStop}%
\bibitem [{\citenamefont {Lee}\ \emph {et~al.}(2019{\natexlab{a}})\citenamefont
  {Lee}, \citenamefont {Malone},\ and\ \citenamefont {Morales}}]{lee_2019_UEG}%
  \BibitemOpen
  \bibfield  {author} {\bibinfo {author} {\bibfnamefont {J.}~\bibnamefont
  {Lee}}, \bibinfo {author} {\bibfnamefont {F.~D.}\ \bibnamefont {Malone}}, \
  and\ \bibinfo {author} {\bibfnamefont {M.~A.}\ \bibnamefont {Morales}},\
  }\href {https://doi.org/10.1063/1.5109572} {\bibfield  {journal} {\bibinfo
  {journal} {J. Chem. Phys.}\ }\textbf {\bibinfo {volume} {151}},\ \bibinfo
  {pages} {064122} (\bibinfo {year} {2019}{\natexlab{a}})}\BibitemShut
  {NoStop}%
\bibitem [{\citenamefont {Lee}\ \emph {et~al.}(2020{\natexlab{a}})\citenamefont
  {Lee}, \citenamefont {Malone},\ and\ \citenamefont
  {Reichman}}]{Lee2020benzene}%
  \BibitemOpen
  \bibfield  {author} {\bibinfo {author} {\bibfnamefont {J.}~\bibnamefont
  {Lee}}, \bibinfo {author} {\bibfnamefont {F.~D.}\ \bibnamefont {Malone}}, \
  and\ \bibinfo {author} {\bibfnamefont {D.~R.}\ \bibnamefont {Reichman}},\
  }\href {\doibase 10.1063/5.0024835} {\bibfield  {journal} {\bibinfo
  {journal} {J. Chem. Phys.}\ }\textbf {\bibinfo {volume} {153}},\ \bibinfo
  {pages} {126101} (\bibinfo {year} {2020}{\natexlab{a}})}\BibitemShut
  {NoStop}%
\bibitem [{\citenamefont {Williams}\ \emph {et~al.}(2020)\citenamefont
  {Williams}, \citenamefont {Yao}, \citenamefont {Li}, \citenamefont {Chen},
  \citenamefont {Shi}, \citenamefont {Motta}, \citenamefont {Niu},
  \citenamefont {Ray}, \citenamefont {Guo}, \citenamefont {Anderson} \emph
  {et~al.}}]{williams2020direct}%
  \BibitemOpen
  \bibfield  {author} {\bibinfo {author} {\bibfnamefont {K.~T.}\ \bibnamefont
  {Williams}}, \bibinfo {author} {\bibfnamefont {Y.}~\bibnamefont {Yao}},
  \bibinfo {author} {\bibfnamefont {J.}~\bibnamefont {Li}}, \bibinfo {author}
  {\bibfnamefont {L.}~\bibnamefont {Chen}}, \bibinfo {author} {\bibfnamefont
  {H.}~\bibnamefont {Shi}}, \bibinfo {author} {\bibfnamefont {M.}~\bibnamefont
  {Motta}}, \bibinfo {author} {\bibfnamefont {C.}~\bibnamefont {Niu}}, \bibinfo
  {author} {\bibfnamefont {U.}~\bibnamefont {Ray}}, \bibinfo {author}
  {\bibfnamefont {S.}~\bibnamefont {Guo}}, \bibinfo {author} {\bibfnamefont
  {R.~J.}\ \bibnamefont {Anderson}},  \emph {et~al.},\ }\href@noop {}
  {\bibfield  {journal} {\bibinfo  {journal} {Phys. Rev. X}\ }\textbf {\bibinfo
  {volume} {10}},\ \bibinfo {pages} {011041} (\bibinfo {year}
  {2020})}\BibitemShut {NoStop}%
\bibitem [{\citenamefont {Malone}\ \emph
  {et~al.}(2020{\natexlab{a}})\citenamefont {Malone}, \citenamefont {Zhang},\
  and\ \citenamefont {Morales}}]{malone2020gpu}%
  \BibitemOpen
  \bibfield  {author} {\bibinfo {author} {\bibfnamefont {F.~D.}\ \bibnamefont
  {Malone}}, \bibinfo {author} {\bibfnamefont {S.}~\bibnamefont {Zhang}}, \
  and\ \bibinfo {author} {\bibfnamefont {M.~A.}\ \bibnamefont {Morales}},\
  }\href {\doibase 10.1021/acs.jctc.0c00262} {\bibfield  {journal} {\bibinfo
  {journal} {J. Chem. Theory Comput.}\ }\textbf {\bibinfo {volume} {16}},\
  \bibinfo {pages} {4286} (\bibinfo {year} {2020}{\natexlab{a}})}\BibitemShut
  {NoStop}%
\bibitem [{\citenamefont {Lee}\ \emph {et~al.}(2020{\natexlab{b}})\citenamefont
  {Lee}, \citenamefont {Malone},\ and\ \citenamefont
  {Morales}}]{lee2020utilizing}%
  \BibitemOpen
  \bibfield  {author} {\bibinfo {author} {\bibfnamefont {J.}~\bibnamefont
  {Lee}}, \bibinfo {author} {\bibfnamefont {F.~D.}\ \bibnamefont {Malone}}, \
  and\ \bibinfo {author} {\bibfnamefont {M.~A.}\ \bibnamefont {Morales}},\
  }\href@noop {} {\bibfield  {journal} {\bibinfo  {journal} {J. Chem. Theory
  Comput.}\ }\textbf {\bibinfo {volume} {16}},\ \bibinfo {pages} {3019}
  (\bibinfo {year} {2020}{\natexlab{b}})}\BibitemShut {NoStop}%
\bibitem [{\citenamefont {Qin}\ \emph {et~al.}(2020)\citenamefont {Qin},
  \citenamefont {Chung}, \citenamefont {Shi}, \citenamefont {Vitali},
  \citenamefont {Hubig}, \citenamefont {Schollw{\"o}ck}, \citenamefont {White},
  \citenamefont {Zhang} \emph {et~al.}}]{qin2020absence}%
  \BibitemOpen
  \bibfield  {author} {\bibinfo {author} {\bibfnamefont {M.}~\bibnamefont
  {Qin}}, \bibinfo {author} {\bibfnamefont {C.-M.}\ \bibnamefont {Chung}},
  \bibinfo {author} {\bibfnamefont {H.}~\bibnamefont {Shi}}, \bibinfo {author}
  {\bibfnamefont {E.}~\bibnamefont {Vitali}}, \bibinfo {author} {\bibfnamefont
  {C.}~\bibnamefont {Hubig}}, \bibinfo {author} {\bibfnamefont
  {U.}~\bibnamefont {Schollw{\"o}ck}}, \bibinfo {author} {\bibfnamefont
  {S.~R.}\ \bibnamefont {White}}, \bibinfo {author} {\bibfnamefont
  {S.}~\bibnamefont {Zhang}},  \emph {et~al.},\ }\href@noop {} {\bibfield
  {journal} {\bibinfo  {journal} {Phys. Rev. X}\ }\textbf {\bibinfo {volume}
  {10}},\ \bibinfo {pages} {031016} (\bibinfo {year} {2020})}\BibitemShut
  {NoStop}%
\bibitem [{\citenamefont {Malone}\ \emph
  {et~al.}(2020{\natexlab{b}})\citenamefont {Malone}, \citenamefont {Benali},
  \citenamefont {Morales}, \citenamefont {Caffarel}, \citenamefont {Kent},\
  and\ \citenamefont {Shulenburger}}]{Malone2020Oct}%
  \BibitemOpen
  \bibfield  {author} {\bibinfo {author} {\bibfnamefont {F.~D.}\ \bibnamefont
  {Malone}}, \bibinfo {author} {\bibfnamefont {A.}~\bibnamefont {Benali}},
  \bibinfo {author} {\bibfnamefont {M.~A.}\ \bibnamefont {Morales}}, \bibinfo
  {author} {\bibfnamefont {M.}~\bibnamefont {Caffarel}}, \bibinfo {author}
  {\bibfnamefont {P.~R.~C.}\ \bibnamefont {Kent}}, \ and\ \bibinfo {author}
  {\bibfnamefont {L.}~\bibnamefont {Shulenburger}},\ }\href {\doibase
  10.1103/PhysRevB.102.161104} {\bibfield  {journal} {\bibinfo  {journal}
  {Phys. Rev. B}\ }\textbf {\bibinfo {volume} {102}},\ \bibinfo {pages}
  {161104} (\bibinfo {year} {2020}{\natexlab{b}})}\BibitemShut {NoStop}%
\bibitem [{\citenamefont {Malone}\ \emph {et~al.}(2019)\citenamefont {Malone},
  \citenamefont {Zhang},\ and\ \citenamefont {Morales}}]{malone_isdf}%
  \BibitemOpen
  \bibfield  {author} {\bibinfo {author} {\bibfnamefont {F.~D.}\ \bibnamefont
  {Malone}}, \bibinfo {author} {\bibfnamefont {S.}~\bibnamefont {Zhang}}, \
  and\ \bibinfo {author} {\bibfnamefont {M.~A.}\ \bibnamefont {Morales}},\
  }\href {\doibase 10.1021/acs.jctc.8b00944} {\bibfield  {journal} {\bibinfo
  {journal} {J. Chem. Theory. Comput.}\ }\textbf {\bibinfo {volume} {15}},\
  \bibinfo {pages} {256} (\bibinfo {year} {2019})}\BibitemShut {NoStop}%
\bibitem [{\citenamefont {Lee}\ and\ \citenamefont
  {Reichman}(2020)}]{lee2020stochastic}%
  \BibitemOpen
  \bibfield  {author} {\bibinfo {author} {\bibfnamefont {J.}~\bibnamefont
  {Lee}}\ and\ \bibinfo {author} {\bibfnamefont {D.~R.}\ \bibnamefont
  {Reichman}},\ }\href@noop {} {\bibfield  {journal} {\bibinfo  {journal} {J.
  Chem. Phys.}\ }\textbf {\bibinfo {volume} {153}},\ \bibinfo {pages} {044131}
  (\bibinfo {year} {2020})}\BibitemShut {NoStop}%
\bibitem [{\citenamefont {Purwanto}\ and\ \citenamefont
  {Zhang}(2004)}]{Purwanto2004Nov}%
  \BibitemOpen
  \bibfield  {author} {\bibinfo {author} {\bibfnamefont {W.}~\bibnamefont
  {Purwanto}}\ and\ \bibinfo {author} {\bibfnamefont {S.}~\bibnamefont
  {Zhang}},\ }\href {\doibase 10.1103/PhysRevE.70.056702} {\bibfield  {journal}
  {\bibinfo  {journal} {Phys. Rev. E}\ }\textbf {\bibinfo {volume} {70}},\
  \bibinfo {pages} {056702} (\bibinfo {year} {2004})}\BibitemShut {NoStop}%
\bibitem [{\citenamefont {Motta}\ and\ \citenamefont
  {Zhang}(2017)}]{motta_back_prop}%
  \BibitemOpen
  \bibfield  {author} {\bibinfo {author} {\bibfnamefont {M.}~\bibnamefont
  {Motta}}\ and\ \bibinfo {author} {\bibfnamefont {S.}~\bibnamefont {Zhang}},\
  }\href {\doibase 10.1021/acs.jctc.7b00730} {\bibfield  {journal} {\bibinfo
  {journal} {J. Chem. Theory Comput.}\ }\textbf {\bibinfo {volume} {13}},\
  \bibinfo {pages} {5367} (\bibinfo {year} {2017})}\BibitemShut {NoStop}%
\bibitem [{\citenamefont {Hedin}(1980)}]{hedin1980effects}%
  \BibitemOpen
  \bibfield  {author} {\bibinfo {author} {\bibfnamefont {L.}~\bibnamefont
  {Hedin}},\ }\href@noop {} {\bibfield  {journal} {\bibinfo  {journal} {Phys.
  Scr.}\ }\textbf {\bibinfo {volume} {21}},\ \bibinfo {pages} {477} (\bibinfo
  {year} {1980})}\BibitemShut {NoStop}%
\bibitem [{sqa()}]{sqa}%
  \BibitemOpen
  \href@noop {} {}\bibinfo {howpublished} {See
  \href{https://github.com/msaitow/SecondQuantizationAlgebra}{https://github.com/msaitow/SecondQuantizationAlgebra}
  for details on how to obtain the source code.}\BibitemShut {Stop}%
\bibitem [{\citenamefont {Neuscamman}\ \emph {et~al.}(2009)\citenamefont
  {Neuscamman}, \citenamefont {Yanai},\ and\ \citenamefont
  {Chan}}]{Neuscamman2009Mar}%
  \BibitemOpen
  \bibfield  {author} {\bibinfo {author} {\bibfnamefont {E.}~\bibnamefont
  {Neuscamman}}, \bibinfo {author} {\bibfnamefont {T.}~\bibnamefont {Yanai}}, \
  and\ \bibinfo {author} {\bibfnamefont {G.~K.-L.}\ \bibnamefont {Chan}},\
  }\href {\doibase 10.1063/1.3086932} {\bibfield  {journal} {\bibinfo
  {journal} {J. Chem. Phys.}\ }\textbf {\bibinfo {volume} {130}},\ \bibinfo
  {pages} {124102} (\bibinfo {year} {2009})}\BibitemShut {NoStop}%
\bibitem [{\citenamefont {Saitow}\ \emph {et~al.}(2013)\citenamefont {Saitow},
  \citenamefont {Kurashige},\ and\ \citenamefont {Yanai}}]{Saitow2013Jul}%
  \BibitemOpen
  \bibfield  {author} {\bibinfo {author} {\bibfnamefont {M.}~\bibnamefont
  {Saitow}}, \bibinfo {author} {\bibfnamefont {Y.}~\bibnamefont {Kurashige}}, \
  and\ \bibinfo {author} {\bibfnamefont {T.}~\bibnamefont {Yanai}},\ }\href
  {\doibase 10.1063/1.4816627} {\bibfield  {journal} {\bibinfo  {journal} {J.
  Chem. Phys.}\ }\textbf {\bibinfo {volume} {139}},\ \bibinfo {pages} {044118}
  (\bibinfo {year} {2013})}\BibitemShut {NoStop}%
\bibitem [{\citenamefont {Zgid}\ \emph {et~al.}(2009)\citenamefont {Zgid},
  \citenamefont {Ghosh}, \citenamefont {Neuscamman},\ and\ \citenamefont
  {Chan}}]{Zgid2009May}%
  \BibitemOpen
  \bibfield  {author} {\bibinfo {author} {\bibfnamefont {D.}~\bibnamefont
  {Zgid}}, \bibinfo {author} {\bibfnamefont {D.}~\bibnamefont {Ghosh}},
  \bibinfo {author} {\bibfnamefont {E.}~\bibnamefont {Neuscamman}}, \ and\
  \bibinfo {author} {\bibfnamefont {G.~K.-L.}\ \bibnamefont {Chan}},\ }\href
  {\doibase 10.1063/1.3132922} {\bibfield  {journal} {\bibinfo  {journal} {J.
  Chem. Phys.}\ }\textbf {\bibinfo {volume} {130}},\ \bibinfo {pages} {194107}
  (\bibinfo {year} {2009})}\BibitemShut {NoStop}%
\bibitem [{\citenamefont {Motta}\ and\ \citenamefont
  {Zhang}(2018)}]{Motta2019}%
  \BibitemOpen
  \bibfield  {author} {\bibinfo {author} {\bibfnamefont {M.}~\bibnamefont
  {Motta}}\ and\ \bibinfo {author} {\bibfnamefont {S.}~\bibnamefont {Zhang}},\
  }\href {\doibase 10.1002/wcms.1364} {\bibfield  {journal} {\bibinfo
  {journal} {WIREs Comput. Mol. Sci.}\ }\textbf {\bibinfo {volume} {8}},\
  \bibinfo {pages} {e1364} (\bibinfo {year} {2018})}\BibitemShut {NoStop}%
\bibitem [{\citenamefont {Hubbard}(1959)}]{Hubbard1959Jul}%
  \BibitemOpen
  \bibfield  {author} {\bibinfo {author} {\bibfnamefont {J.}~\bibnamefont
  {Hubbard}},\ }\href {\doibase 10.1103/PhysRevLett.3.77} {\bibfield  {journal}
  {\bibinfo  {journal} {Phys. Rev. Lett.}\ }\textbf {\bibinfo {volume} {3}},\
  \bibinfo {pages} {77} (\bibinfo {year} {1959})}\BibitemShut {NoStop}%
\bibitem [{\citenamefont {Hirsch}(1983)}]{Hirsch1983}%
  \BibitemOpen
  \bibfield  {author} {\bibinfo {author} {\bibfnamefont {J.~E.}\ \bibnamefont
  {Hirsch}},\ }\href@noop {} {\bibfield  {journal} {\bibinfo  {journal} {Phys.
  Rev. B}\ }\textbf {\bibinfo {volume} {28}},\ \bibinfo {pages} {4059}
  (\bibinfo {year} {1983})}\BibitemShut {NoStop}%
\bibitem [{Mot()}]{MottaPrivate}%
  \BibitemOpen
  \href@noop {} {}\bibinfo {howpublished} {Mario Motta, private
  communication.}\BibitemShut {Stop}%
\bibitem [{\citenamefont {Chen}\ \emph {et~al.}(2020)\citenamefont {Chen},
  \citenamefont {Motta}, \citenamefont {Ma},\ and\ \citenamefont
  {Zhang}}]{ChenSolidsBP2020}%
  \BibitemOpen
  \bibfield  {author} {\bibinfo {author} {\bibfnamefont {S.}~\bibnamefont
  {Chen}}, \bibinfo {author} {\bibfnamefont {M.}~\bibnamefont {Motta}},
  \bibinfo {author} {\bibfnamefont {F.}~\bibnamefont {Ma}}, \ and\ \bibinfo
  {author} {\bibfnamefont {S.}~\bibnamefont {Zhang}},\ }\href@noop {}
  {\bibfield  {journal} {\bibinfo  {journal} {arXiv preprint arXiv:2011.08335}\
  } (\bibinfo {year} {2020})}\BibitemShut {NoStop}%
\bibitem [{\citenamefont {Hohenstein}\ \emph
  {et~al.}(2012{\natexlab{a}})\citenamefont {Hohenstein}, \citenamefont
  {Parrish},\ and\ \citenamefont {Mart{\'\i}nez}}]{thc1}%
  \BibitemOpen
  \bibfield  {author} {\bibinfo {author} {\bibfnamefont {E.~G.}\ \bibnamefont
  {Hohenstein}}, \bibinfo {author} {\bibfnamefont {R.~M.}\ \bibnamefont
  {Parrish}}, \ and\ \bibinfo {author} {\bibfnamefont {T.~J.}\ \bibnamefont
  {Mart{\'\i}nez}},\ }\href {\doibase 10.1063/1.4732310} {\bibfield  {journal}
  {\bibinfo  {journal} {J. Chem. Phys.}\ }\textbf {\bibinfo {volume} {137}},\
  \bibinfo {pages} {044103} (\bibinfo {year} {2012}{\natexlab{a}})}\BibitemShut
  {NoStop}%
\bibitem [{\citenamefont {Parrish}\ \emph {et~al.}(2012)\citenamefont
  {Parrish}, \citenamefont {Hohenstein}, \citenamefont {Mart{\'\i}nez},\ and\
  \citenamefont {Sherrill}}]{thc2}%
  \BibitemOpen
  \bibfield  {author} {\bibinfo {author} {\bibfnamefont {R.~M.}\ \bibnamefont
  {Parrish}}, \bibinfo {author} {\bibfnamefont {E.~G.}\ \bibnamefont
  {Hohenstein}}, \bibinfo {author} {\bibfnamefont {T.~J.}\ \bibnamefont
  {Mart{\'\i}nez}}, \ and\ \bibinfo {author} {\bibfnamefont {C.~D.}\
  \bibnamefont {Sherrill}},\ }\href {\doibase 10.1063/1.4768233} {\bibfield
  {journal} {\bibinfo  {journal} {J. Chem. Phys.}\ }\textbf {\bibinfo {volume}
  {137}},\ \bibinfo {pages} {224106} (\bibinfo {year} {2012})}\BibitemShut
  {NoStop}%
\bibitem [{\citenamefont {Hohenstein}\ \emph
  {et~al.}(2012{\natexlab{b}})\citenamefont {Hohenstein}, \citenamefont
  {Parrish}, \citenamefont {Sherrill},\ and\ \citenamefont
  {Mart{\'\i}nez}}]{thc3}%
  \BibitemOpen
  \bibfield  {author} {\bibinfo {author} {\bibfnamefont {E.~G.}\ \bibnamefont
  {Hohenstein}}, \bibinfo {author} {\bibfnamefont {R.~M.}\ \bibnamefont
  {Parrish}}, \bibinfo {author} {\bibfnamefont {C.~D.}\ \bibnamefont
  {Sherrill}}, \ and\ \bibinfo {author} {\bibfnamefont {T.~J.}\ \bibnamefont
  {Mart{\'\i}nez}},\ }\href {\doibase 10.1063/1.4768241} {\bibfield  {journal}
  {\bibinfo  {journal} {J. Chem. Phys.}\ }\textbf {\bibinfo {volume} {137}},\
  \bibinfo {pages} {221101} (\bibinfo {year} {2012}{\natexlab{b}})}\BibitemShut
  {NoStop}%
\bibitem [{\citenamefont {Lee}\ \emph {et~al.}(2019{\natexlab{b}})\citenamefont
  {Lee}, \citenamefont {Lin},\ and\ \citenamefont
  {Head-Gordon}}]{lee2019systematically}%
  \BibitemOpen
  \bibfield  {author} {\bibinfo {author} {\bibfnamefont {J.}~\bibnamefont
  {Lee}}, \bibinfo {author} {\bibfnamefont {L.}~\bibnamefont {Lin}}, \ and\
  \bibinfo {author} {\bibfnamefont {M.}~\bibnamefont {Head-Gordon}},\ }\href
  {\doibase 10.1021/acs.jctc.9b00820} {\bibfield  {journal} {\bibinfo
  {journal} {J. Chem. Theory Comput.}\ }\textbf {\bibinfo {volume} {16}},\
  \bibinfo {pages} {243} (\bibinfo {year} {2019}{\natexlab{b}})}\BibitemShut
  {NoStop}%
\bibitem [{\citenamefont {Schoof}\ \emph {et~al.}(2015)\citenamefont {Schoof},
  \citenamefont {Groth}, \citenamefont {Vorberger},\ and\ \citenamefont
  {Bonitz}}]{schoof_prl}%
  \BibitemOpen
  \bibfield  {author} {\bibinfo {author} {\bibfnamefont {T.}~\bibnamefont
  {Schoof}}, \bibinfo {author} {\bibfnamefont {S.}~\bibnamefont {Groth}},
  \bibinfo {author} {\bibfnamefont {J.}~\bibnamefont {Vorberger}}, \ and\
  \bibinfo {author} {\bibfnamefont {M.}~\bibnamefont {Bonitz}},\ }\href
  {\doibase 10.1103/PhysRevLett.115.130402} {\bibfield  {journal} {\bibinfo
  {journal} {Phys. Rev. Lett.}\ }\textbf {\bibinfo {volume} {115}},\ \bibinfo
  {pages} {130402} (\bibinfo {year} {2015})}\BibitemShut {NoStop}%
\bibitem [{\citenamefont {Yang}\ \emph {et~al.}(2020)\citenamefont {Yang},
  \citenamefont {Gorelov}, \citenamefont {Pierleoni}, \citenamefont
  {Ceperley},\ and\ \citenamefont {Holzmann}}]{Yang2020Feb}%
  \BibitemOpen
  \bibfield  {author} {\bibinfo {author} {\bibfnamefont {Y.}~\bibnamefont
  {Yang}}, \bibinfo {author} {\bibfnamefont {V.}~\bibnamefont {Gorelov}},
  \bibinfo {author} {\bibfnamefont {C.}~\bibnamefont {Pierleoni}}, \bibinfo
  {author} {\bibfnamefont {D.~M.}\ \bibnamefont {Ceperley}}, \ and\ \bibinfo
  {author} {\bibfnamefont {M.}~\bibnamefont {Holzmann}},\ }\href {\doibase
  10.1103/PhysRevB.101.085115} {\bibfield  {journal} {\bibinfo  {journal}
  {Phys. Rev. B}\ }\textbf {\bibinfo {volume} {101}},\ \bibinfo {pages}
  {085115} (\bibinfo {year} {2020})}\BibitemShut {NoStop}%
\bibitem [{\citenamefont {Sun}\ \emph {et~al.}(2018)\citenamefont {Sun},
  \citenamefont {Berkelbach}, \citenamefont {Blunt}, \citenamefont {Booth},
  \citenamefont {Guo}, \citenamefont {Li}, \citenamefont {Liu}, \citenamefont
  {McClain}, \citenamefont {Sayfutyarova}, \citenamefont {Sharma},
  \citenamefont {Wouters},\ and\ \citenamefont {Chan}}]{Sun2018Jan}%
  \BibitemOpen
  \bibfield  {author} {\bibinfo {author} {\bibfnamefont {Q.}~\bibnamefont
  {Sun}}, \bibinfo {author} {\bibfnamefont {T.~C.}\ \bibnamefont {Berkelbach}},
  \bibinfo {author} {\bibfnamefont {N.~S.}\ \bibnamefont {Blunt}}, \bibinfo
  {author} {\bibfnamefont {G.~H.}\ \bibnamefont {Booth}}, \bibinfo {author}
  {\bibfnamefont {S.}~\bibnamefont {Guo}}, \bibinfo {author} {\bibfnamefont
  {Z.}~\bibnamefont {Li}}, \bibinfo {author} {\bibfnamefont {J.}~\bibnamefont
  {Liu}}, \bibinfo {author} {\bibfnamefont {J.~D.}\ \bibnamefont {McClain}},
  \bibinfo {author} {\bibfnamefont {E.~R.}\ \bibnamefont {Sayfutyarova}},
  \bibinfo {author} {\bibfnamefont {S.}~\bibnamefont {Sharma}}, \bibinfo
  {author} {\bibfnamefont {S.}~\bibnamefont {Wouters}}, \ and\ \bibinfo
  {author} {\bibfnamefont {G.~K.-L.}\ \bibnamefont {Chan}},\ }\href {\doibase
  10.1002/wcms.1340} {\bibfield  {journal} {\bibinfo  {journal} {WIREs Comput.
  Mol. Sci.}\ }\textbf {\bibinfo {volume} {8}},\ \bibinfo {pages} {e1340}
  (\bibinfo {year} {2018})}\BibitemShut {NoStop}%
\bibitem [{\citenamefont {Kent}\ \emph {et~al.}(2020)\citenamefont {Kent},
  \citenamefont {Annaberdiyev}, \citenamefont {Benali}, \citenamefont
  {Bennett}, \citenamefont {Landinez~Borda}, \citenamefont {Doak},
  \citenamefont {Hao}, \citenamefont {Jordan}, \citenamefont {Krogel},
  \citenamefont
  {Kyl{\ifmmode\ddot{a}\else\"{a}\fi}np{\ifmmode\ddot{a}\else\"{a}\fi}{\ifmmode\ddot{a}\else\"{a}\fi}},
  \citenamefont {Lee}, \citenamefont {Luo}, \citenamefont {Malone},
  \citenamefont {Melton}, \citenamefont {Mitas}, \citenamefont {Morales},
  \citenamefont {Neuscamman}, \citenamefont {Reboredo}, \citenamefont
  {Rubenstein}, \citenamefont {Saritas}, \citenamefont {Upadhyay},
  \citenamefont {Wang}, \citenamefont {Zhang},\ and\ \citenamefont
  {Zhao}}]{Kent2020May}%
  \BibitemOpen
  \bibfield  {author} {\bibinfo {author} {\bibfnamefont {P.~R.~C.}\
  \bibnamefont {Kent}}, \bibinfo {author} {\bibfnamefont {A.}~\bibnamefont
  {Annaberdiyev}}, \bibinfo {author} {\bibfnamefont {A.}~\bibnamefont
  {Benali}}, \bibinfo {author} {\bibfnamefont {M.~C.}\ \bibnamefont {Bennett}},
  \bibinfo {author} {\bibfnamefont {E.~J.}\ \bibnamefont {Landinez~Borda}},
  \bibinfo {author} {\bibfnamefont {P.}~\bibnamefont {Doak}}, \bibinfo {author}
  {\bibfnamefont {H.}~\bibnamefont {Hao}}, \bibinfo {author} {\bibfnamefont
  {K.~D.}\ \bibnamefont {Jordan}}, \bibinfo {author} {\bibfnamefont {J.~T.}\
  \bibnamefont {Krogel}}, \bibinfo {author} {\bibfnamefont {I.}~\bibnamefont
  {Kyl{\ifmmode\ddot{a}\else\"{a}\fi}np{\ifmmode\ddot{a}\else\"{a}\fi}{\ifmmode\ddot{a}\else\"{a}\fi}}},
  \bibinfo {author} {\bibfnamefont {J.}~\bibnamefont {Lee}}, \bibinfo {author}
  {\bibfnamefont {Y.}~\bibnamefont {Luo}}, \bibinfo {author} {\bibfnamefont
  {F.~D.}\ \bibnamefont {Malone}}, \bibinfo {author} {\bibfnamefont {C.~A.}\
  \bibnamefont {Melton}}, \bibinfo {author} {\bibfnamefont {L.}~\bibnamefont
  {Mitas}}, \bibinfo {author} {\bibfnamefont {M.~A.}\ \bibnamefont {Morales}},
  \bibinfo {author} {\bibfnamefont {E.}~\bibnamefont {Neuscamman}}, \bibinfo
  {author} {\bibfnamefont {F.~A.}\ \bibnamefont {Reboredo}}, \bibinfo {author}
  {\bibfnamefont {B.}~\bibnamefont {Rubenstein}}, \bibinfo {author}
  {\bibfnamefont {K.}~\bibnamefont {Saritas}}, \bibinfo {author} {\bibfnamefont
  {S.}~\bibnamefont {Upadhyay}}, \bibinfo {author} {\bibfnamefont
  {G.}~\bibnamefont {Wang}}, \bibinfo {author} {\bibfnamefont {S.}~\bibnamefont
  {Zhang}}, \ and\ \bibinfo {author} {\bibfnamefont {L.}~\bibnamefont {Zhao}},\
  }\href {\doibase 10.1063/5.0004860} {\bibfield  {journal} {\bibinfo
  {journal} {J. Chem. Phys.}\ }\textbf {\bibinfo {volume} {152}},\ \bibinfo
  {pages} {174105} (\bibinfo {year} {2020})}\BibitemShut {NoStop}%
\bibitem [{pau()}]{pauxy}%
  \BibitemOpen
  \href@noop {} {}\bibinfo {howpublished} {See
  \href{https://github.com/pauxy-qmc/pauxy}{https://github.com/pauxy-qmc/pauxy}
  for details on how to obtain the source code.}\BibitemShut {Stop}%
\bibitem [{\citenamefont {Holmes}\ \emph {et~al.}(2016)\citenamefont {Holmes},
  \citenamefont {Tubman},\ and\ \citenamefont {Umrigar}}]{Holmes2016Aug}%
  \BibitemOpen
  \bibfield  {author} {\bibinfo {author} {\bibfnamefont {A.~A.}\ \bibnamefont
  {Holmes}}, \bibinfo {author} {\bibfnamefont {N.~M.}\ \bibnamefont {Tubman}},
  \ and\ \bibinfo {author} {\bibfnamefont {C.~J.}\ \bibnamefont {Umrigar}},\
  }\href {\doibase 10.1021/acs.jctc.6b00407} {\bibfield  {journal} {\bibinfo
  {journal} {J. Chem. Theory Comput.}\ }\textbf {\bibinfo {volume} {12}},\
  \bibinfo {pages} {3674} (\bibinfo {year} {2016})}\BibitemShut {NoStop}%
\bibitem [{\citenamefont {Sharma}\ \emph {et~al.}(2017)\citenamefont {Sharma},
  \citenamefont {Holmes}, \citenamefont {Jeanmairet}, \citenamefont {Alavi},\
  and\ \citenamefont {Umrigar}}]{Sharma2017Apr}%
  \BibitemOpen
  \bibfield  {author} {\bibinfo {author} {\bibfnamefont {S.}~\bibnamefont
  {Sharma}}, \bibinfo {author} {\bibfnamefont {A.~A.}\ \bibnamefont {Holmes}},
  \bibinfo {author} {\bibfnamefont {G.}~\bibnamefont {Jeanmairet}}, \bibinfo
  {author} {\bibfnamefont {A.}~\bibnamefont {Alavi}}, \ and\ \bibinfo {author}
  {\bibfnamefont {C.~J.}\ \bibnamefont {Umrigar}},\ }\href {\doibase
  10.1021/acs.jctc.6b01028} {\bibfield  {journal} {\bibinfo  {journal} {J.
  Chem. Theory Comput.}\ }\textbf {\bibinfo {volume} {13}},\ \bibinfo {pages}
  {1595} (\bibinfo {year} {2017})}\BibitemShut {NoStop}%
\bibitem [{\citenamefont {Smith}\ \emph {et~al.}(2017)\citenamefont {Smith},
  \citenamefont {Mussard}, \citenamefont {Holmes},\ and\ \citenamefont
  {Sharma}}]{Smith2017Nov}%
  \BibitemOpen
  \bibfield  {author} {\bibinfo {author} {\bibfnamefont {J.~E.~T.}\
  \bibnamefont {Smith}}, \bibinfo {author} {\bibfnamefont {B.}~\bibnamefont
  {Mussard}}, \bibinfo {author} {\bibfnamefont {A.~A.}\ \bibnamefont {Holmes}},
  \ and\ \bibinfo {author} {\bibfnamefont {S.}~\bibnamefont {Sharma}},\ }\href
  {\doibase 10.1021/acs.jctc.7b00900} {\bibfield  {journal} {\bibinfo
  {journal} {J. Chem. Theory Comput.}\ }\textbf {\bibinfo {volume} {13}},\
  \bibinfo {pages} {5468} (\bibinfo {year} {2017})}\BibitemShut {NoStop}%
\bibitem [{\citenamefont {Wagner}\ \emph {et~al.}(2009)\citenamefont {Wagner},
  \citenamefont {Bajdich},\ and\ \citenamefont {Mitas}}]{wagner_qwalk}%
  \BibitemOpen
  \bibfield  {author} {\bibinfo {author} {\bibfnamefont {L.~K.}\ \bibnamefont
  {Wagner}}, \bibinfo {author} {\bibfnamefont {M.}~\bibnamefont {Bajdich}}, \
  and\ \bibinfo {author} {\bibfnamefont {L.}~\bibnamefont {Mitas}},\ }\href
  {\doibase https://doi.org/10.1016/j.jcp.2009.01.017} {\bibfield  {journal}
  {\bibinfo  {journal} {J. Comput. Phys.}\ }\textbf {\bibinfo {volume} {228}},\
  \bibinfo {pages} {3390} (\bibinfo {year} {2009})}\BibitemShut {NoStop}%
\bibitem [{\citenamefont {Haydock}\ \emph {et~al.}(1975)\citenamefont
  {Haydock}, \citenamefont {Heine},\ and\ \citenamefont
  {Kelly}}]{Haydock1975Aug}%
  \BibitemOpen
  \bibfield  {author} {\bibinfo {author} {\bibfnamefont {R.}~\bibnamefont
  {Haydock}}, \bibinfo {author} {\bibfnamefont {V.}~\bibnamefont {Heine}}, \
  and\ \bibinfo {author} {\bibfnamefont {M.~J.}\ \bibnamefont {Kelly}},\ }\href
  {\doibase 10.1088/0022-3719/8/16/011} {\bibfield  {journal} {\bibinfo
  {journal} {J. Phys. C: Solid State Phys.}\ }\textbf {\bibinfo {volume} {8}},\
  \bibinfo {pages} {2591} (\bibinfo {year} {1975})}\BibitemShut {NoStop}%
\bibitem [{\citenamefont {Dagotto}(1994)}]{DagottoLanczos1994}%
  \BibitemOpen
  \bibfield  {author} {\bibinfo {author} {\bibfnamefont {E.}~\bibnamefont
  {Dagotto}},\ }\href {\doibase 10.1103/RevModPhys.66.763} {\bibfield
  {journal} {\bibinfo  {journal} {Rev. Mod. Phys.}\ }\textbf {\bibinfo {volume}
  {66}},\ \bibinfo {pages} {763} (\bibinfo {year} {1994})}\BibitemShut
  {NoStop}%
\bibitem [{\citenamefont {van Setten}\ \emph {et~al.}(2015)\citenamefont {van
  Setten}, \citenamefont {Caruso}, \citenamefont {Sharifzadeh}, \citenamefont
  {Ren}, \citenamefont {Scheffler}, \citenamefont {Liu}, \citenamefont
  {Lischner}, \citenamefont {Lin}, \citenamefont {Deslippe}, \citenamefont
  {Louie}, \citenamefont {Yang}, \citenamefont {Weigend}, \citenamefont
  {Neaton}, \citenamefont {Evers},\ and\ \citenamefont
  {Rinke}}]{vanSetten2015Dec}%
  \BibitemOpen
  \bibfield  {author} {\bibinfo {author} {\bibfnamefont {M.~J.}\ \bibnamefont
  {van Setten}}, \bibinfo {author} {\bibfnamefont {F.}~\bibnamefont {Caruso}},
  \bibinfo {author} {\bibfnamefont {S.}~\bibnamefont {Sharifzadeh}}, \bibinfo
  {author} {\bibfnamefont {X.}~\bibnamefont {Ren}}, \bibinfo {author}
  {\bibfnamefont {M.}~\bibnamefont {Scheffler}}, \bibinfo {author}
  {\bibfnamefont {F.}~\bibnamefont {Liu}}, \bibinfo {author} {\bibfnamefont
  {J.}~\bibnamefont {Lischner}}, \bibinfo {author} {\bibfnamefont
  {L.}~\bibnamefont {Lin}}, \bibinfo {author} {\bibfnamefont {J.~R.}\
  \bibnamefont {Deslippe}}, \bibinfo {author} {\bibfnamefont {S.~G.}\
  \bibnamefont {Louie}}, \bibinfo {author} {\bibfnamefont {C.}~\bibnamefont
  {Yang}}, \bibinfo {author} {\bibfnamefont {F.}~\bibnamefont {Weigend}},
  \bibinfo {author} {\bibfnamefont {J.~B.}\ \bibnamefont {Neaton}}, \bibinfo
  {author} {\bibfnamefont {F.}~\bibnamefont {Evers}}, \ and\ \bibinfo {author}
  {\bibfnamefont {P.}~\bibnamefont {Rinke}},\ }\href {\doibase
  10.1021/acs.jctc.5b00453} {\bibfield  {journal} {\bibinfo  {journal} {J.
  Chem. Theory Comput.}\ }\textbf {\bibinfo {volume} {11}},\ \bibinfo {pages}
  {5665} (\bibinfo {year} {2015})}\BibitemShut {NoStop}%
\bibitem [{\citenamefont {Dunning}(1989)}]{Dunning1989}%
  \BibitemOpen
  \bibfield  {author} {\bibinfo {author} {\bibfnamefont {T.~H.}\ \bibnamefont
  {Dunning}},\ }\href {\doibase 10.1063/1.456153} {\bibfield  {journal}
  {\bibinfo  {journal} {J. Chem. Phys.}\ }\textbf {\bibinfo {volume} {90}},\
  \bibinfo {pages} {1007} (\bibinfo {year} {1989})}\BibitemShut {NoStop}%
\bibitem [{\citenamefont {Lee}\ \emph {et~al.}(2020{\natexlab{c}})\citenamefont
  {Lee}, \citenamefont {Morales},\ and\ \citenamefont {Malone}}]{Lee2020Dec}%
  \BibitemOpen
  \bibfield  {author} {\bibinfo {author} {\bibfnamefont {J.}~\bibnamefont
  {Lee}}, \bibinfo {author} {\bibfnamefont {M.~A.}\ \bibnamefont {Morales}}, \
  and\ \bibinfo {author} {\bibfnamefont {F.~D.}\ \bibnamefont {Malone}},\
  }\href {https://arxiv.org/abs/2012.12228v2} {\bibfield  {journal} {\bibinfo
  {journal} {arXiv}\ } (\bibinfo {year} {2020}{\natexlab{c}})},\ \Eprint
  {http://arxiv.org/abs/2012.12228} {2012.12228} \BibitemShut {NoStop}%
\bibitem [{\citenamefont {Motta}\ \emph {et~al.}(2019)\citenamefont {Motta},
  \citenamefont {Zhang},\ and\ \citenamefont {Chan}}]{Motta2019Jul}%
  \BibitemOpen
  \bibfield  {author} {\bibinfo {author} {\bibfnamefont {M.}~\bibnamefont
  {Motta}}, \bibinfo {author} {\bibfnamefont {S.}~\bibnamefont {Zhang}}, \ and\
  \bibinfo {author} {\bibfnamefont {G.~K.-L.}\ \bibnamefont {Chan}},\ }\href
  {\doibase 10.1103/PhysRevB.100.045127} {\bibfield  {journal} {\bibinfo
  {journal} {Phys. Rev. B}\ }\textbf {\bibinfo {volume} {100}},\ \bibinfo
  {pages} {045127} (\bibinfo {year} {2019})}\BibitemShut {NoStop}%
\bibitem [{\citenamefont {Goedecker}\ \emph {et~al.}(1996)\citenamefont
  {Goedecker}, \citenamefont {Teter},\ and\ \citenamefont
  {Hutter}}]{Goedecker1996Jul}%
  \BibitemOpen
  \bibfield  {author} {\bibinfo {author} {\bibfnamefont {S.}~\bibnamefont
  {Goedecker}}, \bibinfo {author} {\bibfnamefont {M.}~\bibnamefont {Teter}}, \
  and\ \bibinfo {author} {\bibfnamefont {J.}~\bibnamefont {Hutter}},\ }\href
  {\doibase 10.1103/PhysRevB.54.1703} {\bibfield  {journal} {\bibinfo
  {journal} {Phys. Rev. B}\ }\textbf {\bibinfo {volume} {54}},\ \bibinfo
  {pages} {1703} (\bibinfo {year} {1996})}\BibitemShut {NoStop}%
\bibitem [{\citenamefont {VandeVondele}\ and\ \citenamefont
  {Hutter}(2007)}]{VandeVondele2007Sep}%
  \BibitemOpen
  \bibfield  {author} {\bibinfo {author} {\bibfnamefont {J.}~\bibnamefont
  {VandeVondele}}\ and\ \bibinfo {author} {\bibfnamefont {J.}~\bibnamefont
  {Hutter}},\ }\href {\doibase 10.1063/1.2770708} {\bibfield  {journal}
  {\bibinfo  {journal} {J. Chem. Phys.}\ }\textbf {\bibinfo {volume} {127}},\
  \bibinfo {pages} {114105} (\bibinfo {year} {2007})}\BibitemShut {NoStop}%
\bibitem [{\citenamefont {Vitali}\ \emph {et~al.}(2016)\citenamefont {Vitali},
  \citenamefont {Shi}, \citenamefont {Qin},\ and\ \citenamefont
  {Zhang}}]{VitaliDynamical2016}%
  \BibitemOpen
  \bibfield  {author} {\bibinfo {author} {\bibfnamefont {E.}~\bibnamefont
  {Vitali}}, \bibinfo {author} {\bibfnamefont {H.}~\bibnamefont {Shi}},
  \bibinfo {author} {\bibfnamefont {M.}~\bibnamefont {Qin}}, \ and\ \bibinfo
  {author} {\bibfnamefont {S.}~\bibnamefont {Zhang}},\ }\href {\doibase
  10.1103/PhysRevB.94.085140} {\bibfield  {journal} {\bibinfo  {journal} {Phys.
  Rev. B}\ }\textbf {\bibinfo {volume} {94}},\ \bibinfo {pages} {085140}
  (\bibinfo {year} {2016})}\BibitemShut {NoStop}%
\bibitem [{\citenamefont {Lee}\ \emph {et~al.}(2020{\natexlab{d}})\citenamefont
  {Lee}, \citenamefont {Zhang},\ and\ \citenamefont {Reichman}}]{Lee2020Dec2}%
  \BibitemOpen
  \bibfield  {author} {\bibinfo {author} {\bibfnamefont {J.}~\bibnamefont
  {Lee}}, \bibinfo {author} {\bibfnamefont {S.}~\bibnamefont {Zhang}}, \ and\
  \bibinfo {author} {\bibfnamefont {D.~R.}\ \bibnamefont {Reichman}},\ }\href
  {https://arxiv.org/abs/2012.13473v1} {\bibfield  {journal} {\bibinfo
  {journal} {arXiv}\ } (\bibinfo {year} {2020}{\natexlab{d}})},\ \Eprint
  {http://arxiv.org/abs/2012.13473} {2012.13473} \BibitemShut {NoStop}%
\end{thebibliography}%

%\section{Implementation details}
%\begin{equation}
%\begin{split}
%(\mathbf{F}_{-})_{pq}^{\sigma\sigma} = &-\sum_{j}h_{qj}P_{pj}^{\sigma\sigma}\\ &-\sum_{ikl\tau}(ik|ql)\left[P_{pl}^{\sigma\sigma}P_{ik}^{\tau\tau}-P_{pk}^{\sigma\tau}P_{il}^{\tau\sigma}\right]
%\end{split}
%\end{equation}
%
%\begin{equation}
%\begin{split}
%(\mathbf{F}_{+})_{pq}^{\sigma\sigma} = &\sum_{i}h_{iq}\left(\delta_{pi}-P_{ip}^{\sigma\sigma}\right) \\  &+\sum_{jl\tau} (pq|jl) P_{jl}^{\tau\tau}-\sum_{jl}(jq|pl)P_{jl}^{\sigma\sigma}  \\
%&-\sum_{ijl\tau}(iq|jl)\left[P_{ip}^{\sigma\sigma}P_{jl}^{\tau\tau}-P_{jp}^{\tau\sigma}P_{il}^{\sigma\tau}\right]
%\end{split}
%\end{equation}
%
%In PySCF, the 2-PDM is stored as
%\begin{equation}
%\Gamma(p_\sigma,q_\sigma,r_{\sigma'},s_{\sigma'})
%=
%\langle a_{p_{\sigma}}^\dagger a_{r_{\sigma'}}^\dagger a_{s_{\sigma'}} a_{q_{\sigma}} \rangle
%\end{equation}
\end{document}